%% file: draft.tex
\documentclass[12pt,notitlepage]{article}

\usepackage[T1]{fontenc}
\usepackage{lmodern}
\usepackage[margin=1.05in]{geometry}
\usepackage{setspace}
\AtBeginDocument{\setlength{\abovedisplayskip}{5pt}\setlength{\belowdisplayskip}{5pt}}

\usepackage{mathtools,amssymb,amsthm,bm}

\newcommand{\adjustthreelimits}[9]{%
    \mathop{#1\vphantom{#4#7}}\limits#2{#3\vphantom{#6#9}}%
    \mathop{#4\vphantom{#1#7}}\limits#5{#6\vphantom{#3#9}}%
    \mathop{#7\vphantom{#1#4}}\limits#8{#9\vphantom{#3#6}}%
}

\usepackage{dsfont,bbm}
\usepackage{relsize}
\usepackage[scr]{rsfso}

\usepackage{xcolor}
\usepackage{graphicx}
\usepackage{subcaption}
\usepackage{tikz}
\usetikzlibrary{arrows.meta,positioning,fit,backgrounds,calc,decorations.pathreplacing}
\usepackage{pgfplots}
\usepgfplotslibrary{groupplots}
\pgfplotsset{compat=1.18}
\usepackage[shortlabels]{enumitem}
\newcommand{\figurepanelwidth}{0.49\linewidth}
\usepackage{titlesec}
\titleformat{\subsubsection}[block]
    {\normalfont\normalsize\itshape\filcenter}
    {\thesubsubsection}
    {1em}
    {}

\usepackage{abstract}

\usepackage{natbib}
\usepackage{xurl}

\definecolor{drkred}{rgb}{0.7,0.11,0.11}
\colorlet{drkblue}{blue!61.8!black}

\usepackage[
    hypertexnames=false,
    colorlinks=true,
    linkcolor=drkblue,
    urlcolor=drkblue,
    citecolor=drkblue,
    anchorcolor=drkblue
]{hyperref}

\usepackage{apptools}

\newtheorem{definition}{Definition}
\newtheorem{theorem}{Theorem}
\newtheorem{proposition}{Proposition}
\newtheorem{lemma}{Lemma}

\newtheorem{assumption}{Assumption}
\newtheorem{example}{Example}

\AtAppendix{\counterwithin{claim}{section}\counterwithin{proposition}{section}\counterwithin{lemma}{section}\counterwithin{corollary}{section}\counterwithin{definition}{section}}

\usepackage[nameinlink,capitalize,noabbrev]{cleveref}

\crefname{assumption}{Assumption}{Assumptions}
\crefname{claim}{Claim}{Claims}
\crefname{section}{Section}{Sections}
\crefname{subsection}{Section}{Section}
\crefname{equation}{Eqn.}{Eqns.}
\crefname{figure}{Figure}{Figures}
\crefname{appendix}{Appendix}{Appendices}
\Crefname{appendix}{Appendix}{Appendices}

\usepackage{bookmark}

\DeclarePairedDelimiter\autobracket{(}{)}
\newcommand{\br}[1]{\autobracket*{#1}}

\DeclarePairedDelimiter\autosbracket{[}{]}
\newcommand{\sbr}[1]{\autosbracket*{#1}}

\DeclarePairedDelimiter\autocbracket{\{}{\}}
\newcommand{\cbr}[1]{\autocbracket*{#1}}

\DeclarePairedDelimiterX\autolsbracket[1]{[}{)}{#1}
\newcommand{\lsbr}[1]{\autolsbracket*{#1}}

\DeclarePairedDelimiterX\autorsbracket[1]{(}{]}{#1}
\newcommand{\rsbr}[1]{\autorsbracket*{#1}}

\DeclareMathOperator*{\argmax}{argmax}
\DeclareMathOperator*{\argmin}{argmin}

\DeclareMathOperator*{\esssup}{ess\,sup}

\DeclareMathOperator{\unif}{Uniform}
\DeclareMathOperator{\Binomial}{Binomial}
\DeclareMathOperator{\Bernoulli}{Bernoulli}
\DeclareMathOperator{\Var}{Var}
\DeclareMathOperator{\bgamma}{\boldsymbol{\gamma}}

\newcommand{\MR}{\mathrm{MR}}
\newcommand{\MC}{\mathrm{MC}}

\newcommand{\rbb}{\mathbb{R}}

\newcommand{\p}{\mathbb{P}}
\newcommand{\bP}{\mathbf{P}}

\newcommand{\qcal}{\mathcal Q}

\newcommand{\mcal}{\mathcal{M}}
\newcommand{\ccal}{\mathcal{C}}

\newcommand{\lcal}{\mathcal L}

\newcommand{\pcal}{\mathcal P}

\newcommand{\scal}{\mathcal{S}}

\newcommand{\xcal}{\mathcal X}

\newcommand{\xscr}{\mathscr X}

\newcommand{\E}{\mathds E}

\newcommand{\ltheta}{\underline{\theta}}
\newcommand{\btheta}{\overline{\theta}}

\newcommand{\boldtheta}{\boldsymbol{\theta}}
\newcommand{\crt}{\mathrm{crt}}

\newcommand{\htheta}{\hat{\theta}}
\newcommand{\hatTheta}{\widehat\Theta}

\title{ \vspace{-1em}
    Targeted Advertising Platforms:\\
    Pooling Audiences Despite Poaching Risk\thanks{
        I am grateful to Michael Ostrovsky, Andrzej Skrzypacz and Weijie Zhong for their guidance and support, and Frank Yang for instrumental feedback. I am thankful to Martino Banchio, Heski Bar-Isaac, Alessandro Bonatti, Steven Callander, Roberto Corrao, Ben Davies, Kwabena Donkor, Ellen Muir, Spencer Pantoja, Bing Liu, Ilya Segal, Takuo Sugaya, Eric Tang, and participants at conferences and seminars for helpful comments and suggestions. I owe special thanks to Varanya Chaubey for exceptional editing advice.\\
        \indent The author declares that he has no relevant or material financial interests that relate to the research described in this paper.
    }
}

\author{Klajdi Hoxha
    \thanks{Graduate School of Business, Stanford University. Email: hklajdi@stanford.edu.
    }
}

\date{\today}

\begin{document}

\maketitle

\begin{abstract}
    E-commerce platforms are inviting merchants to contribute their own audiences to a common pool that the platform uses to identify and target customers most likely to click on merchants' ads. While pooling audiences allows a merchant to reach new customers, it also creates a poaching risk. We develop a model of targeted advertising, in which each customer has merchant-specific click-through rates, and study how a platform can incentivize merchants to pool audiences despite poaching risk. We characterize the mechanism that maximizes a weighted sum of platform revenue, merchants' surplus, and customer engagement, measured by expected clicks. On sufficiently large platforms, with merchants grouped by observable tags $k \in K$, the optimal mechanism offers merchants with each tag $k$ three options: (i) receive, at no charge, a campaign built on the pooled audience that delivers their baseline expected clicks, (ii) expand it by paying the posted buying price $p_k^B$ per additional click, or (iii) reduce it and receive the posted selling price $p_k^S \leq p_k^B$ per click forgone. By pooling audiences, the platform can deliver baseline clicks with potentially fewer ads by targeting the customers most likely to click, and sell any additional clicks to other merchants. We show that endogenous, tag-specific virtual prices $\br{\lambda_k: k \in K}$ pin down the posted prices and the supply of clicks across tags. We then characterize when the efficient platform design requires no outside subsidy and extend our analysis to settings in which customers who do not click may be targeted by another merchant.
\end{abstract}

\section{Introduction}
\label{sec:introduction}

E-commerce platforms are rolling out ambitious initiatives to improve ad targeting by pooling audiences of the merchants they host. These initiatives respond to recent privacy regulations and policies\textemdash such as the California Consumer Privacy Act (CCPA), the European Union's General Data Protection Regulation (GDPR), and Apple's App Tracking Transparency (ATT)\textemdash which have limited third-party tracking, the backbone of targeted online advertising.\footnote{Early estimates following the introduction of ATT in iOS 14.5 suggested that about 96\% of U.S. users had not opted into app tracking: \url{https://www.imore.com/96-iphone-users-have-opted-out-app-tracking-ios-145-launched} (accessed October 10, 2026).} Safari and Firefox restrict third-party tracking by default, leaving Chrome and Edge as the only major browsers that still support third-party cookies as they seek alternative solutions.\footnote{In October 2025, Google announced that it would retire several Privacy Sandbox technologies following low adoption; see \url{https://privacysandbox.google.com/blog/update-on-plans-for-privacy-sandbox-technologies} (accessed October 10, 2026).} As third-party tracking declines, pooling audiences within the platform offers a promising alternative. Doing so gives the platform access to a larger audience, allowing it to identify and target customers with ads they are more likely to click than the ads of the merchants who originally reached them.

Yet pooling audiences exposes merchants to customer poaching. Once audiences are pooled, any participating merchant can compete for any customer in the pool. Ads are typically allocated through auctions based on bids and measures of ad quality, such as \textit{click-through rates} (CTRs), the number of clicks an ad receives divided by its number of impressions. Merchants who contribute their audiences may therefore lose some of their existing customers to competitors who bid more aggressively or have higher CTRs. Unless the platform protects merchants from poaching or compensates them for the value of the clicks they forgo, merchants may prefer not to participate and instead advertise only to their own audiences.

Major platforms such as Amazon, Google, and Shopify differ in how much control merchants have over the audiences they build on them.\footnote{Outside e-commerce platforms, data cooperatives such as Epsilon's Abacus and Wiland let members target new audiences using pooled data in exchange for contributing their own customer records. See \url{https://legal.epsilon.com/abacus/services-privacy-policy} and \url{https://wiland.com/blog/data-collaboration-explained/} (accessed October 19, 2026).} At one end, Amazon uses the shopping data it collects to design and run targeted ads for any merchant advertising on its platform.\footnote{See \url{https://advertising.amazon.com/library/guides/display-ads-beyond-amazon} (accessed October 10, 2026).} At the other end, Google's Customer Match lets each advertiser upload his own customer lists to build campaigns, without pooling them across advertisers.\footnote{See \url{https://support.google.com/google-ads/answer/6334160?hl=en-GB} (accessed October 10, 2026).} Between these extremes, Shopify, our running example, allows merchants to participate voluntarily in pooling audiences. In 2022, it launched Shopify Audiences, which uses pooled customer data to build audiences of potential customers for participating merchants.\footnote{See \url{https://www.shopify.com/enterprise/blog/shopify-audiences} (accessed October 10, 2026).}

Shopify's own example makes the poaching risk concrete: a new coffee brand can reach customers ``who have purchased similar specialty coffee products from other Shopify brands.''\footnote{See \url{https://www.shopify.com/enterprise/blog/second-party-vs-third-party-data} (accessed October 19, 2026).} Merchants may therefore prefer to stay out if participation exposes their customers to competitors. This concern is reflected in Shopify's FAQ, which asks whether merchants can exclude competitors from using their data. Shopify confirms that ``the data that you contribute will be used to benefit other participating merchants,''\footnote{See \url{https://help.shopify.com/en/manual/promoting-marketing/shopify-audiences/faq} (accessed October 10, 2026).} which may leave merchants reluctant to participate.

Motivated by this trade-off, this paper develops a model of a targeted advertising platform and characterizes its optimal design. Any merchant $i \in N$ who voluntarily joins the platform contributes his audience and then competes with other participants for the customers in the resulting pool. The model has two sources of heterogeneity. Each merchant is \textit{privately} informed about his value per click (type), $\theta_i$, and his \textit{baseline} expected clicks are those obtained by advertising to his initial audience. Each customer is described by an \textit{observable} vector of CTRs, $\omega = \br{\omega_i: i \in N}$, where $\omega_i$ is the probability that she clicks merchant $i$'s ad.

The platform's problem is to design targeted ads and payments that maximize a weighted sum of platform revenue, merchants' surplus, and customer engagement, with respective normalized weights $\eta_r$, $\eta_v$, and $\eta_w$ summing to $1$, subject to incentive and participation constraints.\footnote{Because the model abstracts from product prices, we use customer engagement, measured by expected clicks, as a proxy for customer welfare. \Cref{subsec:discussion-assumptions} provides a microfoundation for CTRs in which product prices are fixed and publicly observed during the advertising period, under which customer surplus is proportional to expected clicks.} We first study the \textit{finite market}, in which the set of merchants $N$ is finite, and then the \textit{limit market}, in which a continuum of merchants are grouped by finitely many observable tags $\br{k \in K}$ and are subject to a capacity on expected clicks. Tags may describe observable characteristics such as product categories, tiers, or prices, as well as differences in audience contributions.

In the limit market, the first part of our main result shows that unique posted per-click prices $p_k^S \leq p_k^B$ in each tag implement the optimal mechanism. Each merchant with tag $k$ is offered three options: (i) receive, at no charge, a campaign built on the pooled audience that delivers his baseline expected clicks; (ii) expand his campaign by paying $p_k^B$ per additional click; or (iii) monetize his audience by selling access to its customers at $p_k^S$ per baseline click forgone. Because payoffs are linear in expected clicks, merchants with values per click below $p_k^S$ or above $p_k^B$ choose an extreme: those below $p_k^S$ monetize their entire audiences, while those above $p_k^B$ expand their campaigns to capacity. Merchants between the two prices \textit{exchange} access to their audiences for targeted ads that deliver their baseline clicks, neither paying nor receiving a transfer. To the best of our knowledge, this exchange market is a novel screening force.

The optimal mechanism realizes two types of potential gains from pooling: it reallocates clicks toward merchants who value them more, and it generates more clicks from the same customers. The first gain comes from heterogeneous values per click: merchants who value clicks more expand their campaigns while those who value them less forgo clicks, raising the total value of clicks even if their number stays fixed. The second gain comes from heterogeneous CTRs: by targeting the customers most likely to click, the platform can potentially deliver the baseline clicks using fewer customers and monetize the rest through campaign expansions.

The second part shows that unique tag-specific \textit{virtual prices} $\br{\lambda_k:k \in K}$ pin down both the supply and demand of clicks in each tag, where $\lambda_k$ is the designer's marginal value of an additional click supplied to tag $k$. On the supply side, each customer is described by tag-specific CTRs $\xi = \br{\xi_k:k \in K}$, where $\xi_k$ is the best CTR attainable within tag $k$. She is assigned to the tag that maximizes the \textit{customer-specific score} $\lambda_k\xi_k$. Holding other virtual prices fixed, a higher $\lambda_k$ weakly increases the clicks supplied to tag $k$. The rule compares \textit{relative} rather than \textit{absolute} CTRs: tag $k$ wins her over tag $j$ whenever $\xi_k/\xi_j \geq \lambda_j/\lambda_k$, so she need not receive the ad she is most likely to click.\footnote{This is Ricardo's principle of comparative advantage applied to customer attention: as in the allocation of heterogeneous inputs across sectors (e.g., \citealp{roy1951some,costinot2009elementary}), customers are assigned based on relative rather than absolute CTRs, with CTRs in the role of productivities and virtual prices in the role of sector prices.}

On the demand side, virtual prices pin down posted prices in each tag. When both posted prices are interior, each $\lambda_k$ equals the weighted marginal revenue $\MR_{\eta,k}\br{p_k^B}$ from selling an additional click to marginal buyer $\theta = p_k^B$, and the weighted marginal cost $\MC_{\eta,k}\br{p_k^S}$ of compensating the marginal seller $\theta = p_k^S$ for an additional click forgone. A higher virtual price weakly raises both posted prices and weakly reduces the tag's demand for clicks. Virtual prices are set so that supply equals demand in every tag.

Under revenue maximization, the marginal revenue and marginal cost curves are the usual virtual value and virtual cost functions in mechanism design \citep{myerson1981optimal,bulow1989simple}, which transform merchants' values per click to account for the information rents they earn when buying or selling clicks.\footnote{\citet{bulow1989simple} show that, with virtual values in the role of marginal revenue, the optimal auction problem is equivalent to a monopolist's third-degree price discrimination problem. The study of third-degree price discrimination dates back to the classic work of \citet{pigou1920economics} and \citet{robinson1933economics}.} When the platform maximizes revenue $\br{\eta_r = 1}$, it acts as a multiproduct monopolist engaging in third-degree price discrimination across tags. In the classic setting, the monopolist sells additional units in the market with the highest positive marginal revenue until marginal revenue falls to zero or capacity is exhausted (\cref{fig:price-discrimination-comparison}, panel~(a)). In our setting, the clicks the platform sells must first be procured from the merchants themselves, which gives each tag a weighted marginal cost curve alongside its weighted marginal revenue curve (\cref{fig:price-discrimination-comparison}, panel~(b)).

\begin{figure}[t!]
    \centering
    \begin{subfigure}[t]{0.48\textwidth}
        \centering
        \resizebox{\linewidth}{!}{\input{figures/price-discrimination-classic}}
        \caption{Classic third-degree price discrimination with fixed supply}
        \label{fig:price-discrimination-canonical}
    \end{subfigure}
    \hfill
    \begin{subfigure}[t]{0.48\textwidth}
        \centering
        \resizebox{\linewidth}{!}{\input{figures/price-discrimination-platform}}
        \caption{Third-degree price discrimination with endogenous supply}
        \label{fig:price-discrimination-platform}
    \end{subfigure}
    \caption{Third-Degree Price Discrimination.}
    \label{fig:price-discrimination-comparison}
\end{figure}

\paragraph{Finite Market Results.} To build intuition for the limit-market results, we first characterize the optimal design in the finite market and show how the model connects to and generalizes classic settings of bilateral trade and partnership dissolution.

A scoring mechanism that \textit{jointly} determines targeted ads for each customer is optimal: merchant $i$'s report generates a single score $h_i\br{\theta_i}$, and each customer $\omega$ is assigned to a merchant with the highest customer-specific score $\omega_i h_i\br{\theta_i}$, with customer-specific tie-breaking when needed. The scoring function $h_i$ coincides with the weighted virtual cost function for low types and the weighted virtual value function for high types and is flat in between (see \cref{fig:scoring-rule} for an illustration). Because a single score $h_i\br{\theta_i}$ governs merchant $i$'s entire campaign, his countervailing incentives to overstate his type when selling customers and to understate it when buying ads are ``pooled'' into one report. Merchants therefore earn weakly lower information rents than if the mechanism allowed different reports across customers.

The finite market delivers a \textit{direct} scoring implementation, while the limit market offers an \textit{indirect} implementation through posted prices and virtual prices. In the finite market, a merchant's campaign depends on other merchants' realized reports, so a merchant of a given type may forgo baseline clicks in some realizations and expand his campaign in others. To ensure incentive compatibility as merchants switch from net sellers to net buyers, the scoring function is ``ironed'' over an interval between the virtual cost and virtual value regions. Except for boundary cases, the ironing parameter and tie-breaking rule are chosen such that types in the ironed interval receive, \textit{on average}, their baseline expected clicks.

\paragraph{Technical Contributions.} We derive the limit market from large finite markets. As the number of merchants grows, aggregate uncertainty about their types disappears, and merchant-specific CTRs $\omega_{i,k}$, for merchant $i$ with tag $k$, can be summarized by tag-specific CTRs $\xi_k$. The reduction in CTRs follows from a \textit{partial exchangeability} assumption on the countably infinite sequence $\omega = \br{\omega_{i,k}: i \in \mathbb N,\,k \in K}$, of which each finite market observes a finite restriction: relabeling merchants within a tag leaves the joint distribution of the sequence unchanged, while allowing dependence across merchants and tags. Under this assumption, de Finetti's theorem gives a representation in which a customer's CTRs are i.i.d.\ within each tag, conditional on latent CTR distributions. As the number of merchants grows, the highest CTR among merchants with tag $k$ approaches the best CTR attainable within the tag, which becomes the customer's tag-specific CTR $\xi_k$.

We show that every sequence of symmetric optimal finite-market mechanisms converges to the optimal limit-market mechanism in normalized interim expected clicks and transfers. The main step is to show that every sequence of symmetric, feasible, nondecreasing normalized expected clicks in finite markets has a subsequence that converges to a feasible expected-clicks allocation in the limit market, and conversely, that every feasible, nondecreasing expected-clicks allocation in the limit market can be approximated by such finite-market allocations. These two directions imply convergence of optimal values, so every limit point of optimal finite-market expected clicks attains the limit-market optimum and, by (essential) uniqueness, coincides with the optimal limit-market expected clicks. We also construct incentive-compatible and individually rational finite-market mechanisms that approximate the optimal limit-market mechanism and are, by convergence of optimal values, asymptotically optimal.

\paragraph{Related Literature.} Our framework introduces merchant-customer specific CTRs, the defining feature of targeted advertising that the classic position-auction models abstract from. In the latter, CTRs are multiplicatively separable into merchant and position effects: the CTR of merchant $i$ in position $s$ is $e_i x_s$, where $e_i$ is the ad's quality and $x_s$ is the position effect \citep{varian2007position}, or depends on position alone \citep{edelman2007internet}.\footnote{Cascade models relax this separability by allowing an ad's CTR to depend on the ads shown before it \citep{kempe2008cascade,aggarwal2008sponsored}; for more details, see \cref{subsec:sequential-targeting}.} As a result, the relative ranking of merchants does not vary across positions. Merchants are therefore ranked by their quality-weighted bids, with higher-ranked merchants receiving better ad positions. In our model, by contrast, each customer represents a distinct advertising opportunity and induces a potentially different ranking of merchants' ads. Ads are instead allocated customer by customer: each customer goes to the merchant with the highest customer-specific score $\omega_ih_i\br{\theta_i}$, and the same merchant may rank first for one customer and last for another.

Our finite-market model generalizes classic settings of bilateral trade \citep{myerson1983efficient} and partnership dissolution \citep{cgk,loertscher2019optimal} to trade over multiple heterogeneous items. More generally, it fits within a class of combinatorial exchange markets in which agents have privately known heterogeneous values for a bundle of items.\footnote{Following \citet{babaioff2008approximability}, a combinatorial exchange generalizes combinatorial auctions by having agents act as both buyers and sellers: an agent can offer a bundle of items, provided he receives a different bundle and/or an appropriate side payment in return.}\ Finding optimal allocations in general combinatorial exchanges can be computationally intractable.\footnote{E.g., \citet{babaioff2008approximability} study exchanges in which buyers and sellers trade bundles of indivisible products and show that maximizing surplus is generally NP-hard. Tractable settings typically assume complete information and rule out side payments, as in the housing market of \citet{shapley1974cores}.}\ We gain considerable tractability from one-dimensional private information (values per click) and additive preferences over multiple heterogeneous and divisible items (customers) in a quasi-linear environment with side payments.

Our optimal finite-market mechanism determines each merchant's targeted ads jointly through a single score, which can be viewed as \textit{bundling} customer assignments into a campaign. This connects our analysis to the literature on optimal bundling, and in particular to \citet{yang2025nested}, who studies nested bundling in a setting with one-dimensional heterogeneity and non-additive preferences. In our setting, preferences are additive across customers, and with one-dimensional heterogeneity alone, bundling would not benefit the designer. Initial audiences, however, create countervailing incentives across customers, making joint campaign design beneficial.

Our setting can be viewed as a matching problem in which a mediator proposes an alternative to the initial matching between merchants and customers that maximizes her objective, subject to merchants' incentive and participation constraints. This connects our analysis to the literature on platform-mediated matching. \citet{gomes2016many, gomes2024price} study optimal many-to-many matching between a continuum of privately informed agents on two sides of a market, and ask how cross-subsidization and price discrimination shape matching patterns, targeting, and welfare. The main difference is that, in our setting, merchants start from an initial matching, which gives them type-dependent outside options and requires the mediator to compensate those whose matches she takes away. In addition, the matching is many-to-one between customers and merchants, and only merchants are privately informed.

Our convergence result relates to the literature that derives continuum models as limits of large finite markets. In matching models, \citet{azevedo2016supply} and \citet{greinecker2021pairwise} provide conditions under which continuum stable matchings are limits of stable matchings in approximating finite economies. In assignment problems, \citet{che2010asymptotic} show that random priority and probabilistic serial converge to the same continuum assignment, and \citet{hashimoto2018generalized} constructs finite-market mechanisms that approximate continuum mechanisms while preserving feasibility, ex post individual rationality, and ex post incentive compatibility. Our approximation result is closest to the latter, although we impose incentive and participation constraints at the interim stage. We derive the limit market's pooled audience distribution and feasible allocation set from finite-market primitives under partial exchangeability of merchant-specific CTRs. We then establish convergence of normalized interim expected clicks and transfers for every sequence of symmetric optimal finite-market mechanisms.

On the practical side, the paper studies a platform that must procure audiences from merchants before it can sell them targeted ads, bringing together the supply of customer data and the design of targeted ads. The literature on targeted advertising studies how platforms use customer data to allocate and price ads \citep{bergemann2011targeting,eliaz2020incentive,bergemann2024data,bergemann2024digital}, while the literature on data intermediation studies how intermediaries acquire and monetize personal data supplied by consumers \citep{bergemann2022economics,galperti2022competitive,galperti2024value}.\footnote{Related work studies the pricing of customer data \citep{bergemann2015selling} and how information disclosure shapes competition \citep{elliott2026market}. See \citet{farboodi2023data} for a review on data and markets.} In our setting, the suppliers of customer data and the buyers of ads are the same merchants, who contribute their audiences to the platform and purchase targeted ads designed from the pooled audience.

\paragraph{Roadmap.} \Cref{sec:finite-market} introduces the finite market and characterizes its optimal scoring mechanism. \Cref{sec:limit-market} presents the limit market and characterizes its optimal mechanism through posted prices and virtual prices. \Cref{sec:limit-market-convergence} derives the limit market from large finite markets. \cref{sec:efficiency} studies efficiency and sequential targeting in the limit market. \Cref{sec:conclusion} concludes.

\section{Finite Market}
\label{sec:finite-market}
We formally introduce the model with finitely many merchants (henceforth, the \textit{finite market}) in \cref{subsec:model-finite} and characterize the optimal \textit{scoring mechanism} in \cref{subsec:optimal-mechanism-finite}. \Cref{subsec:proof-overview-finite} presents a proof overview, and the complete proofs are in \cref{app:finite}.

\subsection{Model}
\label{subsec:model-finite}
A designer operates an advertising platform serving a finite set $N$ of merchants, each of whom has exclusive targeting access to an audience. These audiences are pairwise disjoint and collectively constitute a unit mass of customers.

Each merchant $i \in N$ has a private value per click (henceforth, \textit{type}) $\theta_i \in \Theta_i = \sbr{\ltheta_i,\btheta_i} \subset \rbb_+$. Types are independently drawn from distributions $F_i$, each admitting a strictly positive, continuous density $f_i$. We assume that each $F_i$ is \textit{regular}, meaning that its virtual value and cost functions (also referred to as \textit{marginal revenue} and \textit{cost} curves)
\[
    \phi_i^B\br{\theta_i}
    :=
    \theta_i - \frac{1 - F_i\br{\theta_i}}{f_i\br{\theta_i}}
    \quad \text{and} \quad
    \phi_i^S\br{\theta_i}
    :=
    \theta_i + \frac{F_i\br{\theta_i}}{f_i\br{\theta_i}},
\]
respectively, are strictly increasing. Let $\Theta := \prod_{i \in N}\Theta_i$ denote the product type space, and let $F := \bigotimes_{i \in N}F_i$ denote the joint type distribution.

Each customer is characterized by a vector of \textit{click-through rates} (CTRs)
\[
    \omega = \br{\omega_i}_{i \in N} \in \sbr{0,1}^{N},
\]
where the CTR $\omega_i$ is the probability that the customer clicks on merchant $i$'s ad.\footnote{Depending on context, $N$ denotes the set of merchants or its size.} CTRs are \textit{observable}: the realization $\omega_i$ is commonly observed by the designer and merchant $i$. Let $\Omega \subseteq \sbr{0,1}^{N}$ be the (Borel) set of possible CTR realizations.

Each merchant $i$ initially has exclusive access to an audience of mass $\alpha_i > 0$, where $\sum_{i \in N}\alpha_i = 1$. Conditional on belonging to merchant $i$'s audience, customers are distributed according to Borel measures $\gamma_i \in \Delta\br{\Omega}$. Let $\alpha = \br{\alpha_i}_{i \in N}$ and $\bgamma = \br{\gamma_i}_{i \in N}$ denote the audience shares and conditional CTR distributions, respectively. Finally, let $\gamma_\alpha \in \Delta\br{\Omega}$ denote the \textit{pooled audience}, defined as the mixture distribution
\[
    \gamma_\alpha\br{B} := \sum_{i \in N}\alpha_i\,\gamma_i\br{B} \quad \text{for all Borel sets } B \subseteq \Omega.
\]

Each merchant has the option to show an ad to the customers in his own audience, which generates total expected clicks\footnote{When referring to campaign quantities, we use ``clicks'' as shorthand for expected clicks.}
\[
    a_i := \alpha_i \int_\Omega \omega_i\,d\gamma_i\br{\omega}.
\]
Showing an ad to a customer with CTR $\omega_i$ generates expected value $\omega_i\theta_i$ for merchant $i$, so his type-dependent \textit{outside option} is $\theta_i a_i$.

\subsubsection{Targeted Advertising Campaigns}

A targeted advertising campaign (henceforth, \textit{campaign}) specifies the customers each merchant targets and the transfer between him and the platform. Each customer can be targeted by at most one merchant.

Specifically, let $\mcal_+\br{\Omega}$ denote the space of finite nonnegative Borel measures on $\Omega$. A campaign consists of targeted ads $x_i \in \mcal_+\br{\Omega}$ and a transfer $t_i \in \rbb$ for each merchant $i \in N$.\footnote{Negative transfers $t_i < 0$ are compensation to merchants who forgo clicks and can be paid in cash or in kind, such as fee rebates or future advertising credits.} For a Borel set $B \subseteq \Omega$, let $x_i\br{B}$ denote the mass of customers with CTRs in $B$ targeted by merchant $i$'s ads.

Since the designer can replicate each merchant's baseline campaign by targeting his initial audience within the pooled audience and assigning zero transfers, we can restrict attention, without loss of generality, to campaigns in which every merchant participates. With all audiences pooled, targeted ads must satisfy the feasibility constraints
\[
    \sum_{i \in N} x_i\br{B} \leq \gamma_\alpha\br{B} \quad
    \text{for all Borel sets } B \subseteq \Omega.
\]
That is, for every set of CTR realizations $B \subseteq \Omega$, the total mass of customers assigned across merchants cannot exceed the mass $\gamma_\alpha\br{B}$ available in the pooled audience.

\paragraph{Direct Mechanism.} By the revelation principle, we can restrict attention without loss of generality to direct mechanisms that assign a campaign for each $\boldtheta \in \Theta$.

Specifically, a direct mechanism $\br{x,t}$ consists of measurable targeted ads $x = \br{x_i}_{i \in N}$ and transfers $t = \br{t_i}_{i \in N}$ such that each $x_i:\Theta \to \mcal_+\br{\Omega}$ and $t_i:\Theta \to \rbb$ satisfy the usual \textit{interim} incentive compatibility (I-IC) and individual rationality (I-IR) constraints
\begin{align}
    \theta_i\br{Q_i\br{\theta_i} - a_i} - T_i\br{\theta_i}
    &\geq
    \theta_i\br{Q_i\br{\widetilde\theta_i} - a_i} - T_i\br{\widetilde\theta_i}
    & & \text{for all } \theta_i,\widetilde\theta_i \in \Theta_i \text{ and } i \in N,
    \label{eqn:finite-IC}
    \tag{I-IC}\\
    \theta_i\br{Q_i\br{\theta_i} - a_i} - T_i\br{\theta_i}
    &\geq 0
    & & \text{for all } \theta_i \in \Theta_i \text{ and } i \in N,
    \label{eqn:finite-IR}
    \tag{I-IR}
\end{align}
where $Q_i\br{\theta_i}$ and $T_i\br{\theta_i}$ denote the expected clicks and expected transfers provided to type $\theta_{i}$, defined respectively by
\[
    Q_i\br{\theta_i} :=
    \E_{F_{-i}}
    \sbr{
        \int_\Omega
        \omega_i\,x_i\br{\theta_i,\boldtheta_{-i};d\omega}
    }
    \quad \text{and} \quad
    T_i\br{\theta_i} :=
    \E_{F_{-i}}
    \sbr{
        t_i\br{\theta_i,\boldtheta_{-i}}
    },
\]
and targeted ads are subject to \textit{ex post} feasibility constraints
\begin{equation}
    \label{eqn:targeted-ads-feasibility}
    \tag{$\gamma_\alpha$-Fsb}
    \sum_{i \in N}
    x_i\br{\boldtheta;B}
    \leq
    \gamma_\alpha\br{B}
    \quad
    \text{for all Borel sets } B \subseteq \Omega
    \text{ and } \boldtheta \in \Theta.
\end{equation}

\subsubsection{Designer's Problem}
\label{subsec:designers-problem}

The designer chooses a direct mechanism to maximize a weighted sum of platform revenue, merchants' surplus, and customer engagement.

\paragraph{Platform Revenue.} Let $R\br{t}$ denote the platform's expected revenue, defined by
\[
    R\br{t} := \E_F\sbr{\sum_{i \in N}t_i\br{\boldtheta}},
\]
where positive transfers are payments from merchants and negative transfers are compensation to them. The platform may need to raise revenue to finance the technological infrastructure required to pool audiences and run targeted advertising campaigns.

\paragraph{Merchants' Surplus.} Let $V\br{x}$ denote merchants' surplus, defined by
\[
    V\br{x} := \E_F\sbr{\sum_{i \in N}\theta_i\br{Q_i\br{\theta_i} - a_i}}.
\]
$V\br{x}$ measures the total value of merchants' expected clicks under the mechanism net of the value generated by their baseline campaigns. Merchants' total expected payoff,
\[
    U\br{x,t} := V\br{x} - R\br{t},
\]
measures their surplus net of transfers to the platform. The platform may value merchants' payoff because satisfied merchants are more likely to continue participating and increase their future advertising expenditures.

\paragraph{Customer Engagement.} Let $W\br{x}$ denote customer engagement, defined by
\[
    W\br{x} := \E_F\sbr{\sum_{i \in N}\br{Q_i\br{\theta_i} - a_i}}.
\]
$W\br{x}$ measures total expected clicks under the mechanism net of those generated by the baseline campaigns. Because the setting abstracts from product prices, we use clicks as a measure of customer's welfare from the platform.\footnote{See \cref{subsec:discussion-assumptions} for a detailed discussion on the relationship between customer engagement and surplus when product prices are fixed and publicly observed.} The platform may value better targeting because it improves the customer experience and keeps customers engaged.

\paragraph{Designer's Objective.} The designer's objective is the weighted sum
\[
    \Psi_\eta\br{x,t} := \eta_r R\br{t} + \eta_v V\br{x} + \eta_w W\br{x},
\]
with welfare weights $\eta = \br{\eta_r,\eta_v,\eta_w} \in \rbb_+^3$ normalized so that $\eta_r + \eta_v + \eta_w = 1$ and $\eta_r > 0$.

The designer may care about merchants' payoff $U\br{x,t}$ rather than their surplus $V\br{x}$. Since $U\br{x,t} = V\br{x} - R\br{t}$, however, a weighted sum of platform revenue, merchants' payoff, and customer engagement reduces to the form above whenever the weight on revenue exceeds the weight on payoff, so that transfers carry a positive net weight.\footnote{Starting from $\widetilde\eta_rR\br{t} + \widetilde\eta_vU\br{x,t} + \widetilde\eta_wW\br{x}$ with nonnegative weights and $\widetilde\eta_r > \widetilde\eta_v$, set $\br{\eta_r,\eta_v,\eta_w} = \br{\widetilde\eta_r - \widetilde\eta_v,\widetilde\eta_v,\widetilde\eta_w}/\br{\widetilde\eta_r + \widetilde\eta_w}$. If $\widetilde\eta_r = \widetilde\eta_v$, transfers cancel from the objective, and if $\widetilde\eta_r < \widetilde\eta_v$, the objective is unbounded from above.}

The designer's problem is therefore given by
\begin{equation}
    \label{eqn:designer-problem-finite}
    \tag{DP-F}
    \begin{aligned}
        \sup_{\br{x,t}} \quad & \Psi_\eta\br{x,t}\\
        \text{subject to} \quad & \eqref{eqn:finite-IC},\ \eqref{eqn:finite-IR},\ \text{and }\eqref{eqn:targeted-ads-feasibility}.
    \end{aligned}
\end{equation}
Let $\Pi_\eta$ denote the optimal value of \eqref{eqn:designer-problem-finite}.

\subsubsection{Discussion of Assumptions}
\label{subsec:discussion-assumptions}

\paragraph{Audience Pooling.} Audience pooling requires participating merchants to waive their exclusive targeting rights over their audiences.\footnote{For example, Shopify's terms (\url{https://www.shopify.com/legal/terms}, accessed October 22, 2026) state in Section~8.1 that Shopify does not take ownership of merchants' data and in Section~9.2 that it may use customer data to deliver advertising for other merchants.} Moreover, our assumption of disjoint audiences is a modeling simplification: customers may have visited several stores, but we assign each merchant a separate share of the pool.

\paragraph{Observable CTRs.} The platform's estimates for CTRs may reflect product prices and categories, ad quality, or customers' shopping history. The assumption that CTRs are observable is motivated by platforms' sharing of campaign performance metrics with merchants.\footnote{Shopify Audiences, for example, provides benchmark metrics that include CTRs; see \textit{Advertising Campaign Benchmarks} on \url{https://help.shopify.com/en/manual/promoting-marketing/shopify-audiences/benchmarks} (accessed October 10, 2026).}

\paragraph{Microfoundation for CTRs.} Suppose the value per click $\theta_i$ reflects merchant $i$'s profit margin, and that each click results in a purchase. Suppose each merchant $i$ posts publicly observable price $p_i$ and has private constant marginal cost $c_i \geq 0$, so his value per click is the profit margin $\theta_i = p_i - c_i$. A customer is interested in his product with probability $\widetilde\omega_i$. Conditional on interest, willingness to pay $v_i$ has the same commonly known distribution $H_i$ across customers. The customer clicks and purchases if she is interested and $v_i > p_i$, so her CTR for merchant $i$'s ad is given by
\[
    \omega_i := \widetilde\omega_i \,\br{1 - H_i\br{p_i}}.
\]

\paragraph{Customer Surplus.} Given the preceding microfoundation for CTRs, for fixed $\mathbf{p} = \br{p_i}_{i \in N}$, in \cref{app:customer-surplus} we show that customer surplus from targeted ads $x$ relative to baseline campaigns is given by
\[
    \mathrm{CS}\br{\mathbf{p},x}
    = \sum_{i \in N} \E_{H_i}\sbr{v - p_i \mid v > p_i}
    \br{\E_{F_i}\sbr{Q_i\br{\theta_i}} - a_i}.
\]
If we further suppose that $H_i = H$ and $p_i = p$ for all $i \in N$, then
\[
    \mathrm{CS}\br{p,x} = \E_{H}\sbr{v - p \mid v > p} W\br{x},
\]
so customer surplus and customer engagement are equivalent measures of customer welfare, differing only by a positive constant factor.

\paragraph{One Ad per Customer.} \eqref{eqn:targeted-ads-feasibility} requires that each customer be targeted at most once. The single-ad benchmark captures settings in which repeated exposure is costly. Advertising can build brand awareness \citep{percy1992model}, while repeated exposure may become less effective as customers stop paying attention or react negatively to repeated ads \citep{craig1976advertising}. To capture settings in which customers may still respond to an ad for a different product, \cref{subsec:sequential-targeting} studies sequential targeting in the limit market, where a second ad from a different tag is shown when the customer does not click on the first.

\paragraph{Pooling and Competition.} Audience pooling may affect competition among merchants. \citet{iyer2005targeting}, for example, show that targeted advertising can soften price competition and increase firms' equilibrium profits. If nonparticipation changes a merchant's expected profits by a fixed, known proportion, we can incorporate this effect by scaling his outside option accordingly.

Specifically, suppose nonparticipation reduces merchant $i$'s value per click to $\rho_i\theta_i$, compared with $\theta_i$ upon participation, for a known constant $\rho_i \in \br{0,1}$. His outside option is then $\rho_i\theta_i a_i$, which we can accommodate by replacing $a_i$ with $\widetilde a_i := \rho_i a_i$.

\subsection{Optimal Mechanism}
\label{subsec:optimal-mechanism-finite}

We show that the optimal mechanism assigns targeted ads using merchant-specific \textit{scoring} functions. To build intuition, we first draw parallels with bilateral trade and partnership dissolution when all customers have the same CTRs.

\begin{example}[Homogeneous Audiences and Revenue Maximization]
    \label{ex:homogeneous}
    Consider two merchants $i = 1,2$ with types drawn independently from $F_i = \unif\sbr{0,1}$, and suppose that the designer maximizes revenue. Merchant $i$ initially has exclusive access to an audience of mass $\alpha_i$, with $\alpha_1 + \alpha_2 = 1$. All customers have CTRs $\omega = \br{1,1}$, so merchant $i$'s baseline expected clicks are $a_i = \alpha_i$.

    If $\alpha_1 = 0$, the problem reduces to bilateral trade with a mediator, with merchant $1$ as the buyer and merchant $2$ as the seller \citep[Section~5]{myerson1983efficient}. The scoring functions are the buyer's virtual value, $2\theta_1 - 1$, and the seller's virtual cost, $2\theta_2$. The optimal mechanism assigns merchant $1$ exclusive access to merchant $2$'s audience if and only if $2\theta_1 - 1 \geq 2\theta_2$, or equivalently $\theta_1 - \theta_2 \geq 1/2$.

    If $\alpha_i \in \br{0,1}$, the problem reduces to partnership dissolution with a mediator, with buyer and seller roles determined endogenously \citep{cgk,loertscher2019optimal}. This creates countervailing incentives, with incentive constraints binding in opposite directions: high types would understate their values per click, whereas low types would overstate them. The optimal mechanism therefore uses ``ironed'' virtual functions: for suitable \textit{ironing parameters} $z_i$, merchant $i$'s score equals his virtual cost $2\theta_i$ for low types, his virtual value $2\theta_i - 1$ for high types, and has constant value $z_i$ in between:
    \begin{equation}
        \label{eqn:ironed-virtual-type-example}
        \overline\phi_i\br{\theta_i;z_i}
        =
        \begin{cases}
            2\theta_i & \text{if } 2\theta_i \leq z_i,\\
            z_i & \text{if } 2\theta_i - 1 < z_i < 2\theta_i,\\
            2\theta_i - 1 & \text{if } 2\theta_i - 1 \geq z_i.
        \end{cases}
    \end{equation}
    The optimal mechanism assigns the entire pooled audience to the merchant with the higher score, with tie-breaking chosen so that types with score $z_i$ receive their baseline expected clicks $\alpha_i$ and earn zero rents at the interim stage.
\end{example}

With homogeneous audiences, gains from trade come from reallocating clicks toward merchants with higher values per click. Heterogeneous audiences create additional opportunities to generate more clicks through better targeting, even when merchants have the same value per click. The designer must therefore determine which merchant should target each customer.

Although different customers may receive ads from different merchants, the designer need not run a separate single-good mechanism for each customer as in the classic single-good settings. This motivates the following definition.

\begin{definition}[Scoring Mechanism]
    \label{def:scoring-mechanism}
    We say that $\br{h,\pi}$ is a scoring mechanism if it consists of merchant-specific, nondecreasing scoring functions $h_i$ and a customer-specific tie-breaking rule $\pi$, defined by (Borel) measurable functions $h_i:\Theta_i \to \mathbb R_+$ for all $i \in N$ and $\pi: \Theta\times\Omega \to \Delta\br{N\cup\cbr{0}}$ such that
    \[
        \begin{aligned}
            x_i\br{\boldtheta;B} &= \int_B
            \pi_i\br{\boldtheta,\omega}\,d\gamma_\alpha\br{\omega} \quad \text{for all Borel sets } B \subseteq \Omega,\\
            \text{where} \quad \pi_i\br{\boldtheta,\omega} &= 0
            \;\text{ if }\;
            i \notin
            \argmax_{j \in N\cup\{0\}}
            \cbr{
                \omega_jh_j\br{\theta_j}
            } \quad \text{for all } i \in N\cup\cbr{0} \text{ and } \boldtheta \in \Theta,
        \end{aligned}
    \]
    where $\omega_0h_0\br{\theta_0} \equiv 0$ by convention and $\pi_0\br{\boldtheta,\omega}$ is the share of customers with CTRs $\omega$ left unassigned.
\end{definition}
In a scoring mechanism, each customer is assigned to a merchant with the highest customer-specific \textit{score} $\omega_ih_i\br{\theta_i}$. A single score $h_i\br{\theta_i}$ therefore determines merchant $i$'s entire campaign.

\begin{figure}[t!]
    \centering

    \begin{subfigure}[t]{\figurepanelwidth}
        \centering
        \resizebox{\linewidth}{!}{\input{figures/scoring-negative-intercept.tex}}
        \caption{$\phi^{B}_{\eta,i}\br{0} < 0$.}
        \label{fig:scoring-rule-negative}
    \end{subfigure}
    \hfill
    \begin{subfigure}[t]{\figurepanelwidth}
        \centering
        \resizebox{\linewidth}{!}{\input{figures/scoring-positive-intercept.tex}}
        \caption{$\phi^{B}_{\eta,i}\br{0} > 0$.}
        \label{fig:scoring-rule-positive}
    \end{subfigure}

    \caption{Examples of (optimal) scoring and weighted virtual functions.}
    \label{fig:scoring-rule}
\end{figure}

We show that a scoring mechanism is optimal when its scores are constructed by ironing the weighted virtual value and cost functions, defined respectively by
\[
    \phi^B_{\eta,i}\br{\theta_i} := \eta_r \phi^B_i\br{\theta_i} + \eta_v\theta_i + \eta_w
    \quad \text{and} \quad
    \phi^S_{\eta,i}\br{\theta_i} := \eta_r \phi^S_i\br{\theta_i} + \eta_v\theta_i + \eta_w.
\]
\Cref{fig:scoring-rule} illustrates these functions together with their ironed counterpart, denoted by $\overline\phi_{\eta,i}\br{\theta_i;z_i}$, which equals the nonnegative ironing parameter $z_i$ on the ironing interval and coincides with the weighted virtual cost function to its left and virtual value function to its right.

To state the results, for each $j \in \cbr{S,B}$ and $i \in N$, we use the inverse convention that $\br{\phi_{\eta,i}^j}^{-1}\br{z} = \ltheta_i$ whenever $z < \phi_{\eta,i}^j\br{\ltheta_i}$ and $\br{\phi_{\eta,i}^j}^{-1}\br{z} = \btheta_i$ whenever $z > \phi_{\eta,i}^j\br{\btheta_i}$.

\subsubsection{Finite-Support CTRs}
\label{subsec:finite-support-CTRs}

Let $\Omega$ be a finite set. For ease of notation, we write $\gamma_\alpha\br{\omega} := \gamma_\alpha\br{\cbr{\omega}}$, $x_i^\omega\br{\boldtheta} := x_i\br{\boldtheta;\cbr{\omega}}$, and $\pi_i^\omega\br{\boldtheta} := \pi_i\br{\boldtheta,\omega}$ for all $\omega \in \Omega$ and $\boldtheta \in \Theta$.\footnote{ Note that \eqref{eqn:targeted-ads-feasibility} becomes equivalent to
\[
    \sum_{i \in N} x_i^\omega\br{\boldtheta} \leq \gamma_\alpha\br{\omega} \quad \text{for all } \omega \in \Omega \text{ and } \boldtheta \in \Theta.
\]
}

\begin{theorem}[Optimal Scoring Mechanism]
    \label{thm:optimal-finite-mechanism}
    Suppose $a_i > 0$ for all $i \in N$. There exists an optimal mechanism that can be represented as a scoring mechanism in which, for each $i \in N$, the scoring function $h_i$ is given by $h_i \equiv \overline\phi_{\eta,i}\br{\cdot \,;z_i}$ for some ironing parameter $\br{\phi_{\eta,i}^{B}\br{\ltheta_i}}_+\leq z_i \leq \phi_{\eta,i}^{S}\br{\btheta_i}$, where
    \begin{equation}
        \label{eqn:scoring-function}
        \overline\phi_{\eta,i}\br{\theta_i;z_i} :=
        \begin{cases}
            \phi^S_{\eta,i}\br{\theta_i} & \text{if } \theta_i < \theta_i^S,\\
            z_i & \text{if } \theta_i^S \leq \theta_i \leq \theta_i^B,\\
            \phi^B_{\eta,i}\br{\theta_i} & \text{if } \theta_i > \theta_i^B,
        \end{cases}
    \end{equation}
    with cutoff types $\theta_i^j = \br{\phi_{\eta,i}^j}^{-1}\br{z_i}$ for each $j \in \cbr{S,B}$. Moreover, the corresponding customer-specific tie-breaking rule $\pi$ is chosen such that the critical types $\htheta_i$, defined uniquely for each $i \in N$ by
    \begin{equation}
        \label{eqn:worst-off-type-ironing-parameter-identity}
        \htheta_i\br{z_i} := \E_{F_i} \sbr{\overline\phi_{\eta,i}\br{\theta_i;z_i} - \eta_v \theta_i - \eta_w} / \eta_r \,\in \sbr{\theta_i^S,\theta_i^B},
    \end{equation}
    are worst-off types \`{a} la \citet{cgk}: $Q_i\br{y} \leq a_i$ for all $y < \htheta_i$ and $Q_i\br{y} \geq a_i$ for all $y > \htheta_i$. Lastly, interim transfers are recovered by
    \[
        T_i\br{\theta_i} = \theta_i\br{Q_i\br{\theta_i} - a_i} - \int_{\htheta_i}^{\theta_i} \br{Q_i\br{y} - a_i}dy.
    \]
\end{theorem}

Relative to baseline clicks $a_i$, expected clicks $Q_i$ determine whether a merchant is on average a net seller or a net buyer. As we show later, \eqref{eqn:finite-IC} requires expected clicks to be nondecreasing in type, so in general low types are net sellers and high types are net buyers. The scoring function in \eqref{eqn:scoring-function} reflects this distinction: low-type scores follow the weighted virtual cost function $\phi^S_{\eta,i}\br{\theta_i}$, whereas high-type scores follow the weighted virtual value function $\phi^B_{\eta,i}\br{\theta_i}$.

Ironing pools ``intermediate'' types and pins down each critical worst-off type $\htheta_i\br{z_i}$ as in \eqref{eqn:worst-off-type-ironing-parameter-identity}. When the critical worst-off type is interior, all types on the ironing interval are worst-off types and receive baseline expected clicks $Q_i\br{\htheta_i} = a_i$. If $\htheta_i = \underline\theta_i$, higher types receive at least their baseline clicks, and if $\htheta_i = \overline\theta_i$, lower types receive at most their baseline clicks. Importantly, the designer may be able to deliver these baseline clicks using fewer customers with higher CTRs, possibly from other merchants' initial audiences.

The critical worst-off type equals $\ltheta_i$ when, relative to merchant $i$'s initial audience, the pooled audience contains many customers with high CTRs for his ad relative to other merchants' ads. It can also equal $\ltheta_i$ when the designer prefers targeting every customer to leaving some unassigned, for instance when the weight on customer engagement is strictly positive. Conversely, the critical worst-off type equals $\btheta_i$ when merchant $i$ is assigned relatively few ads, making him a net seller regardless of his type. These endpoint cases do not arise in the partnership dissolution setting of \citet{loertscher2019optimal} when types have common support: because the good is homogeneous, the allocation depends only on types, so each merchant is a net buyer at high types and a net seller at low types.\footnote{For example, if $\omega = \br{1/2,1}$ for every customer and $\theta_i\sim\unif\sbr{1/2,1}$, then $\theta_1/2 \leq 1/2 \leq \theta_2$, so merchant $1$ never values a customer more than merchant $2$, and the critical worst-off types are $\btheta_1$ and $\ltheta_2$.}

With finite-support CTRs, the flat regions of the scoring functions can produce ties with positive probability, and their resolution affects each critical type's expected clicks. We show that the designer chooses the ironing parameters and a tie-breaking rule so that each critical type is a worst-off type. The optimal tie-breaking rule, however, is defined only implicitly by these worst-off-type conditions. In the next subsection, we show that when CTRs are continuously distributed, ties occur only on null sets, which allows us to characterize the optimal ironing parameters as the solution to a system of equations.

\subsubsection{Continuously Distributed CTRs}
\label{subsec:tie-free-CTRs}

In this subsection, we assume that the pooled audience $\gamma_\alpha$ has a density on $\sbr{0,1}^N$. We further require that at least one merchant's baseline clicks are large relative to what he would receive under a scoring mechanism that treats all merchants as buyers:

\begin{assumption}[Audience Scarcity (Finite Market)]
    \label{asm:scarcity-finite}
    There exists $i \in N$ such that
    \begin{equation}
        \label{eqn:audience-scarcity-finite}
        a_i \geq
        \left(
        \prod_{j \neq i}
        F_j\br{
            \br{\phi_{\eta,j}^{B}}^{-1}\br{0}
        }
        \right)
        \int_\Omega \omega_i\,d\gamma_\alpha\br{\omega}.
    \end{equation}
\end{assumption}

Together with $a_i > 0$ for every $i \in N$, we show that \cref{asm:scarcity-finite} ensures that some merchant's score is strictly positive almost surely under the optimal scoring mechanism. Since CTRs are continuously distributed, the highest score is then attained by a unique merchant for almost every customer, so ties occur only on null sets. The following proposition formalizes these arguments and gives the system of equations that pins down the optimal ironing parameters.

\begin{proposition}
    \label{prop:ironing-system}
    Suppose \cref{asm:scarcity-finite} holds and $a_i > 0$ for all $i \in N$. There exists an optimal scoring mechanism as in \cref{thm:optimal-finite-mechanism} with a nonzero vector of ironing parameters $z = \br{z_i}_{i \in N} \neq \mathbf{0}$. For almost every $\boldtheta \in \Theta$ and $\omega \in \Omega$, $\pi_i\br{\boldtheta,\omega} = 1$ for some $i \in N$. If the critical worst-off type $\htheta_i\br{z_i}$ is interior for some $i \in N$, then $Q_i\br{\htheta_i} = a_i$, so $z$ satisfies
    \[
        a_i
        = \int_\Omega
        \omega_i\,
        \p_{F_{-i}}\br{
            \omega_i z_i
            > \max_{j \neq i}\omega_j\overline\phi_{\eta,j}\br{\theta_j;z_j}
        }\,d\gamma_\alpha\br{\omega}.
    \]

\end{proposition}

The next example applies the proposition to the two-merchant setting of \cref{ex:homogeneous} with heterogeneous CTRs and shows how the ironing parameters change as CTRs become more or less dispersed.

\begin{figure}[t!]
    \centering
    \begin{subfigure}[t]{\figurepanelwidth}
        \centering
        \input{figures/ironing-equal-audiences}
        \caption{$\alpha_1 = \alpha_2 = 1/2$.}
        \label{fig:example-N2-continuum-equal}
    \end{subfigure}
    \hfill
    \begin{subfigure}[t]{\figurepanelwidth}
        \centering
        \input{figures/ironing-unequal-audiences}
        \caption{$\alpha_1 = 4/5$ and $\alpha_2 = 1/5$.}
        \label{fig:example-N2-continuum-asymmetric}
    \end{subfigure}
    \caption{Optimal Ironing Parameters $z_1\br{\varepsilon}$ and $z_2\br{\varepsilon}$.}
    \label{fig:example-N2-continuum}
\end{figure}

\begin{example}[Heterogeneous Audiences and Revenue Maximization]
    \label{ex:heterogeneous}
    Return to \cref{ex:homogeneous}, but now allow CTRs to vary across customers.

    Specifically, for $\varepsilon \in \lsbr{1/2,1}$, suppose that customers in either initial audience have independently distributed CTRs $\omega_1,\omega_2\overset{\mathrm{i.i.d.}}{\sim}\unif\sbr{2\varepsilon - 1,1}$ with $\E\sbr{\omega_i} = \varepsilon$. The pooled audience has this same distribution regardless of the initial shares, and merchant $i$'s baseline expected clicks are $a_i = \alpha_i\varepsilon$. Increasing $\varepsilon$ raises average CTRs and reduces their variance; as $\varepsilon \longrightarrow 1$, CTRs converge to $\omega = \br{1,1}$, recovering the homogeneous benchmark.

    If $\alpha_1 = 0$, merchant $1$ is the buyer and merchant $2$ is the seller. Their scoring functions are again the buyer's virtual value, $2\theta_1 - 1$, and the seller's virtual cost, $2\theta_2$, respectively, but the optimal mechanism now compares customer-specific scores: a customer with CTRs $\omega$ receives merchant $1$'s ad if
    \[
        \omega_1\br{2\theta_1 - 1} > \omega_2\br{2\theta_2}.
    \]
    Thus, different customers may receive ads from different merchants. As $\varepsilon \longrightarrow 1$, this condition reduces to $\theta_1 - \theta_2 \geq 1/2$ as in bilateral trade.

    If $\alpha_i \in \br{0,1}$, buyer and seller roles are again determined endogenously. The optimal mechanism uses the ironed scoring functions in \eqref{eqn:ironed-virtual-type-example} and assigns each customer to the merchant with the higher score $\omega_i\overline\phi_i\br{\theta_i;z_i\br{\varepsilon}}$. \Cref{fig:example-N2-continuum} illustrates the optimal ironing parameters $z\br{\varepsilon} = \br{z_1\br{\varepsilon},z_2\br{\varepsilon}}$, where the parameters converge to their homogeneous-benchmark values as $\varepsilon \longrightarrow 1$.

    The ironing parameters reflect two ways the designer can address poaching concerns: compensate merchants for the clicks they forgo or give them higher priority in the scoring mechanism. When CTRs are high and similar across customers, the gains from better targeting are small, so the designer relies more on compensation by assigning higher ironing parameters, making merchants more likely to be net sellers who are paid for their forgone baseline clicks. Across merchants, those with larger baseline campaigns have more to lose from pooling, so the designer protects them with higher ironing parameters, giving their intermediate types higher targeting priority.
\end{example}

\subsection{Proof Overview}
\label{subsec:proof-overview-finite}

We provide a proof overview for \cref{thm:optimal-finite-mechanism,prop:ironing-system}. The main steps are standard. We reduce the designer's objective to a virtual objective as in \citet{myerson1981optimal} and solve a minimax formulation of the problem as in \citet{loertscher2019optimal}.

\paragraph{Virtual Objective.} First, we apply standard Myersonian techniques and characterize \eqref{eqn:finite-IC} in terms of nondecreasing expected clicks $Q_i$, and by the envelope theorem also pin down transfers $T_i$ up to a constant and rewrite the designer's objective as
\[
    \Psi_\eta\br{x,t}
    = \sum_{i \in N}\E_{F_i}\sbr{
        \br{Q_i\br{\theta_i} - a_i}
        \phi_{\eta,i}\br{\theta_i;\theta_i^\crt}
    } - \eta_r \sum_{i \in N} U_i\br{\theta_i^\crt}
\]
for any critical type $\boldtheta^{\crt} \in \Theta$, with payoff $U_i\br{\theta_i^\crt} := \theta_i^\crt \br{Q_i\br{\theta_i^\crt} - a_i} - T_i\br{\theta_i^\crt}$ and weighted virtual type function
\[
    \phi_{\eta,i}\br{\theta_i;\theta_i^{\crt}}
    :=
    \begin{cases}
        \phi^S_{\eta,i}\br{\theta_i} & \text{if }\theta_i < \theta_i^{\crt},\\
        \phi^B_{\eta,i}\br{\theta_i} & \text{if }\theta_i \geq \theta_i^{\crt}.
    \end{cases}
\]

Because $Q_i\br{\theta_i} - a_i$ can be negative if a merchant is a net seller and positive if  he is a net buyer, the worst-off type is determined endogenously by the mechanism $\br{x,t}$.

\paragraph{Minimax Characterization.} Let $\xscr$ denote feasible targeted ads with nondecreasing interim expected clicks. Following \citet{loertscher2019optimal} and using Fan's minimax theorem \citep[Theorem~1]{fan1953minimax}, we can express the optimal value of \eqref{eqn:designer-problem-finite} as
\begin{equation}
    \label{eqn:finite-minimax}
    \Pi_\eta
    = \adjustlimits\min_{\boldtheta^{\crt} \in \Theta}\max_{x \in \xscr}
    \psi_\eta\br{x,\boldtheta^{\crt}},
\end{equation}
where
\[
    \psi_\eta\br{x,\boldtheta^\crt}
    := \sum_{i \in N}\E_{F_i}\sbr{
        \br{Q_i\br{\theta_i} - a_i}
        \phi_{\eta,i}\br{\theta_i;\theta_i^\crt}}.
\]

We show that \eqref{eqn:finite-minimax} has a saddle point $\br{x^*,\boldsymbol{\htheta}}$ such that $x^*$ is optimal and $\boldsymbol{\htheta}$ is a worst-off critical type with payoff $U_i\br{\htheta_i} = 0$ for all $i \in N$.

\paragraph{Ironing.} For a fixed worst-off critical type $\boldsymbol{\htheta}$, we characterize $x^*$ as the solution to the inner maximization problem in \eqref{eqn:finite-minimax}. Since $\phi_{\eta,i}\br{\theta_i;\theta_i^\crt}$ has a downward jump at an interior $\theta_i^\crt$, maximizing pointwise for $x_i\br{\boldtheta;\omega}$ at each $\boldtheta \in \Theta$ and $\omega \in \Omega$ need not yield nondecreasing expected clicks $Q_i$, and may violate \eqref{eqn:finite-IC}.

To overcome this, we apply the ironing technique of \citet{myerson1981optimal} and instead maximize pointwise the ironed virtual objective
\begin{equation}
    \label{eqn:finite-ironing-bound}
    \overline\psi_\eta\br{x,z} := \sum_{i \in N}\E_{F_i}\sbr{
        \br{Q_i\br{\theta_i} - a_i}
        \overline\phi_{\eta,i}\br{\theta_i;z_i}
    },
\end{equation}
where $\overline\phi_{\eta,i}\br{\theta_i;z_i}$ is nondecreasing and $z = z\br{\boldsymbol{\htheta}}$ is pinned down by \eqref{eqn:worst-off-type-ironing-parameter-identity}. Since $\psi_\eta\br{x,\boldsymbol{\htheta}} \leq \overline\psi_\eta\br{x,z}$ for all $x \in \xscr$, with equality whenever each $Q_i$ is constant on its ironing interval, any maximizer of the ironed objective whose expected clicks are constant on these intervals solves the inner maximization in \eqref{eqn:finite-minimax}.

\paragraph{Pointwise Maximization.} Substituting the definition of $Q_i$ into \eqref{eqn:finite-ironing-bound}, for each $\boldtheta$ and $\omega$ the pointwise maximization problem is
\begin{equation}
    \label{eqn:pointwise-problem-scoring}
    \begin{aligned}
        \max_{\pi \in \rbb_+^N} \quad
        & \sum_{i \in N}\pi_i\,\omega_i\overline\phi_{\eta,i}\br{\theta_i;z_i}\\
        \text{subject to} \quad & \sum_{i \in N}\pi_i \leq 1.
    \end{aligned}
\end{equation}
The optimal value in \eqref{eqn:pointwise-problem-scoring} equals $\max\cbr{0,\max_{i \in N}\omega_i\overline\phi_{\eta,i}\br{\theta_i;z_i}}$. Each customer is assigned to a merchant with the highest customer-specific score, with probability one whenever this score is strictly positive. If ties are broken by a rule that depends on types only through the scores $\overline\phi_{\eta,i}\br{\theta_i;z_i}$, $Q_i$ is nondecreasing and constant on the ironing interval. We construct such a rule using the saddle point $\br{x^*,\boldsymbol{\htheta}}$. For continuously distributed CTRs, \cref{asm:scarcity-finite} and positive baseline clicks imply $z \neq \mathbf{0}$, so ties occur only on null sets and the worst-off type characterization gives $Q_i\br{\htheta_i\br{z_i}} = a_i$ at interior critical types.

\section{Limit Market}
\label{sec:limit-market}
In this section, we describe a model with a continuum of merchants (henceforth, the \textit{limit market}) and show in \cref{sec:limit-market-convergence} that it is the limit of a sequence of finite markets as the number of merchants grows. The limit market preserves the trade-off between better targeting and customer poaching of the finite market while eliminating aggregate uncertainty about merchants' types, thereby yielding a simple implementation of the optimal mechanism through posted prices.

We formally introduce the model in \cref{subsec:model-limit} and characterize the optimal \textit{posted-price mechanism} in \cref{subsec:optimal-mechanism-limit}. \Cref{subsec:proof-sketch-limit} presents a proof sketch, and the complete proofs are in \cref{app:limit}.

\subsection{Model}
\label{subsec:model-limit}

A designer operates an advertising platform serving a unit mass of merchants, each with an observable tag $k$ from a finite set $K$. There is also a unit mass of customers, initially distributed evenly across merchants: any group of merchants with mass $\beta \in \sbr{0,1}$ collectively contributes audience mass $\beta$.

Merchants with tag $k \in K$ have mass $\beta_k > 0$, with $\sum_{k \in K}\beta_k = 1$. Each merchant has a private value per click (type) $\theta \in \Theta_k = \sbr{\ltheta_k,\btheta_k} \subset \rbb_+$. Within tag $k$, types are distributed according to $F_k$, which admits a strictly positive, continuous density $f_k$ and is regular as in the finite market.

Each merchant initially has exclusive targeting access to his audience. Advertising only to his own audience gives a merchant with tag $k$ baseline expected clicks $\mu_k \in \br{0,1}$ and an outside option $\mu_k\theta$, both per unit of merchant mass. Once audiences are pooled, each customer is represented by observable \textit{tag-specific} CTRs
\[
    \xi = \br{\xi_k}_{k \in K} \in \sbr{0,1}^K,
\]
where $\xi_k$ is the best CTR attainable by targeting the customer within tag $k$.\footnote{Depending on context, similarly $K$ denotes the set of tags or its cardinality.} CTRs $\xi$ are distributed according to $G \in \Delta\br{\sbr{0,1}^K}$, which has a density $g$ that is strictly positive on $\br{0,1}^K$. We require $G$ to be \textit{consistent}, in the sense that the pooled audience can reproduce every merchant's baseline expected clicks, which need not come from the same customers.\footnote{We provide a formal treatment of consistency in \cref{sec:limit-market-convergence}.}

Lastly, we impose a capacity normalized to one expected click per unit of merchant mass.\footnote{The capacity is a stylized form of diminishing returns: the marginal value of a click is $\theta$ up to capacity and zero beyond it. Our framework can be extended to settings with diminishing returns to clicks.} A merchant with tag $k$ can therefore expand his campaign from its baseline $\mu_k$ by at most $1 - \mu_k$ clicks.

\subsubsection{Direct Mechanism}
Let $\mcal_+\br{\sbr{0,1}^K}$ denote the space of finite nonnegative Borel measures on $\sbr{0,1}^K$. As in the finite market, we can restrict attention to direct mechanisms $\br{x,t}$ that consist of measurable targeted ads $x = \br{x_k}_{k \in K}$ and transfers $t = \br{t_k}_{k \in K}$ such that each $x_k:\Theta_k \to \mcal_+\br{\sbr{0,1}^K}$ and $t_k:\Theta_k \to \rbb$ satisfy the usual incentive compatibility (IC) and individual rationality (IR) constraints
\begin{align}
    \theta\br{q_k\br{\theta} - \mu_k} - t_k\br{\theta}
    &\geq \theta\br{q_k\br{\tilde\theta} - \mu_k} - t_k\br{\tilde\theta}
    &&\text{for all }\theta,\tilde\theta \in \Theta_k\text{ and }k \in K,
    \label{eqn:ICk}
    \tag{IC$_k$}\\
    \theta\br{q_k\br{\theta} - \mu_k} - t_k\br{\theta}
    &\geq 0
    &&\text{for all }\theta \in \Theta_k\text{ and }k \in K,
    \label{eqn:IRk}
    \tag{IR$_k$}
\end{align}
where $q_k: \Theta_k \to \rbb_+$ denotes expected clicks per unit of merchant mass for a merchant with tag $k$ and type $\theta$, defined by
\begin{equation}
    \label{eqn:expected-clicks-limit}
    q_k\br{\theta} := \int_{\sbr{0,1}^K}\xi_k\,x_k\br{\theta;d\xi}
    \quad \text{for all }\theta \in \Theta_k,
\end{equation}
and targeted ads are subject to capacity constraints
\begin{equation}
    \label{eqn:Ck}
    \tag{Cap$_k$}
    0 \leq q_k\br{\theta} \leq 1
    \quad \text{for all }\theta \in \Theta_k\text{ and }k \in K,
\end{equation}
and the aggregate feasibility constraints
\begin{equation}
    \label{eqn:aggregate-feasibility-limit}
    \tag{G-Fsb}
    \sum_{k \in K}\beta_k\int_{\Theta_k}x_k\br{\theta;B}\,dF_k\br{\theta}
    \leq G\br{B}
    \quad \text{for all Borel sets }B \subseteq \sbr{0,1}^K.
\end{equation}

Here, $x_k\br{\theta;B}$ denotes the mass of customers with CTRs in $B$ assigned to a merchant with tag $k$ and type $\theta$, per unit of merchant mass. Thus, \eqref{eqn:aggregate-feasibility-limit} requires that the total mass of customers assigned across tags and types cannot exceed the mass $G\br{B}$ available in the pooled audience, for every Borel set $B \subseteq \sbr{0,1}^K$.

\subsubsection{Designer's Problem}
As in the finite market, the designer chooses a direct mechanism to maximize a weighted sum of platform revenue, merchants' surplus, and customer engagement. For given welfare weights $\eta = \br{\eta_r,\eta_v,\eta_w}$ the designer's problem is
\begin{equation}
    \label{eqn:designer-problem-limit}
    \tag{DP-L}
    \begin{aligned}
        \sup_{\br{x,t}} \quad & \sum_{k \in K}\beta_k\E_{F_k}\sbr{
            \eta_r t_k\br{\theta}
            + \br{\eta_v\theta + \eta_w}\br{q_k\br{\theta} - \mu_k}
        }\\
        \text{subject to} \quad & \eqref{eqn:aggregate-feasibility-limit},\ \eqref{eqn:Ck},\ \eqref{eqn:ICk},\ \text{and }\eqref{eqn:IRk}\ \text{for all }k \in K,
    \end{aligned}
\end{equation}
where $q$ are implemented by $x$ through \eqref{eqn:expected-clicks-limit}. Let $\Pi_\eta^\infty$ denote the optimal value of \eqref{eqn:designer-problem-limit}.

\subsection{Optimal Mechanism}
\label{subsec:optimal-mechanism-limit}

We show that the optimal mechanism can be implemented with tag-specific posted prices for buying and selling clicks, pinned down by virtual prices that also determine the supply of clicks across tags.

For each tag $k$, let $\phi_{\eta,k}^B$ and $\phi_{\eta,k}^S$ denote the weighted virtual value and cost functions of $F_k$, defined as in \cref{subsec:optimal-mechanism-finite}, which give the weighted marginal revenue $\MR_{\eta,k}\br{\theta} := \phi_{\eta,k}^B\br{\theta}$ and marginal cost $\MC_{\eta,k}\br{\theta} := \phi_{\eta,k}^S\br{\theta}$ when the marginal buyer or seller has type $\theta$.\footnote{When $\eta_r = 1$, they coincide with the usual marginal revenue and marginal cost curves of \citet{bulow1989simple}.}

Analogous to \cref{asm:scarcity-finite} in the finite market, we impose the following audience-scarcity condition, where the inverses $\MC_{\eta,k}^{-1}$ and $\MR_{\eta,k}^{-1}$ follow the convention in \cref{subsec:optimal-mechanism-finite}.

\begin{assumption}[Audience Scarcity (Limit Market)]
    \label{asm:scarcity-limit}
    The tag-specific type distributions $\br{F_k}_{k \in K}$ and the pooled audience distribution $G$ jointly satisfy
    \[
        \sum_{k \in K}\beta_k\sbr{1 - \br{1 - \mu_k}F_k\br{\MR_{\eta,k}^{-1}\br{0}}}
        > \E_G\sbr{\max_{k \in K}\xi_k}.
    \]
\end{assumption}

The assumption requires that the total expected clicks demanded when no merchant sells exceed what the pooled audience can supply. Specifically, the left-hand side gives the demand for expected clicks under the three-option mechanism at virtual prices $\lambda_k = 0$ for all $k$, when all merchants either receive their baseline clicks or expand their campaigns to capacity. The right-hand side gives the maximum expected clicks the pooled audience can generate. Together with the density assumption on $G$, audience scarcity ensures that every tag has a strictly positive virtual price, as shown in \cref{thm:optimal-limit-mechanism}.

\begin{theorem}[Optimal Posted-Price Mechanism]
    \label{thm:optimal-limit-mechanism}
    Suppose \cref{asm:scarcity-limit} holds. Then an optimal mechanism exists, and every optimal mechanism induces the same expected clicks and transfers $dF_k$-almost everywhere as the mechanism implemented by per-click posted prices $p_k^S \leq p_k^B$ in $\Theta_k$ that segment each tag $k \in K$ into
    \begin{enumerate}[label=\roman*)]
        \item a selling market, with types $\theta \leq p_k^S$ who receive no targeted ads and are paid $p_k^S$ per baseline click forgone: $q_k\br{\theta} = 0$ and $t_k\br{\theta} = - \mu_kp_k^S$;

        \item an exchange market, with types $\theta \in \rsbr{p_k^S,p_k^B}$ who receive their baseline clicks at no charge: $q_k\br{\theta} = \mu_k$ and $t_k\br{\theta} = 0$;

        \item a buying market, with types $\theta > p_k^B$ who expand their campaigns to capacity and pay $p_k^B$ per additional click: $q_k\br{\theta} = 1$ and $t_k\br{\theta} = \br{1 - \mu_k}p_k^B$.
    \end{enumerate}
    There exists a unique vector of tag-specific \textit{virtual prices} $\lambda = \br{\lambda_k}_{k \in K}$, with $0 < \lambda_k < \MC_{\eta,k}\br{\btheta_k}$ for every $k \in K$, that pins down the posted prices through
    \begin{equation}
        \label{eqn:posted-price-cutoffs}
        \tag{MR$_k{=}$MC$_k$}
        p_k^S = \MC_{\eta,k}^{-1}\br{\lambda_k} \quad \text{and} \quad p_k^B = \MR_{\eta,k}^{-1}\br{\lambda_k},
    \end{equation}
    and solves the market-clearing conditions
    \begin{equation}
        \label{eqn:market-clearing-limit}
        \tag{S$_k{=}$D$_k$}
        \E_G\sbr{
            \xi_k
            \mathbf{1}\cbr{
                k = \argmax_{j \in K}\lambda_j\xi_j
            }
        }
        =
        \beta_k \sbr{
            1 - F_k\br{p_k^B} +
            \mu_k \br{F_k\br{p_k^B} - F_k\br{p_k^S}}
        }.
    \end{equation}
\end{theorem}

The optimal mechanism is determined by market-clearing virtual prices $\lambda = \br{\lambda_k}_{k \in K}$ that segment each tag into at most three markets: selling, buying and exchange. By \cref{thm:optimal-limit-mechanism}, $\lambda_k < \MC_{\eta,k}\br{\btheta_k}$ and hence $p_k^S < \btheta_k$, so every tag contains a positive mass of merchants in the exchange or buying market.

Within tag $k$, the designer’s problem reduces to a \textit{screening problem} in which every click she promises to merchants costs her the \textit{shadow cost} $\lambda_k$. The designer acquires baseline clicks from sellers as long as their weighted marginal cost is below the shadow cost of supplying them from the pooled audience, and sells additional clicks to buyers as long as their weighted marginal revenue exceeds it. At an interior optimum, both margins therefore equal the shadow cost:
\[
    \MC_{\eta,k}\br{p_k^S}
    = \lambda_k
    = \MR_{\eta,k}\br{p_k^B}.
\]
Importantly, the \textit{qualitative} structure of the optimal menu within each tag does not depend on the level of $\lambda_k$, which affects only the posted prices and the sizes of the three markets.

The optimal menu gives sellers and buyers positive information rents, while merchants in the exchange market earn none. Sellers forgo their baseline clicks and earn rents $\mu_k\br{p_k^S - \theta}$, which decrease in the value per click $\theta$. Buyers expand to capacity and earn rents $\br{1 - \mu_k}\br{\theta - p_k^B}$, which increase in $\theta$. Intermediate types receive their baseline clicks by targeting customers in the pooled audience, potentially different from those in their initial audiences, and make no transfers, so they earn no rents. The exchange market parallels the finite-market ironing interval: in both, merchants receive their baseline clicks and earn zero rents, but here each merchant’s type alone determines his expected clicks and transfer, whereas in the finite market they also depend on other merchants’ reports.

Across tags, the designer's problem reduces to a \textit{targeting problem} in which every click she supplies to tag $k$ earns her the \textit{shadow value} $\lambda_k$. We show that the targeting problem is essentially a ``production problem'' with customers as inputs, clicks as outputs, CTRs as productivities, and shadow values $\lambda_k$ as output prices. As in the allocation of heterogeneous inputs across sectors (e.g., \citealp{roy1951some,costinot2009elementary}), the designer maximizes the value of production by assigning each customer to the tag with the highest price times productivity, $\lambda_k\xi_k$. That is, the optimal targeting rule compares relative CTRs with relative virtual prices: tag $k$ wins a customer over tag $j$ whenever it has the higher score for her, or equivalently, a \textit{comparative advantage} at those prices:
\[
    \lambda_k\xi_k \geq \lambda_j\xi_j
    \quad \Leftrightarrow \quad
    \xi_k / \xi_j \geq \lambda_j /\lambda_k.
\]
A customer therefore need not receive the ad she is most likely to click, but the one whose click the designer values most.

Market clearing \eqref{eqn:market-clearing-limit} links the two problems by equating supply from targeting with demand from screening in every tag. The left-hand side of \eqref{eqn:market-clearing-limit} is the supply of clicks from targeting. On the right-hand side, the buying and exchange markets demand $1 - F_k\br{p_k^B}$ and $\mu_k\sbr{F_k\br{p_k^B} - F_k\br{p_k^S}}$ expected clicks per unit of merchant mass, respectively, which after scaling by the tag’s mass $\beta_k$ yields the total demand for expected clicks in tag $k$. At the market-clearing $\lambda_k$ (the virtual price), the shadow cost of promising a click to merchants with tag $k$ equals the shadow value of supplying one from the pooled audience.

The virtual prices pin down merchants' expected clicks and transfers across tags. The following example illustrates how to construct the mechanism and certify its optimality from candidate shadow prices.

\begin{example}[Certifying Optimality with Candidate Shadow Prices]
    \label{ex:certifying}
    Consider a revenue-maximizing designer $\br{\eta_r = 1}$ and two observable tags $k \in K = \cbr{1,2}$, corresponding to road-running and trail-running shoes. Each tag has merchant mass $\beta_k = 1/2$ and values per click drawn from $F_k = \unif\sbr{0,1}$, so $\MC_{\eta,k}\br{\theta} = 2\theta$ and $\MR_{\eta,k}\br{\theta} = 2\theta - 1$. CTRs in the pooled audience satisfy $\xi_1,\xi_2\overset{\mathrm{i.i.d.}}{\sim}\unif\sbr{0,1}$. Suppose that $60\%$ of customers initially come from the higher-CTR tag and $40\%$ from the other. Baseline expected clicks are therefore $\mu_1 = \mu_2 = (3/5)(2/3) + (2/5)(1/3) = 8/15$. Lastly, at zero shadow prices, total demand is $23/30$, exceeding the maximum $2/3$ expected clicks the pooled audience can generate, so \cref{asm:scarcity-limit} holds.

    Now, consider the candidate shadow prices $\lambda_1 = \lambda_2 = 1/5 < \MC_{k}\br{1}$. The marginal revenue and marginal cost identities in \eqref{eqn:posted-price-cutoffs} give
    \[
        p_k^S = \frac{\lambda_k}{2} = \frac{1}{10}
        \quad \text{and} \quad
        p_k^B = \frac{1 + \lambda_k}{2} = \frac{3}{5}.
    \]
    Since shadow prices are equal, each customer is assigned to the tag with the higher CTR, which supplies each tag with $(1/2)(2/3) = 1/3$ total expected clicks. On the demand side, at posted prices $p_k^S = 1/10$ and $p_k^B = 3/5$, each tag demands $(1/2)\sbr{1 - 3/5 + (8/15)(3/5 - 1/10)} = 1/3$. Both markets therefore clear, and \cref{thm:optimal-limit-mechanism} certifies $\lambda_1 = \lambda_2 = 1/5$ as the unique virtual prices.

    The optimal mechanism redirects the $40\%$ of customers initially belonging to the lower-CTR tag to the higher-CTR tag, raising total expected clicks from $8/15$ to $2/3$, a $25\%$ increase. This maximizes customer engagement, so the same posted prices remain optimal for $\eta_w > 0$ when $\eta_v = 0$ and $\eta_r > 0$, and only the virtual prices need to adjust.

    If the designer instead weights revenue and merchants' surplus equally $\br{\eta_r = \eta_v = 1/2}$, similar steps give virtual prices $\lambda_1 = \lambda_2 = 4/15$. Customer engagement remains at its maximum, but the exchange market shrinks from one half to one third in each tag: the price paid to sellers rises from $1/10$ to $8/45$, and the price charged to buyers falls from $3/5$ to $23/45$.
\end{example}

\subsection{Proof Sketch}
\label{subsec:proof-sketch-limit}

We provide a proof sketch for \cref{thm:optimal-limit-mechanism}. The key step is a reduced-form characterization of \eqref{eqn:aggregate-feasibility-limit} through linear constraints on $q$, one for each vector of shadow prices $\lambda \in \rbb_+^K$. A pseudo-Lagrangian then bounds the designer's optimal value by the sum of screening and targeting values. The unique market-clearing shadow prices (the virtual prices) attain this bound.

\paragraph{Pooling and Supplying Clicks.} We first characterize the set of aggregate expected clicks feasible under the pooled audience distribution $G$.

Specifically, let $m_k:\sbr{0,1}^K \to \rbb_+$ be the share of customers with CTRs $\xi$ assigned to tag $k$, and write $m = \br{m_k}_{k \in K}$. Each customer can be assigned at most once, so $\sum_{k \in K}m_k\br{\xi} \leq 1$ $dG$-almost everywhere. Then $\E_G\sbr{\xi_km_k\br{\xi}}$ gives the aggregate expected clicks supplied to tag $k$. Since the designer can leave customers unassigned (\textit{free disposal}), we define the feasible set of aggregate expected clicks by
\begin{equation}
    \label{eqn:pooling-feasible-outputs}
    \scal_G
    := \cbr{
        \begin{aligned}
            s \in \rbb_+^K:\;&\text{there exist measurable shares }
            m:\sbr{0,1}^K \to \rbb_+^K\\
            &\text{such that } \sum_{k \in K}m_k\br{\xi} \leq 1
            \quad dG\text{-almost everywhere}\\
            &\hspace{2.5em}\text{and }s_k \leq \E_G\sbr{\xi_km_k\br{\xi}}
            \quad \text{for all }k \in K
        \end{aligned}
    }
\end{equation}
which is compact and convex and has free disposal (\cref{lem:feasible-set}).

\begin{figure}[t!]
    \centering
    \resizebox{0.5\textwidth}{!}{\input{figures/feasible-set-uniform-ctrs.tex}}
    \caption{Feasible Set $\scal_G$ for $K = \{1,2\}$ and $\xi_1,\xi_2\overset{\mathrm{i.i.d.}}{\sim}\unif\sbr{0,1}$.}
    \label{fig:uniform-feasible-set}
\end{figure}
\Cref{fig:uniform-feasible-set} illustrates $\scal_G$ for \cref{ex:certifying}. Initial audiences provide aggregate expected clicks $s^0 := \br{\beta_1\mu_1,\beta_2\mu_2} = \br{4/15,4/15}$ in the interior of $\scal_G$, while the revenue-maximizing mechanism supplies $s^\lambda = \br{1/3,1/3}$ on its frontier at virtual prices $\lambda = \br{1/5,1/5}$.

More generally, the frontier of $\scal_G$ is characterized by its support function $\sigma:\rbb_+^K \to \rbb_+$,
\begin{equation}
    \label{eqn:pooling-support-function}
    \sigma\br{\lambda}
    := \max_{s \in \scal_G}\lambda \cdot s
    = \E_G\sbr{\max_{k \in K}\cbr{\lambda_k\xi_k}}
    \quad \text{for all }\lambda \in \rbb_+^K,
\end{equation}
where the second equality holds because assigning each customer to a tag maximizing $\lambda_k\xi_k$ attains the maximum. At given shadow prices $\lambda$, the support function $\sigma\br{\lambda}$ gives the maximum shadow value of aggregate expected clicks that the pooled audience can supply.

\paragraph{Reduced-Form Feasibility.} We refer to $q = \br{q_k}_{k \in K}$ as an \textit{allocation of expected clicks}, or simply an \textit{allocation}. We next characterize which allocations can be implemented by targeted ads $x = \br{x_k}_{k \in K}$.
\begin{definition}
    \label{def:implementable}
    We say that an allocation $q = \br{q_k}_{k \in K}$ is implementable if there exist targeted ads $x = \br{x_k}_{k \in K}$ that satisfy \eqref{eqn:aggregate-feasibility-limit} and
    \[
        q_k\br{\theta}
        = \int_{\sbr{0,1}^K}\xi_k\,x_k\br{\theta;d\xi}
        \quad \text{for all }\theta \in \Theta_k\text{ and }k \in K.
    \]
\end{definition}
\begin{proposition}[Reduced-Form Feasibility]
    \label{prop:reduced-form-feasibility}
    An allocation $q = \br{q_k}_{k \in K}$ is implementable if and only if
    \begin{equation}
        \label{eqn:R-Fsb}
        \tag{R-Fsb}
        \sum_{k \in K}\lambda_k\beta_k\E_{F_k}\sbr{q_k\br{\theta}}
        \leq \E_G\sbr{\max_{k \in K}\cbr{\lambda_k\xi_k}}
        \quad \text{for all }\lambda \in \rbb_+^K.
    \end{equation}
\end{proposition}

Let $d\br{q} = \br{d_k\br{q}}_{k \in K} \in \rbb_+^K$ denote the aggregate expected clicks induced by $q = \br{q_k}_{k \in K}$, where
\[
    d_k\br{q} := \beta_k\E_{F_k}\sbr{q_k\br{\theta}}
    \quad \text{for all }k \in K.
\]
We prove \cref{prop:reduced-form-feasibility} by first showing that implementability is equivalent to $d\br{q} \in \scal_G$ (\cref{lem:implementable}).

Next, using the strict separating hyperplane theorem we show that
\[
    d\br{q} \in \scal_G
    \quad \Leftrightarrow \quad
    \lambda \cdot d\br{q} \leq \max_{s \in \scal_G}\lambda \cdot s
    \quad \text{for all }\lambda \in \rbb^K.
\]
By free disposal of the feasible set $\scal_G$, we can restrict attention to $\lambda \geq 0$. The right-hand side then equals $\sigma\br{\lambda}$, giving \eqref{eqn:R-Fsb}.

Lastly, note that by applying \cref{prop:reduced-form-feasibility} to baseline expected clicks, a necessary condition for $G$ to be consistent is
\begin{equation}
    \label{eqn:G-consistency}
    \tag{C1}
    \sum_{k \in K}\lambda_k\beta_k\mu_k
    \leq \E_G\sbr{\max_{k \in K}\cbr{\lambda_k\xi_k}}
    \quad \text{for all }\lambda \in \rbb_+^K.
\end{equation}
That is, consistency requires the aggregate expected clicks $\br{\beta_k\mu_k}_{k \in K}$ to lie in $\scal_G$. With a single tag, this reduces to $\E_G\sbr{\xi_1} \geq \mu_1$.

\paragraph{Virtual Objective.} Following similar steps as in the proof of \cref{thm:optimal-finite-mechanism}, we reduce the designer's problem to a max-min problem over allocations and critical types.

Let $\qcal_k$ be the set of nondecreasing functions $q_k:\Theta_k \to \sbr{0,1}$. Define $\qcal := \prod_{k \in K}\qcal_k$ and let $\qcal_R \subseteq \qcal$ consist of allocations $q$ satisfying \eqref{eqn:R-Fsb}. Next define $\Theta_K := \prod_{k \in K}\Theta_k$ and denote a critical type by $\boldtheta^{\crt} = \br{\theta_k^{\crt}}_{k \in K} \in \Theta_K$. Then we show that the designer's problem in \eqref{eqn:designer-problem-limit} is equivalently given by
\begin{equation}
    \label{eqn:limit-reduced-problem}
    \Pi_\eta^\infty
    = \adjustlimits\sup_{q \in \qcal_R}
    \min_{\boldtheta^{\crt} \in \Theta_K}
    \psi_\eta\br{q,\boldtheta^{\crt}},
\end{equation}
where the virtual objective is defined by
\begin{equation}
    \label{eqn:virtual-objective-limit}
    \psi_\eta\br{q,\boldtheta^{\crt}}
    := \sum_{k \in K}\beta_k\E_{F_k}\sbr{
        \br{q_k\br{\theta} - \mu_k}
        \phi_{\eta,k}\br{\theta;\theta_k^{\crt}}
    }.
\end{equation}

\paragraph{Pseudo-Lagrangian and Weak Duality.} Next we incorporate \eqref{eqn:R-Fsb} through the pseudo-Lagrangian
\begin{equation}
    \label{eqn:limit-pseudo-lagrangian}
    \lcal\br{q,\boldtheta^{\crt};\lambda}
    := \sigma\br{\lambda} - \lambda \cdot d\br{q}
    + \psi_\eta\br{q,\boldtheta^{\crt}}
\end{equation}
and show that
\begin{equation}
    \label{eqn:pseudo-lagrangian-equality}
    \Pi_\eta^\infty
    = \adjustthreelimits\sup_{q \in \qcal}
    \min_{\boldtheta^{\crt} \in \Theta_K}
    \inf_{\lambda \geq 0}
    \lcal\br{q,\boldtheta^{\crt};\lambda}
    = \adjustthreelimits\sup_{q \in \qcal}
    \inf_{\lambda \geq 0}
    \min_{\boldtheta^{\crt} \in \Theta_K}
    \lcal\br{q,\boldtheta^{\crt};\lambda}.
\end{equation}
Then, by weak duality we get
\begin{equation}
    \label{eqn:dual-upper-bound}
    \Pi_\eta^\infty
    \leq \inf_{\lambda \geq 0}
    \cbr{\sigma\br{\lambda} + \ell\br{\lambda}},
\end{equation}
where $\ell:\rbb_+^K \to \rbb$ is defined by
\begin{equation}
    \label{eqn:screening-value}
    \ell\br{\lambda}
    := \adjustlimits\sup_{q \in \qcal}\min_{\boldtheta^{\crt} \in \Theta_K}
    \cbr{
        \psi_\eta\br{q,\boldtheta^{\crt}}
        - \lambda \cdot d\br{q}
    }
    \quad \text{for all }\lambda \in \rbb_+^K.
\end{equation}

We refer to $\sigma\br{\lambda}$ as the \textit{targeting value} and to $\ell\br{\lambda}$ as the \textit{screening value} at shadow prices $\lambda$. We next characterize the solution to each problem.

\paragraph{The Screening Problem.} Fix shadow prices $\lambda \in \rbb_+^K$. \eqref{eqn:screening-value} gives the optimal value from screening merchants when clicks can be supplied to each tag $k$ at shadow cost $\lambda_k$. For any fixed allocation $q \in \qcal$, we show that monotonicity implies
\begin{equation}
    \label{eqn:reduced-screening-objective}
    \min_{\boldtheta^{\crt} \in \Theta_K}
    \psi_\eta\br{q,\boldtheta^{\crt}}
    =
    \sum_{k \in K}\beta_k\E_{F_k}\sbr{
        \min\cbr{
            \br{q_k\br{\theta} - \mu_k}\phi_{\eta,k}^S\br{\theta},\,
            \br{q_k\br{\theta} - \mu_k}\phi_{\eta,k}^B\br{\theta}
        }
    }.
\end{equation}
Then, subtracting the shadow cost $\lambda \cdot d\br{q}$ and maximizing pointwise over $q_k\br{\theta} \in \sbr{0,1}$ gives $q_k^\lambda$ (see \cref{lem:screening}), which is defined for all $\theta \in \Theta_k$ and $k \in K$ by
\begin{equation}
    \label{eqn:screening-clicks}
    q_k^\lambda\br{\theta}
    :=
    \begin{cases}
        0
        & \text{if }\phi_{\eta,k}^S\br{\theta} \leq \lambda_k,\\
        \mu_k
        & \text{if }\phi_{\eta,k}^B\br{\theta} \leq \lambda_k < \phi_{\eta,k}^S\br{\theta},\\
        1
        & \text{if }\lambda_k < \phi_{\eta,k}^B\br{\theta}.
    \end{cases}
\end{equation}
Taking expectations with respect to $F_k$ gives the aggregate demand for clicks in tag $k$ (right-hand side of \eqref{eqn:market-clearing-limit}):
\begin{equation}
    \label{eqn:screening-demand}
    d_k\br{q^\lambda}
    = \beta_k\sbr{
        1 - F_k\br{p_k^B\br{\lambda_k}}
        + \mu_k\br{F_k\br{p_k^B\br{\lambda_k}} - F_k\br{p_k^S\br{\lambda_k}}}
    }.
\end{equation}

Regularity of each distribution $F_k$ implies $q_k^\lambda$ is nondecreasing, so no additional ironing is required, unlike in the proof of \cref{thm:optimal-finite-mechanism}. Regularity also makes the pointwise maximizer unique except at the two cutoff types, so $q_k^\lambda$ is the unique maximizer $dF_k$-almost everywhere. Substituting $q^\lambda$ into the virtual objective gives the screening value
\begin{equation}
    \label{eqn:pooling-dual-decomposition}
    \ell\br{\lambda}
    = \sum_{k \in K}\beta_k\E_{F_k}\sbr{
        \max\cbr{
            - \mu_k\phi_{\eta,k}^S\br{\theta},\,
            - \mu_k\lambda_k,\,
            \br{1 - \mu_k}\phi_{\eta,k}^B\br{\theta} - \lambda_k
        }
    }.
\end{equation}

\paragraph{The Targeting Problem.} Fix shadow prices $\lambda \in \rbb_+^K\setminus\cbr{\mathbf{0}}$. The targeting problem is given by \eqref{eqn:pooling-support-function}. We show in \cref{lem:support-function} that the density assumption on $G$ implies that
\begin{equation}
    \label{eqn:optimal-s-frontier}
    s_k^\lambda
    := \E_G\sbr{
        \xi_k\,
        \mathbf{1}\cbr{
            k = \argmax_{j \in K}\lambda_j\xi_j
        }
    }
    \quad \text{for all }k \in K,
\end{equation}
is the unique solution to \eqref{eqn:pooling-support-function}, which gives the supply of clicks on the left-hand side of \eqref{eqn:market-clearing-limit}. The optimal $s^\lambda$ attains $\lambda \cdot s^\lambda = \sigma\br{\lambda}$, so the hyperplane $\cbr{s:\lambda \cdot s = \sigma\br{\lambda}}$ supports $\scal_G$ at $s^\lambda$ (see \cref{fig:uniform-feasible-set}).

\paragraph{Market Clearing and Optimality.} Finally, we minimize the right-hand side $\sigma + \ell$ in \eqref{eqn:dual-upper-bound} and show that the weak duality bound is attained. By \cref{lem:dual-compact}, this minimization can be restricted to the compact set defined by $0 \leq \lambda_k \leq \MC_{\eta,k}\br{\btheta_k}$ for every $k \in K$. The sum is convex and continuous, so it attains a minimum at some $\lambda^*$.

We then show that the first-order conditions at a strictly positive candidate $\lambda$ are
\[
    0
    = \frac{\partial}{\partial\lambda_k}
    \sbr{\sigma\br{\lambda} + \ell\br{\lambda}}
    = s_k^\lambda - \beta_k\E_{F_k}\sbr{q_k^\lambda\br{\theta}}
    \quad \text{for all }k \in K.
\]
These conditions equate supply and demand in every tag and, by convexity, are sufficient for global optimality.

We then show in \cref{lem:market-clearing} that \cref{asm:scarcity-limit} and the density assumption on $G$ imply that every minimizer is strictly positive. Since positive shadow prices generate positive supply in every tag, whereas demand is zero when $\lambda_k \geq \MC_{\eta,k}\br{\btheta_k}$, market clearing also implies $\lambda_k^* < \MC_{\eta,k}\br{\btheta_k}$ for every $k \in K$.

Lastly, market clearing gives $d\br{q^{\lambda^*}} = s^{\lambda^*} \in \scal_G$, so $q^{\lambda^*} \in \qcal_R$ is feasible. Since $s^{\lambda^*}$ attains the targeting value,
\[
    \sigma\br{\lambda^*} - \lambda^* \cdot d\br{q^{\lambda^*}}
    = \sigma\br{\lambda^*} - \lambda^* \cdot s^{\lambda^*}
    = 0.
\]
Therefore,
\[
    \min_{\boldtheta^{\crt} \in \Theta_K}
    \psi_\eta\br{q^{\lambda^*},\boldtheta^{\crt}}
    \leq \Pi_\eta^\infty
    \leq \sigma\br{\lambda^*} + \ell\br{\lambda^*}
    = \min_{\boldtheta^{\crt} \in \Theta_K}
    \psi_\eta\br{q^{\lambda^*},\boldtheta^{\crt}},
\]
where the first inequality follows from the definition of $\Pi_\eta^\infty$ and feasibility of $q^{\lambda^*}$, the second from weak duality, and the equality from the preceding market-clearing identity. Both inequalities therefore hold with equality, proving optimality of the virtual prices $\lambda^*$.

\paragraph{Uniqueness.} Let $q^* \in \qcal_R$ be any other optimal allocation. Feasibility and the definition of the screening problem give
\[
    \begin{aligned}
        \Pi_\eta^\infty
        &= \min_{\boldtheta^{\crt} \in \Theta_K}
        \psi_\eta\br{q^*,\boldtheta^{\crt}}\\
        &\leq \sigma\br{\lambda^*}
        + \min_{\boldtheta^{\crt} \in \Theta_K}
        \psi_\eta\br{q^*,\boldtheta^{\crt}}
        - \lambda^* \cdot d\br{q^*}\\
        &\leq \sigma\br{\lambda^*} + \ell\br{\lambda^*}
        = \Pi_\eta^\infty.
    \end{aligned}
\]
Equality throughout implies that $q^*$ solves the screening problem at $\lambda^*$. Essential uniqueness of its maximizer gives
\[
    q_k^*\br{\theta} = q_k^{\lambda^*}\br{\theta}
    \quad dF_k\text{-almost everywhere,} \quad \text{for all } k \in K.
\]
Since $\eta_r > 0$, every optimal mechanism gives a worst-off type in each tag zero payoff, so by the envelope formula we also pin down transfers $dF_k$-almost everywhere.

Lastly, virtual prices are unique in both relative terms and level. The intuition for relative prices is the usual \textit{gross substitutes} argument: raising $\lambda_k$ increases supply to tag $k$ and weakly reduces its demand, while weakly reducing supply to other tags. Since $G$ has a strictly positive density, supply in \eqref{eqn:optimal-s-frontier} responds strictly to relative prices. In our setting, demand also depends on the level of virtual prices.\footnote{In a Walrasian economy, excess demand is homogeneous of degree zero, so prices are determined only up to a normalization.} Raising all virtual prices proportionately leaves supply unchanged but strictly reduces demand in at least one tag with an interior posted price.

\section{Finite-to-Limit Market Convergence}
\label{sec:limit-market-convergence}

We consider a sequence of finite markets with $n$ merchants, each with a capacity of $1/n$ expected clicks. We assign each merchant mass $1/n$ based on his audience contribution, so that his size relative to the market becomes negligible as $n \longrightarrow \infty$. Dividing expected clicks by merchant mass gives normalized expected clicks $q^n_i\br{\theta_i} := nQ_i\br{\theta}$ and a normalized capacity of one, as in the limit market. Our main result shows that the optimal mechanism in the limit market is the limit of symmetric optimal mechanisms in these finite markets, and that it can be approximated by asymptotically optimal finite-market mechanisms. We introduce the finite markets \textit{with capacity constraints} in \cref{subsec:convergence-model} and take the limit in \cref{subsec:convergence-taking-limit}. \Cref{subsec:convergence-overview} presents a proof overview, and the complete proofs are in \cref{app:symmetric-large-markets}.

\subsection{Large Markets with Capacity}
\label{subsec:convergence-model}

We introduce tags into the finite market model of \cref{subsec:model-finite}, assume symmetric type distributions and audience contributions within each tag, and impose capacity constraints on each merchant's expected clicks. In doing so, we fix the tag primitives $\br{F_k,\mu_k,\beta_k}_{k \in K}$ from the limit market model of \cref{subsec:model-limit}.

Specifically, fix $n \in \mathbb N$ and let the $n$th market consist of a finite set of merchants $N^n := \sqcup_{k \in K}N_k$, partitioned into tags indexed by $K$, where tag $k$ contains $n_k := |N_k|$ merchants, $\sum_{k \in K}n_k = n$, and $\beta_k^n := n_k/n$ is its share of merchants. Types are drawn independently across merchants with $\theta_{i,k}\sim F_k$ for each $i \in N_k$ and $k \in K$. The product type space and joint type distribution are $\Theta^n := \prod_{k \in K}\Theta_k^{n_k}$ and $F^n := \bigotimes_{k \in K}F_k^{\otimes n_k}$.

Our key assumption for obtaining a tractable limit is that the vector of CTRs
\[
    \omega^n := \br{\omega_{i,k}:i \in N_k,\,k \in K}\sim\gamma^n \in \Delta\br{\sbr{0,1}^n}
\]
is a finite restriction of a countably infinite CTR vector $\omega = \br{\omega_{i,k}}_{i \in \mathbb N,k \in K}$ with measure $\gamma \in \Delta\br{\sbr{0,1}^{\mathbb N\times K}}$ that is partially exchangeable.

To formally state the assumption, for each collection of finite tag sets $\br{N_k}_{k \in K}$, let $\pcal$ be the set of permutations $\varsigma = \br{\varsigma_k}_{k \in K}$ such that $\varsigma_k:N_k \to N_k$ permutes merchants' indices within tag $k$.

\begin{assumption}[Partial Exchangeability]
    \label{asm:partial-exchangeability}
    The countably infinite sequence of random variables $\br{\omega_{i,k}}_{i \in \mathbb N,k \in K} \in [0,1]^{\mathbb N \times K}$ is partially exchangeable with respect to the index set $K$. That is, for all finite sets $N_k \subset \mathbb N$ with $k \in K$, the measure $\gamma^n$ is invariant under every permutation $\varsigma \in \pcal$:
    \[
        \br{\omega_{i,k}:i \in N_k,k \in K}
        \,\overset{\mathrm{dist.}}{=} \,
        \br{\omega_{\varsigma_k\br{i},k}:i \in N_k,k \in K}.
    \]
\end{assumption}

Partial exchangeability implies that relabeling merchants within each tag leaves the pooled audience's joint CTR measure unchanged, while allowing CTR measures to differ across tags and CTRs to remain dependent across merchants.

Next, for all $i \in N_k$ and $k \in K$, suppose the audience shares are uniform across merchants, $\alpha_{i,k}^n = 1/n$, and the conditional CTR measures $\gamma^n_{i,k}$ of the initial audiences satisfy
\begin{equation}
    \label{eqn:convergence-ownership}
    \int_{\sbr{0,1}^n}\omega^n_{i,k}\,d\gamma_{i,k}^n\br{\omega^n} = \mu_k.
\end{equation}
Then, the mixture measure of the pooled audience, denoted by $\gamma^n$, is given by
\begin{equation}
    \label{eqn:convergence-mixture-dist}
    \gamma^n\br{B} = \frac{1}{n}\sum_{i \in N_k,k \in K}\gamma_{i,k}^n\br{B}
    \quad \text{for all Borel sets } B \subseteq [0,1]^n.
\end{equation}
The baseline expected clicks of merchant $i$ with tag $k$ are therefore $\mu_k/n$, and his outside option is $\theta_{i,k}\mu_k/n$.

\paragraph{Direct Mechanisms.} Let $\br{x^n,t^n} = \br{\br{x^n_{i,k}},\br{t^n_{i,k}}}$ be a direct mechanism defined as in \cref{subsec:model-finite}. Suppose that, in addition to ex post feasibility constraints, it satisfies capacity constraints on expected clicks: for all $i \in N_k$ and $k \in K$,
\begin{align}
    \label{eqn:convergence-feasibility}
    \tag{$\gamma^n$-Fsb}
    \sum_{i \in N_k,k \in K}x_{i,k}^n\br{\boldtheta;B}
    &\leq \gamma^n\br{B}
    & & \text{for all Borel } B \subseteq \sbr{0,1}^{n}
    \text{ and } \boldtheta \in \Theta^{n},\\
    \label{eqn:convergence-capacity}
    \tag{Cap$^n_{i,k}$}
    q_{i,k}^n\br{\boldtheta} :=
    \int_{[0,1]^n}\omega^n_{i,k}x_{i,k}^n\br{\boldtheta;d\omega^n}
    &\leq 1/n
    & & \text{for all } \boldtheta \in \Theta^{n}.
\end{align}
Thus, \eqref{eqn:convergence-capacity} implies that a merchant with tag $k$ can expand his campaign from its baseline $\mu_k/n$ by at most $\br{1 - \mu_k}/n$ expected clicks. Next, define interim expected clicks and transfers by
\[
    Q_{i,k}^n\br{\theta_{i,k}}
    := \E_{F^n_{-\br{i,k}}}\sbr{q_{i,k}^n\br{\theta_{i,k},\boldtheta_{-\br{i,k}}}}
    \quad \text{and} \quad
    T_{i,k}^n\br{\theta_{i,k}}
    := \E_{F^n_{-\br{i,k}}}\sbr{t_{i,k}^n\br{\theta_{i,k},\boldtheta_{-\br{i,k}}}}.
\]

\paragraph{Symmetric Direct Mechanisms.} A direct mechanism is symmetric within tags if its interim expected clicks and transfers do not depend on a merchant's identity within his tag:
\[
    Q^n_{i,k}\br{\theta} = Q^n_{j,k}\br{\theta}
    \quad \text{and} \quad
    T^n_{i,k}\br{\theta} = T^n_{j,k}\br{\theta}
    \quad \text{for all }\theta \in \Theta_k,\ i,j \in N_k\text{ and }k \in K.
\]
By \cref{asm:partial-exchangeability} and the symmetry of type distributions and outside options within tags, we restrict attention to symmetric direct mechanisms without loss of generality. Since symmetry makes expected clicks and transfers independent of the merchant's identity within each tag, define their normalized counterparts by
\[
    q_k^n\br{\theta} := nQ_{i,k}^n\br{\theta}
    \quad \text{and} \quad
    t_k^n\br{\theta} := nT_{i,k}^n\br{\theta}
    \quad \text{for all }\theta \in \Theta_k,\ i \in N_k\text{ and }k \in K.
\]
We call $q^n = \br{q_k^n}_{k \in K}$ the \textit{normalized interim allocation}. Since each merchant contributes an audience of mass $1/n$, we interpret $1/n$ as the merchant's size in the market and refer to $q_k^n\br{\theta}$ as expected clicks per unit of merchant mass. Taking expectations over other merchants' types in \eqref{eqn:convergence-capacity} gives
\[
    q_k^n\br{\theta} \leq 1
    \quad \text{for all }\theta \in \Theta_k\text{ and }k \in K,
\]
as in the limit market. Using this normalization, the interim IC and IR constraints are equivalent to the following \textit{normalized} constraints: for all $k \in K$,
\begin{align}
    \theta\br{q_k^n\br{\theta} - \mu_k} - t_k^n\br{\theta}
    &\geq \theta\br{q_k^n\br{\widetilde\theta} - \mu_k} - t_k^n\br{\widetilde\theta}
    & & \text{for all } \theta,\widetilde\theta \in \Theta_k,
    \label{eqn:convergence-IC}
    \tag{IC$^n_k$}\\
    \theta\br{q_k^n\br{\theta} - \mu_k} - t_k^n\br{\theta} &\geq 0
    & & \text{for all } \theta \in \Theta_k,
    \label{eqn:convergence-IR}
    \tag{IR$^n_k$}
\end{align}
which take the same form as \eqref{eqn:ICk} and \eqref{eqn:IRk}.

\paragraph{Designer's Problem.} Restricting attention to symmetric direct mechanisms, the designer's problem in the $n$th market is
\begin{equation}
    \label{eqn:designer-problem-finite-n}
    \tag{DP-F$^n$}
    \begin{aligned}
        \sup_{\br{x^n,t^n}} \quad & \Psi_\eta^n\br{x^n,t^n}\\
        \text{subject to} \quad
        & \eqref{eqn:convergence-feasibility},\
        \eqref{eqn:convergence-capacity},\
        \eqref{eqn:convergence-IC},\
        \text{and }\eqref{eqn:convergence-IR} \quad \text{for all } i \in N_k \text{ and } k \in K,
    \end{aligned}
\end{equation}
where, by the law of iterated expectations, the objective can be written in normalized form
\[
    \begin{aligned}
        \Psi_\eta^n\br{x^n,t^n}
        &:= \sum_{i \in N_k,\,k \in K}\E_{F_k}\sbr{
            \eta_rT^n_{i,k}\br{\theta}
            + \br{\eta_v\theta + \eta_w}\br{Q^n_{i,k}\br{\theta} - \mu_k/n}}\\
        &= \sum_{k \in K}\beta_k^n\E_{F_k}\sbr{
            \eta_rt_k^n\br{\theta}
            + \br{\eta_v\theta + \eta_w}\br{q_k^n\br{\theta} - \mu_k}}.
    \end{aligned}
\]
It has the same form as the objective in \eqref{eqn:designer-problem-limit}, with tag shares $\beta_k^n$ and normalized expected clicks $q_k^n$ and expected transfers $t_k^n$. Let $\Pi_\eta^n$ denote the optimal value of \eqref{eqn:designer-problem-finite-n}.

\subsection{Taking the Limit}
\label{subsec:convergence-taking-limit}

The following de Finetti representation for partially exchangeable sequences follows from \citet[Corollary~1.7, p.~30]{kallenberg2005symmetries}.

\begin{proposition}[De Finetti Representation]
    \label{prop:de-finetti}
    Under \cref{asm:partial-exchangeability}, there exists a vector of random probability measures $\bP = \br{P_k}_{k \in K} \in \br{\Delta\br{\sbr{0,1}}}^K$, with joint measure $\Gamma$, such that, conditional on $\bP$, CTRs $\omega_{i,k}$ are independent across all $i \in \mathbb N$ and $k \in K$, and identically distributed for fixed $k \in K$ according to $P_k$:
    \[
        \br{\omega_{i,k}:i \in \mathbb N \mid \bP}
        \,\overset{\mathrm{i.i.d.}}{\sim}\,
        P_k
        \quad \text{for all } k \in K, \quad \Gamma\text{-almost everywhere}.
    \]
\end{proposition}

As the number of merchants with tag $k$ grows, the highest CTR among them approaches the upper endpoint of the support of $P_k$. For a given realization $\bP$, define the conditional CTR \textit{frontiers} $\xi\br{\bP} = \br{\xi_k\br{P_k}}_{k \in K} \in \sbr{0,1}^K$ and their measure $\xi\sim G \in \Delta\br{\sbr{0,1}^K}$ by
\begin{equation}
    \label{eqn:convergence-frontier}
    \xi_k\br{P_k} := \esssup_{\omega_{i,k} \sim P_k}\omega_{i,k}
    \quad \text{and} \quad G := \Gamma\circ\xi^{-1}.
\end{equation}
We show that these frontiers are the limit market's tag-specific CTRs, and $G$ is its pooled audience distribution.

Since $\omega_{i,k} \leq \xi_k$ $\gamma$-almost everywhere for all $i \in \mathbb N$ and $k \in K$, our first observation is that $G$ is consistent, i.e., it satisfies \eqref{eqn:G-consistency}.

\begin{proposition}[Consistency]
    \label{prop:consistency}
    Let $n \longrightarrow \infty$ with $\beta_k^n \longrightarrow \beta_k$ for all $k \in K$. If the initial and pooled audience measures satisfy \eqref{eqn:convergence-ownership} and \eqref{eqn:convergence-mixture-dist} for all $n$, then $G$ satisfies \eqref{eqn:G-consistency}.
\end{proposition}

The limit market's assumption that $G$ has a density strictly positive on $\br{0,1}^K$ is an additional restriction on the CTR frontiers that partial exchangeability alone does not imply it. Accordingly, Part~I) of the convergence result (\cref{thm:finite-limit-convergence}) holds for general distributions $G$, while Part~II) invokes the assumptions of \cref{thm:optimal-limit-mechanism}.

Lastly, for each $n \in \mathbb N$, let $\xcal^n$ denote the set of targeted ads satisfying \eqref{eqn:convergence-feasibility} and \eqref{eqn:convergence-capacity} whose interim allocations are symmetric within each tag. Using $\qcal$ from \cref{subsec:proof-sketch-limit}, define the set of feasible and nondecreasing normalized interim allocations in the $n$th market by
\[
    \qcal_{\gamma^n} := \cbr{q \in \qcal:q\text{ is induced by some }x^n \in \xcal^n},
\]
where its limit-market counterpart is $\qcal_R$. We show that these symmetric finite markets converge, in terms of feasible normalized interim allocations $\qcal_{\gamma^n}$ and optimal values $\Pi_\eta^n$, to the limit market's feasible allocations $\qcal_R$ and optimal value $\Pi_\eta^\infty$, and that optimal normalized interim allocations and transfers converge to their essentially unique limit-market counterparts.

\begin{theorem}[Finite-to-Limit Market Convergence]
    \label{thm:finite-limit-convergence}
    Fix limit-market primitives $\br{F_k,\mu_k,\beta_k}_{k \in K}$ and pooled audience distribution $G$ defined by \eqref{eqn:convergence-frontier}, and suppose $\beta_k^n \longrightarrow \beta_k > 0$ for all $k \in K$.
    \begin{enumerate}[label=\Roman*)]
        \item For any sequence of normalized interim allocations $\br{q^n}_{n \geq 1}$ in $\qcal_{\gamma^n}$, there exists a subsequence $\br{q^{n_j}}_{j \geq 1}$ and an allocation $q^\infty \in \qcal_R$ such that, for all $k \in K$,
        \[
            q_k^{n_j}\br{\theta} \longrightarrow q_k^\infty\br{\theta}
            \quad dF_k\text{-almost everywhere.}
        \]
        Conversely, for any allocation $q \in \qcal_R$ there exists a sequence of normalized interim allocations $\br{q^n}_{n \geq 1}$ in $\qcal_{\gamma^n}$ such that, for all $k \in K$,
        \[
            q_k^n\br{\theta} \longrightarrow q_k\br{\theta}
            \quad dF_k\text{-almost everywhere.}
        \]
        \item Under the assumptions of \cref{thm:optimal-limit-mechanism}, let $q^*$ and $t^*$ denote the essentially unique optimal allocation and transfers in the limit market. There exists a sequence of symmetric incentive-compatible and individually rational mechanisms $\br{x^n,t^n}$, with $x^n \in \xcal^n$, such that
        \[
            \Pi_\eta^n \longrightarrow \Pi_\eta^\infty
            \quad \text{and} \quad
            \Pi_\eta^n - \Psi_\eta^n\br{x^n,t^n} \longrightarrow 0.
        \]
        Their normalized interim allocations and transfers satisfy, for every $k \in K$,
        \begin{equation}
            \label{eqn:convergence-optimal-allocations-transfers}
            \br{q_k^n\br{\theta},t_k^n\br{\theta}}
            \longrightarrow \br{q_k^*\br{\theta},t_k^*\br{\theta}}
            \quad dF_k\text{-almost everywhere}.
        \end{equation}
        The same convergence holds for every sequence of symmetric optimal finite-market mechanisms.
    \end{enumerate}
\end{theorem}

Part~I) gives both directions of the approximation: limits of feasible finite-market interim allocations are feasible in the limit market, and every feasible limit allocation can be approximated by finite-market interim allocations. Part~II) gives the main takeaway from \cref{thm:finite-limit-convergence}: finite approximations of the optimal limit mechanism are asymptotically optimal in large finite markets, and every sequence of symmetric optimal finite-market mechanisms converges to the optimal limit mechanism in normalized interim allocations and transfers.

\subsection{Proof Overview}
\label{subsec:convergence-overview}

We now give a proof overview of \cref{thm:finite-limit-convergence}.

\paragraph{Finite-to-Limit Feasibility.} Since $\omega_{i,k} \leq \xi_k$ $\gamma$-almost everywhere, taking expectations in \eqref{eqn:convergence-feasibility} gives, for every $q^n \in \qcal_{\gamma^n}$ and $\lambda \in \rbb_+^K$,
\[
    \sum_{k \in K}\lambda_k\beta_k^n\E_{F_k}\sbr{q_k^n\br{\theta}}
    \leq \E_G\sbr{\max_{k \in K}\lambda_k\xi_k},
\]
which gives \eqref{eqn:R-Fsb} but with $\beta_k^n$ in place of $\beta_k$. Because each $q_k^n$ is nondecreasing with values in $\sbr{0,1}$, Helly's selection theorem gives a subsequence converging $dF_k$-almost everywhere, and passing to the limit in the inequality shows that the limit lies in $\qcal_R$.

\paragraph{Limit-to-Finite Approximation.} For the converse, fix $q \in \qcal_R$ with limit audience shares $m$ from \cref{lem:implementable}. Since the finite-market designer only observes $\omega^n$ but not the frontier $\xi$, we use the feasible \textit{tentative} audience shares
\[
    m_k^n\br{\omega^n} := \E_\gamma\sbr{m_k\br{\xi} \mid \omega^n}.
\]
Within tag $k$, we then make each merchant $i$ eligible independently with probability $q_k\br{\theta_{i,k}}$, assign the tentative share to the eligible merchant with the highest CTR, breaking ties uniformly, and scale his targeted ads down whenever his expected clicks exceed his capacity $1/n$. By \cref{prop:de-finetti}, conditional on $\bP$ the CTRs within a tag are i.i.d., so the highest CTR among the eligible merchants converges to $\xi_k$ as their number grows. For tags with positive demand, the total tentative clicks and the mass of eligible merchants therefore both converge to $d_k\br{q}$, and by symmetry each eligible merchant's normalized tentative clicks approach one.

To state this formally, let $b^n_{i,k}$ denote the number of other eligible merchants in tag $k$ conditional on merchant $i$ being eligible. It is independent of merchant $i$'s report and is distributed as $1 + \Binomial\br{n_k - 1,\E_{F_k}\sbr{q_k\br{\theta}}}$. For $h$ eligible merchants, let $s_k^n\br{h}$ denote their total tentative expected clicks divided by their total capacity $h/n$. After imposing the capacity constraint, an eligible merchant therefore receives $\min\cbr{1,s_k^n\br{b^n_{i,k}}}$ normalized expected clicks. We show that the normalized interim allocation satisfies
\[
    q_k^n\br{\theta}
    = \E\sbr{\min\cbr{1,s_k^n\br{b^n_{i,k}}}}\,q_k\br{\theta}
    \longrightarrow q_k\br{\theta},
\]
where the expectation is over $b^n_{i,k}$ (see \cref{prop:limit-to-finite}).

\paragraph{Upper and Lower Bounds on Optimal Values.} For Part~II), \cref{lem:objective-convergence} shows that if $q_k^n \longrightarrow q_k^\infty$ $dF_k$-almost everywhere for every $k \in K$, the designer's reduced objective converges to
\[
    \min_{\boldtheta^{\crt} \in \Theta_K}\psi_\eta^n\br{q^n,\boldtheta^{\crt}}
    \longrightarrow
    \min_{\boldtheta^{\crt} \in \Theta_K}\psi_\eta\br{q^\infty,\boldtheta^{\crt}}.
\]
Then, by Part~I), finite-market allocations whose values approach $\Pi_\eta^n$ have a subsequence converging to a feasible limit allocation, whose value is at most $\Pi_\eta^\infty$. Hence, $\limsup_n\Pi_\eta^n \leq \Pi_\eta^\infty$.

For the lower bound, we apply the construction above to the optimal allocation $q^*$ and set transfers whose normalized interim values are
\[
    t_k^n\br{\theta}
    = \E\sbr{\min\cbr{1,s_k^n\br{b^n_{i,k}}}}\,t_k^*\br{\theta}
    - \br{1 - \E\sbr{\min\cbr{1,s_k^n\br{b^n_{i,k}}}}}\mu_k\btheta_k.
\]
Scaling his clicks and transfers by the same factor preserves incentive compatibility, and compensating him at $\btheta_k$ per forgone baseline click preserves individual rationality. We then show that the value of these mechanisms converges to $\Pi_\eta^\infty$, so $\liminf_n\Pi_\eta^n \geq \Pi_\eta^\infty$.

\paragraph{Convergence of Optimal Mechanisms.} Every limit point of symmetric optimal finite-market allocations is feasible by Part~I) and attains $\Pi_\eta^\infty$ by convergence of optimal values and \cref{lem:objective-convergence}, so essential uniqueness in \cref{thm:optimal-limit-mechanism} identifies it with $q^*$. Since all limit points coincide and allocations are nondecreasing, the full sequence converges to $q^*$ $dF_k$-almost everywhere. The transfers then converge as well, since the envelope formula pins them down from the allocations once the worst-off type's payoff is zero.

\section{Efficiency and Sequential Targeting}
\label{sec:efficiency}

In \cref{subsec:efficiency-limit}, we characterize efficient platform design in the limit market and show when it can be achieved without an outside subsidy. In \cref{subsec:sequential-targeting}, we extend the framework to sequential targeting. The complete proofs are in \cref{app:extensions}.

\subsection{Efficiency}
\label{subsec:efficiency-limit}

We consider the problem of a designer who maximizes a weighted sum of merchants' surplus and customer engagement. That is, take welfare weights $\eta = \br{\eta_r,\eta_v,\eta_w}$ with $\eta_r = 0$ and $\eta_v > 0$.

\subsubsection{Efficient Targeting}

Since the reduced-form feasibility characterization in \cref{prop:reduced-form-feasibility} does not depend on welfare weights, expected clicks $q^E = \br{q^E_k}_{k \in K}$ are \textit{efficient} if they solve
\begin{equation}
    \label{eqn:designer-problem-efficiency}
    \tag{DP-E}
    \begin{aligned}
        \sup_{q} \quad & \sum_{k \in K}\beta_k\E_{F_k}\sbr{\br{\eta_v\theta + \eta_w}\br{q_k\br{\theta} - \mu_k}}\\
        \text{subject to} \quad & \eqref{eqn:R-Fsb}\ \text{and }\eqref{eqn:Ck}\ \text{for all }k \in K,
    \end{aligned}
\end{equation}
and let $\Pi_\eta^E$ denote the optimal value.

Within a tag, assigning clicks from one merchant to another who values them more raises merchants' surplus without changing customer engagement. An efficient allocation therefore eliminates the exchange market.

\begin{lemma}[Efficient Expected Clicks]
    \label{lem:efficient-clicks}
    Suppose $\eta_r = 0$ and $\eta_v > 0$. An efficient allocation exists, and every efficient allocation gives the same expected clicks $dF_k$-almost everywhere in each tag: for each $k \in K$ there exists a cutoff type $p_k^E$ such that
    \begin{equation}
        \label{eqn:efficient-clicks-q-E}
        q_k^E\br{\theta}
        =
        \begin{cases}
            0 & \text{if } \theta \leq p_k^E,\\
            1 & \text{if } \theta > p_k^E,
        \end{cases}
        \quad dF_k\text{-almost everywhere.}
    \end{equation}
    There exists a unique vector of tag-specific \emph{efficient} virtual prices $\lambda^E = \br{\lambda_k^E}_{k \in K}$, with $0 < \lambda_k^E < \eta_v\btheta_k + \eta_w$ for every $k \in K$, that pins down the posted prices through
    \begin{equation}
        \label{eqn:efficient-prices}
        p_k^E
        =
        \min\cbr{
            \overline{\theta}_k,\,
            \max\cbr{
                \underline{\theta}_k,\,
                \frac{\lambda_k^E - \eta_w}{\eta_v}
            }
        },
    \end{equation}
    and satisfies the market-clearing conditions
    \begin{equation}
        \label{eqn:market-clearing-efficiency}
        \tag{S$^E_k{=}$D$^E_k$}
        \E_G\sbr{
            \xi_k
            \mathbf{1}\cbr{
                k = \argmax_{j \in K}\lambda_j^E\xi_j
            }
        }
        = \beta_k\sbr{1 - F_k\br{p_k^E}} \quad \text{for all } k \in K.
    \end{equation}
\end{lemma}

Since $\br{\lambda_k^E - \eta_w} / \eta_v < \btheta_k$, we have $p_k^E < \btheta_k$, so each tag contains a positive mass of buyers.

\subsubsection{Efficient Platform Design}

\begin{definition}[Efficient Platform Design]
    \label{def:efficient-design}
    The efficient platform design can be achieved without an outside subsidy if there exists a (direct) mechanism $\br{x,t}$ that is incentive compatible and individually rational, where targeted ads $x$ induce efficient expected clicks $q^E$ as in \eqref{eqn:efficient-clicks-q-E} and transfers $t$ satisfy the \emph{no-deficit} constraint:
    \begin{equation}
        \label{eqn:no-deficit}
        \sum_{k \in K}\beta_k \E_{F_k}\sbr{t_k\br{\theta}} \geq 0.
        \tag{ND}
    \end{equation}
    We say that the platform design is \textit{merchant-efficient} when $\eta_v = 1$ and \textit{customer- and merchant-efficient} when $0 < \eta_v < 1$, so that $\eta_w = 1 - \eta_v > 0$.
\end{definition}

The no-deficit condition \eqref{eqn:no-deficit} need not hold separately in every tag: deficits in some tags may be financed by surpluses in others, provided that the aggregate net payments are nonnegative.

The cutoff type $p_k^E$ is a worst-off type, since lower types forgo their baseline clicks and higher types expand to capacity. Setting his rent to zero and applying the envelope theorem therefore gives the largest net payments consistent with incentive compatibility and individual rationality. Each seller receives $\mu_kp_k^E$ for his forgone baseline clicks, while each buyer pays $\br{1 - \mu_k}p_k^E$ to expand to capacity. After multiplying by the respective masses of buyers and sellers, net payments from tag $k$ are
\[
    \beta_k\br{1 - \mu_k}p_k^E\sbr{1 - F_k\br{p_k^E}}
    - \beta_k\mu_kp_k^E F_k\br{p_k^E} = \beta_kp_k^E\sbr{1 - \mu_k - F_k\br{p_k^E}}.
\]
This gives the following characterization.

\begin{proposition}[Customer- and Merchant-Efficient Platform Design]
    \label{prop:cm-efficient-design}
    Suppose $\eta_r = 0$ and $\eta_v \in \br{0,1}$. Then, the customer- and merchant-efficient platform design can be achieved without an outside subsidy if and only if
    \begin{equation}
        \label{eqn:efficiency-no-deficit-condition}
        \sum_{k \in K} \beta_k p_k^E \sbr{1 - \mu_k - F_k\br{p_k^E}} \geq 0,
    \end{equation}
    where the posted prices $p_k^E$ are given by \eqref{eqn:efficient-prices}.
\end{proposition}

Under customer- and merchant-efficient design, the efficient virtual price at an interior cutoff is $\lambda_k^E = \eta_vp_k^E + \eta_w$. Although $\eta_w$ is common across tags, its contribution to a customer's score is $\eta_w\xi_k$, which is larger for tags in which she has higher CTRs. The efficient allocation may therefore assign a customer to tag $k$ rather than tag $j$ because $\xi_k > \xi_j$, even when $p_k^E < p_j^E$. A positive weight on customer engagement may therefore favor a tag with a higher CTR even when its cutoff type is lower, so its buyers pay less per additional click. Buyers' total payments may then fall short of sellers' compensation, so the customer- and merchant-efficient platform design requires an outside subsidy when \eqref{eqn:efficiency-no-deficit-condition} fails.

Under merchant-efficient design, by contrast, $\lambda_k^E = p_k^E$ at an interior cutoff. When all cutoff types are interior, the efficient allocation therefore assigns each customer to the tag $k$ with the highest $\xi_kp_k^E$. The targeting and screening problems are thus aligned: each customer is assigned where her click is worth most to the cutoff types, who buy and sell clicks at exactly their value. This yields the following possibility result.

\begin{proposition}[Merchant-Efficient Platform Design]
    \label{prop:m-efficient-design}
    Suppose $\eta_v = 1$. Then, the merchant-efficient platform design can be achieved without an outside subsidy.
\end{proposition}

As merchants become negligible in large markets, the empirical distribution of types converges to its population distribution, and aggregate uncertainty about demand disappears. Efficient virtual prices then decentralize the efficient allocation, coordinating posted prices within tags with targeting across tags, much as competitive prices do in the Debreu--Scarf limit \citep{debreu1963limit}. The efficiency result also parallels \citet[Proposition~4]{cgk}, who show that \textit{any} partnership can be dissolved efficiently after sufficiently many replications, regardless of initial ownership shares, where replication adds copies of each partner and divides his share equally among them. Our large market plays the role of their replicated economy: they study the reallocation of ownership of a single asset, whereas we study the reallocation of advertising access to heterogeneous customers.

\subsection{Sequential Targeting}
\label{subsec:sequential-targeting}

We conclude the paper with an extension in which a customer who does not click the first ad may be shown a second from a different tag. Targeting stops after a click or the second ad. We restrict analysis to at most two ads, which capture the main economic trade-offs introduced by sequential targeting. The analysis to more than two ads extends readily. Lastly, we keep the primitives of the limit market with $K \geq 3$ and $\eta_r > 0$.

Consider a customer with CTRs $\xi = \br{\xi_k}_{k \in K}$. Conditional on $\xi$, the probability of clicking on tag $j$ after not clicking on tag $k$ remains $\xi_j$. Targeting the customer with an ad from tag $k$ first and, if she does not click, with an ad from tag $j$ therefore generates $\xi_k$ expected clicks for tag $k$ and $\br{1 - \xi_k}\xi_j$ for tag $j$. For fixed shadow prices $\lambda \in \rbb_+^K$, we show that the targeting value under sequential targeting is given by
\begin{equation}
    \label{eqn:sequential-support}
    \sigma^{(2)}\br{\lambda} := \E_G\sbr{\max_{k \neq j \in K}\cbr{\lambda_k\xi_k + \br{1 - \xi_k}\lambda_j\xi_j}},
\end{equation}
while the screening value remains $\ell\br{\lambda}$ as defined in \eqref{eqn:screening-value}. We obtain the following extension of \cref{thm:optimal-limit-mechanism}.

\begin{proposition}[Optimal Sequential Targeting]
    \label{prop:sequential-targeting}
    There exists a vector of virtual prices $\lambda^{(2)} \in \rbb_{++}^K$ such that the following mechanism is optimal. In each tag, it offers the selling, exchange, and buying menu of \cref{thm:optimal-limit-mechanism}, with posted prices given by \eqref{eqn:posted-price-cutoffs} at $\lambda_k = \lambda_k^{(2)}$. For $G$-almost every $\xi \in \br{0,1}^K$, it selects a tag $r$ with the highest score,
    \begin{equation}
        \label{eqn:sequential-first-tag}
        r \in \argmax_{k \in K}\lambda_k^{(2)}\xi_k,
    \end{equation}
    and a tag $s \neq r$ satisfying
    \begin{equation}
        \label{eqn:sequential-second-tag}
        s \in \argmax_{j \neq r}\cbr{\lambda_j^{(2)}\xi_j - \min\cbr{\lambda_r^{(2)},\lambda_j^{(2)}}\xi_r\xi_j}.
    \end{equation}
    Lastly, the optimal targeting rule shows the ad from the tag with the higher virtual price first and the other if the customer does not click, with ties resolved such that markets clear.
\end{proposition}

The optimal targeting rule first selects a maximizing pair of tags $\cbr{r,s}$, and then determines the order in which to show their ads. The customer's CTRs affect the pair, whereas the order depends only on the virtual prices.

To understand the optimal ordering, fix shadow prices $\lambda \in \rbb_+^K$ and any two tags $k \neq j$. Showing tag $k$ first yields $\lambda_k\xi_k + \br{1 - \xi_k}\lambda_j\xi_j$, and showing tag $j$ first yields $\lambda_j\xi_j + \br{1 - \xi_j}\lambda_k\xi_k$. Because targeting stops after a click, the two orders differ only in the event that she would click on both ads if shown separately, which has probability $\xi_k\xi_j$. In that event, the click goes to whichever tag is shown first, so showing tag $k$ first rather than $j$ changes the value by $\br{\lambda_k - \lambda_j}\xi_k\xi_j$. The designer therefore gives priority to the tag with the higher shadow price, regardless of which ad the customer is more likely to click on its own, and so the pair $\cbr{k,j}$ has targeting value
\begin{equation}
    \label{eqn:sequential-pair-value}
    \lambda_k\xi_k + \lambda_j\xi_j - \min\cbr{\lambda_k,\lambda_j}\xi_k\xi_j.
\end{equation}

To see \eqref{eqn:sequential-first-tag}, note that some maximizing pair contains the tag $r$ with the highest score $\lambda_r\xi_r$: if a mechanism shows some tag $k$ followed by tag $j$ and omits $r$, replacing the second ad with tag $r$'s changes the value by $\br{1 - \xi_k}\br{\lambda_r\xi_r - \lambda_j\xi_j} \geq 0$. To see \eqref{eqn:sequential-second-tag}, fix $r$ and choose the second tag $j \neq r$ to maximize the value \eqref{eqn:sequential-pair-value} with $k = r$. The first term $\lambda_r\xi_r$ does not depend on $j$, so the second tag $s$ maximizes $\lambda_j\xi_j - \min\cbr{\lambda_r,\lambda_j}\xi_r\xi_j$.

Sequential targeting connects our analysis to models of sponsored search in which consumers scan ads in order and stop once an ad meets their need. \citet{athey2011position} endogenize the value of positions from consumers who search optimally from the top down and stop when a click succeeds, so an ad's expected clicks depend on the ads above it. In the cascade models of \citet{kempe2008cascade} and \citet{aggarwal2008sponsored}, each ad is clicked and scanned past with ad-specific probabilities, and the allocation problem is to order ads on a page given these externalities.

Our setting shares these externalities, with two important differences. First, CTRs are customer-specific rather than ad-specific, so the optimal pair of ads varies across customers. Second, the order within the pair depends only on the value of a click to each tag, not on CTRs. In cascade models, by contrast, an ad with a high value but a low continuation probability may be shown later, because a customer who stops scanning is lost to all later ads. Here, the platform targets customers directly, so the opportunity to show a second ad is forgone only if the first, the more valuable one, is clicked.

\section{Conclusions}
\label{sec:conclusion}

We study how platforms can incentivize merchants to pool their audiences despite concerns about customer poaching. Pooling audiences can improve targeting, but it also exposes merchants to customer poaching, and the platform must resolve this trade-off to incentivize voluntary pooling.

We characterize the optimal mechanism in the limit market through posted prices and show that its finite-market approximations are asymptotically optimal. The posted-price mechanism offers merchants in each observable tag three options: receive, at no charge, a campaign that delivers their baseline clicks; expand it by paying the buying price per additional click; or forgo it and receive the selling price per baseline click. By targeting the customers most likely to click, the platform can deliver baseline clicks with fewer customers and sell the additional clicks to merchants who expand their campaigns. Unique tag-specific virtual prices pin down both the supply and demand of clicks in each tag: within each tag, the virtual price determines the posted selling and buying prices, and across tags, virtual prices determine which customers are assigned to each tag through the scores $\lambda_k\xi_k$.

We also characterize when the efficient platform design can be achieved without an outside subsidy. This is always possible under merchant-efficient design, and we give a necessary and sufficient condition for customer- and merchant-efficient design. Lastly, we extend the analysis to sequential targeting, in which a customer who does not click may be targeted by a merchant from another tag. A customer's CTRs determine which pair of tags she sees, while virtual prices alone determine their order, so a merchant from the tag with the higher virtual price targets her first.

A key implication of our findings is that merchants are valuable to the platform both as buyers of ads and as contributors to the pooled audience. Merchants with low values per click can still supply valuable targeting access, while others can contribute their audiences in exchange for campaigns that deliver their baseline clicks. In practice, pooled audience programs can therefore incentivize voluntary pooling by promising merchants their baseline clicks rather than exclusive access to their audiences, and by compensating merchants for the audiences they contribute even when they are not interested in advertising in a given period.

\bibliography{references}
\bibliographystyle{ecta}

\clearpage
\appendix

\begin{center}
    \Large\bfseries Appendix
\end{center}

\addcontentsline{toc}{section}{Appendix}

\section{Proofs for Finite Market}
\label[appendix]{app:finite}

\subsection{Customer Surplus and Expected Clicks}
\label[appendix]{app:customer-surplus}

Consider the microfoundation in \cref{subsec:discussion-assumptions}, with fixed, publicly observed prices $\mathbf{p} = \br{p_i}_{i \in N}$. For each merchant $i$, assume that willingness to pay is nonnegative, $H_i$ has finite mean, and $H_i\br{p_i} < 1$. The expected customer surplus per click from merchant $i$ is
\[
    S_i\br{p_i}
    := \E_{H_i}\sbr{v - p_i \mid v > p_i}
    = \frac{\int_{p_i}^\infty\br{1 - H_i\br{v}}\,dv}{1 - H_i\br{p_i}},
\]
which is finite and strictly positive.

Now, a customer is interested in merchant $i$'s product with probability $\widetilde\omega_i$, so expected surplus from showing his ad is
\[
    \widetilde\omega_i\E_{H_i}\sbr{\br{v - p_i}_+}
    = \widetilde\omega_i\br{1 - H_i\br{p_i}}S_i\br{p_i}
    = \omega_i S_i\br{p_i}.
\]
Thus, total expected customer surplus from targeted ads $x$ is given by
\[
    \begin{aligned}
        \mathrm{CS}\br{\mathbf{p},x}
        &=
        \E_F\sbr{
            \sum_{i \in N}S_i\br{p_i}
            \br{
                \int_\Omega \omega_i\,x_i\br{\boldtheta;d\omega}
                - a_i
            }
        }
        =
        \sum_{i \in N}S_i\br{p_i}
        \br{\E_{F_i}\sbr{Q_i\br{\theta_i}} - a_i}.
    \end{aligned}
\]

\subsection{Proof of \texorpdfstring{\cref{thm:optimal-finite-mechanism}}{Theorem \ref*{thm:optimal-finite-mechanism}}}
\label{proof:main-theorem}

The following arguments apply both to finite-support and continuously distributed CTRs. The density and audience-scarcity assumptions are used only in the proof of \cref{prop:ironing-system} below.

\paragraph{Characterizing \eqref{eqn:finite-IC}.} Define (interim) payoffs $U_i\br{\theta_i}$ by
\[
    U_i\br{\theta_i}
    := \theta_i\br{Q_i\br{\theta_i} - a_i} - T_i\br{\theta_i}.
\]
By the envelope theorem, \eqref{eqn:finite-IC} holds if and only if
\begin{align}
    U_i\br{\theta_i}
    &=
    U_i\br{\theta_i^{\crt}}
    + \int_{\theta_i^{\crt}}^{\theta_i}
    \br{Q_i\br{y} - a_i}\,dy
    \quad \text{for all }\theta_i,\theta_i^{\crt} \in \Theta_i,
    \notag\\
    Q_i\br{\theta_i}
    &\text{ is nondecreasing in }\theta_i.
    \label{eqn:monotonicity}
    \tag{M}
\end{align}
Hence, for any critical type $\theta_i^{\crt}$, transfers are pinned down by
\begin{equation}
    \label{eqn:transfers}
    T_i\br{\theta_i}
    = \theta_i\br{Q_i\br{\theta_i} - a_i}
    - \int_{\theta_i^{\crt}}^{\theta_i}\br{Q_i\br{y} - a_i}\,dy
    - U_i\br{\theta_i^{\crt}}.
\end{equation}

\begin{definition}[Worst-Off Types]
    \label{def:worst-off-types}
    We say that $\htheta_i \in \Theta_i$ is a \textit{worst-off type} of merchant $i \in N$ for a given direct mechanism $\br{x,t}$ if $U_i\br{\htheta_i} \leq U_i\br{\theta_i}$ for all $\theta_i \in \Theta_i$. Let $\hatTheta_i\br{x,t}$ denote the set of worst-off types, and set $\hatTheta\br{x,t} := \mathlarger{\prod}_{i \in N}\hatTheta_i\br{x,t}$.
\end{definition}

\paragraph{Characterizing Worst-Off Types \`{a} la \citet{cgk}.} By the envelope theorem, $U_i'\br{\theta_i} = Q_i\br{\theta_i} - a_i$ almost everywhere. Together with absolute continuity of $U_i$, \eqref{eqn:monotonicity} therefore implies that $U_i$ is convex. The characterization in \citet{cgk} then applies in terms of expected clicks: $\htheta_i \in \hatTheta_i\br{x,t}$ if and only if $Q_i\br{y} \leq a_i$ for all $y < \htheta_i$ and $Q_i\br{y} \geq a_i$ for all $y > \htheta_i$.

\paragraph{Virtual Objective and IR.} With $\psi_\eta$ as in \cref{subsec:proof-overview-finite}, substituting \eqref{eqn:transfers} into the designer's objective and integrating by parts gives
\begin{equation}
    \label{eqn:virtual-identity}
    \Psi_\eta\br{x,t}
    = \psi_\eta\br{x,\boldtheta^{\crt}}
    - \eta_r\sum_{i \in N}U_i\br{\theta_i^{\crt}}
    \quad \text{for all }\boldtheta^{\crt} \in \Theta.
\end{equation}
The left-hand side does not depend on $\boldtheta^{\crt}$. Since $\eta_r > 0$, the minimizers of $\psi_\eta\br{x,\cdot}$ are therefore exactly the worst-off types. Following \citet{loertscher2019optimal}, this implies
\[
    \hatTheta\br{x}
    = \argmin_{\boldtheta^{\crt} \in \Theta}
    \psi_\eta\br{x,\boldtheta^{\crt}}
\]
and hence the optimal value of \eqref{eqn:designer-problem-finite} given by
\[
    \Pi_\eta = \adjustlimits
    \sup_{x \in \xscr}\min_{\boldtheta^{\crt} \in \Theta}
    \psi_\eta\br{x,\boldtheta^{\crt}}.
\]

\paragraph{Minimax and Saddle Point.} By \eqref{eqn:targeted-ads-feasibility}, each measure $x_i\br{\boldtheta;\cdot}$ is absolutely continuous with respect to $\gamma_\alpha$. Hence, by the Radon--Nikodym theorem, there exist measurable functions $\pi_i\br{\boldtheta,\omega}$ such that
\[
    x_i\br{\boldtheta;B}
    = \int_B\pi_i\br{\boldtheta,\omega}\,d\gamma_\alpha\br{\omega}
    \quad \text{for all Borel sets }B \subseteq \Omega,
\]
where $\pi_i\br{\boldtheta,\omega}$ is the share of customers with CTRs $\omega$ assigned to merchant $i$, with $\pi_i \geq 0$ and $\sum_{i \in N}\pi_i \leq 1$ almost everywhere. With finite-support CTRs, this becomes $x_i^\omega\br{\boldtheta} = \gamma_\alpha\br{\omega}\pi_i\br{\boldtheta,\omega}$.

We identify targeted ads $x$ with their Radon--Nikodym derivatives $\pi$ in $\br{L^\infty\br{dF\otimes\gamma_\alpha}}^N$, endowed with the weak$^*$ topology. Since \eqref{eqn:targeted-ads-feasibility} and \eqref{eqn:monotonicity} are weak$^*$-closed linear constraints, $\xscr$ is convex and weak$^*$-closed in a finite product of unit balls, hence weak$^*$-compact by the Banach--Alaoglu theorem.

For every $\boldtheta^{\crt} \in \Theta$, $\psi_\eta\br{\cdot,\boldtheta^{\crt}}$ is weak$^*$-continuous and affine. For every $x \in \xscr$, $\psi_\eta\br{x,\cdot}$ is separable and absolutely continuous in each coordinate, with
\[
    \frac{\partial\psi_\eta\br{x,\boldtheta^{\crt}}}{\partial\theta_i^{\crt}}
    = \eta_r\br{Q_i\br{\theta_i^{\crt}} - a_i}
    \quad \text{almost everywhere}.
\]
It is therefore continuous and convex by \eqref{eqn:monotonicity}. Then, we can apply Fan's minimax theorem \citep[Theorem~1]{fan1953minimax} which gives
\[
    \adjustlimits\max_{x \in \xscr\vphantom{\boldtheta^{\crt} \in \Theta}}\min_{\boldtheta^{\crt} \in \Theta}
    \psi_\eta\br{x,\boldtheta^{\crt}}
    =
    \adjustlimits\min_{\boldtheta^{\crt} \in \Theta}\max_{x \in \xscr\vphantom{\boldtheta^{\crt} \in \Theta}}
    \psi_\eta\br{x,\boldtheta^{\crt}},
\]
yielding \eqref{eqn:finite-minimax} and the existence of a saddle point $\br{x^*,\boldsymbol{\htheta}}$.

\paragraph{Ironing and Pointwise Maximization.} Fix a saddle point $\br{x^*,\boldsymbol{\htheta}}$. Then $x^*$ maximizes $\psi_\eta\br{x,\boldsymbol{\htheta}}$ over $\xscr$. Since $\phi_{\eta,i}\br{\cdot \,;\htheta_i}$ may be nonmonotone at an interior $\htheta_i$, we iron it following \citet{myerson1981optimal}. For now, allow negative ironing parameters in \eqref{eqn:finite-ironing-bound}. For each $i \in N$, there exists a unique $z_i \in \sbr{\phi_{\eta,i}^B\br{\ltheta_i},\phi_{\eta,i}^S\br{\btheta_i}}$ such that
\[
    \E_{F_i}\sbr{
        \overline\phi_{\eta,i}\br{\theta_i;z_i}
    }
    =
    \eta_r\htheta_i + \eta_v\E_{F_i}\sbr{\theta_i} + \eta_w.
\]
For such $z = \br{z_i}_{i \in N}$, by \citet{myerson1981optimal} ironing implies
\[
    \psi_\eta\br{x,\boldsymbol{\htheta}}
    \leq \overline\psi_\eta\br{x,z}
    \quad \text{for every }x \in \xscr,
\]
with equality if and only if each $Q_i$ is constant almost everywhere on the nondegenerate ironing interval.

Maximizing the ironed objective pointwise requires solving \eqref{eqn:pointwise-problem-scoring}. Let $\widetilde x$ be such a pointwise maximizer, using a fixed priority order whenever ties occur. Its expected clicks $\widetilde Q_i$ are nondecreasing and constant on ironing intervals, so
\[
    \overline\psi_\eta\br{x^*,z}
    \leq
    \overline\psi_\eta\br{\widetilde x,z}
    =
    \psi_\eta\br{\widetilde x,\boldsymbol{\htheta}}
    \leq
    \psi_\eta\br{x^*,\boldsymbol{\htheta}}
    \leq
    \overline\psi_\eta\br{x^*,z}.
\]
All inequalities are therefore equalities: $Q_i^*$ is constant almost everywhere on its ironing interval, and $x^*$ maximizes the ironed objective pointwise almost everywhere.

\paragraph{Nonnegative Ironing Parameters.} Suppose $z_i < 0$. Since $\htheta_i \leq \theta_i^B < \btheta_i$ and $\overline\phi_{\eta,i}\br{\htheta_i;z_i} = z_i < 0$, continuity implies that the score remains negative immediately to the right of $\htheta_i$. Pointwise optimality gives $Q_i^* = 0$ almost everywhere on this interval, so the envelope formula implies
\[
    U_i\br{\theta_i} - U_i\br{\htheta_i}
    = - a_i\br{\theta_i - \htheta_i} < 0
\]
for $\theta_i > \htheta_i$ sufficiently close to $\htheta_i$, contradicting that $\htheta_i$ is a worst-off type. Hence $z_i \geq 0$ for every $i \in N$.

\paragraph{Score-Based Tie-Breaking.} Let $\pi^*$ denote the Radon--Nikodym derivatives representing $x^*$. By pointwise optimality, $\pi^*$ is almost everywhere a tie-breaking rule as in \cref{def:scoring-mechanism}, assigning customers only to alternatives with the highest customer-specific score, including the designer with score $h_0\equiv0$.

Pointwise optimality alone does not ensure monotonicity of $Q_i$ at every type since tie-breaking may treat types in the ironed interval differently even though they have the same score. To overcome this, first average $\pi^*$ over type profiles with the same scores while holding CTRs fixed. This preserves feasibility and pointwise optimality almost everywhere. Since types are independent and $Q_i^*$ is constant almost everywhere on ironing intervals, averaging also leaves expected clicks unchanged almost everywhere, hence preserving the saddle point property. On the null sets, break ties according to some fixed priority order. The resulting rule is then feasible everywhere and depends on reports only through their scores, thus ensuring monotonicity of $Q_i$ everywhere.

Lastly, we recover transfers from \eqref{eqn:transfers} at $\boldtheta^{\crt} = \boldsymbol{\htheta}$, setting $U_i\br{\htheta_i} = 0$ for every $i \in N$.

\subsection{Proof of \texorpdfstring{\cref{prop:ironing-system}}{Proposition \ref*{prop:ironing-system}}}
\label{proof:optimal-mechanism-continuum}

The preceding proof also applies to continuously distributed CTRs. We now use the density of $\gamma_\alpha$ and \cref{asm:scarcity-finite} to obtain explicit characterizations of the ironing parameters.

\paragraph{Nonzero Ironing Parameters.} Suppose, toward a contradiction, that $z = \mathbf{0}$. Since $\phi_{\eta,i}^S \geq 0$, the inverse convention gives $\theta_i^S = \ltheta_i$, so merchant $i$'s score is
\[
    \overline\phi_{\eta,i}\br{\theta_i;0}
    = \max\cbr{\phi_{\eta,i}^B\br{\theta_i},0},
\]
which equals zero on $\sbr{\ltheta_i,\theta_i^B}$ and hence with probability
\[
    \tau_i
    := F_i\br{\theta_i^B}
    = F_i\br{\br{\phi_{\eta,i}^B}^{-1}\br{0}}.
\]
We consider two cases.

\textit{Case 1: $\tau_j = 0$ for some $j \in N$.} Then $\theta_j^B = \ltheta_j$, so $\phi_{\eta,j}^B\br{\theta_j} > 0$ for all $\theta_j > \ltheta_j$. Since the density assumption gives $\omega_j > 0$ almost everywhere, merchant $j$'s customer-specific score $\omega_j\overline\phi_{\eta,j}\br{\theta_j;0}$ is strictly positive almost surely.

Now, fix $i \neq j$. Merchant $i$'s score is continuous and equals zero on $\sbr{\ltheta_i,\theta_i^B}$, so it converges to zero as $\theta_i\downarrow\htheta_i \in \sbr{\ltheta_i,\theta_i^B}$. By pointwise optimality, merchant $i$ receives a customer only when his customer-specific score is at least merchant $j$'s, which is strictly positive, so $Q_i\br{\theta_i} \longrightarrow 0$ as $\theta_i\downarrow\htheta_i$. Because $\htheta_i \leq \theta_i^B < \btheta_i$ and $a_i > 0$, there exist types $\theta_i > \htheta_i$ with $Q_i\br{\theta_i} < a_i$, contradicting the worst-off-type characterization.

\textit{Case 2: $\tau_i > 0$ for every $i \in N$.} Fix $i \in N$. By \eqref{eqn:worst-off-type-ironing-parameter-identity}, $\htheta_i < \theta_i^B$. Since expected clicks are constant on the ironing interval, the worst-off-type characterization implies
\[
    Q_i\br{\htheta_i}
    = Q_i\br{y}
    \geq a_i
    \quad \text{for } y \in \br{\htheta_i,\theta_i^B}.
\]
Let $\pi^0\br{\omega}$ denote the tie-breaking rule when all merchants' scores are zero, which depends on $\omega$ alone because tie-breaking is score-based. A merchant with score zero receives clicks only when all other merchants also have score zero, an event of probability $\prod_{j \neq i}\tau_j$ by independence of types. Hence,
\[
    a_i
    \leq Q_i\br{\htheta_i}
    =
    \left(\prod_{j \neq i}\tau_j\right)
    \int_\Omega\omega_i\pi_i^0\br{\omega}\,
    d\gamma_\alpha\br{\omega}
    <
    \left(\prod_{j \neq i}\tau_j\right)
    \int_\Omega\omega_i\,d\gamma_\alpha\br{\omega}.
\]
The last inequality is strict because the same argument applied to any $j \neq i$ gives $Q_j\br{\htheta_j} \geq a_j > 0$, so $\pi_j^0 > 0$ and hence $\pi_i^0 < 1$ on a set of positive $\gamma_\alpha$-measure, while $\omega_i > 0$ almost everywhere and $\prod_{j \neq i}\tau_j > 0$. Since $i$ was arbitrary, the strict inequality holds for every $i \in N$, contradicting \cref{asm:scarcity-finite}. Thus, $z \neq \mathbf{0}$.

\paragraph{Unique Targeting and Solving for $z$.} Since $z_i > 0$ for some $i$ and $\phi_{\eta,i}^S$ is nonnegative and strictly increasing, the highest customer-specific score is positive for almost every type profile $\boldtheta$ and customer $\omega$, with ties only in a null set given that $\gamma_\alpha$ has a density. Thus, almost every customer is assigned to the merchant with the unique highest score.

Suppose that $\htheta_i$ is interior. By \eqref{eqn:worst-off-type-ironing-parameter-identity}, it lies in the interior of its ironing interval and has score $z_i$. Since expected clicks are constant on this interval, the worst-off type characterization implies $Q_i\br{\htheta_i} = a_i$. With ties occurring only on null sets, Fubini's theorem and independence of CTRs and types give
\[
    a_i
    = Q_i\br{\htheta_i}
    = \int_\Omega
    \omega_i\,
    \p_{F_{-i}}\br{
        \omega_i z_i
        > \max_{j \neq i}\omega_j\overline\phi_{\eta,j}\br{\theta_j;z_j}
    }\,d\gamma_\alpha\br{\omega}.
\]

\section{Proofs for Limit Market}
\label[appendix]{app:limit}

\subsection{Reduced-Form Feasibility}
\label{app:pooling-feasibility}

We first prove the properties of $\scal_G$ and its support function used in \cref{subsec:proof-sketch-limit}. We then show that an allocation $q$ is implementable if and only if $d\br{q} \in \scal_G$ and apply the strict separating hyperplane theorem to prove \cref{prop:reduced-form-feasibility}.

\paragraph{Properties of Feasible Set $\scal_G$ and Its Support Function.}
\begin{lemma}
    \label{lem:feasible-set}
    The feasible set $\scal_G$ is nonempty, convex, compact, and has free disposal, that is, if $s \in \scal_G$ and $0 \leq \tilde s \leq s$, then $\tilde s \in \scal_G$.
\end{lemma}
\begin{proof}[Proof of \cref{lem:feasible-set}]
    The zero vector belongs to $\scal_G$, so the set is nonempty. Moreover, for any $c \in [0,1]$ and points $s,\tilde s \in \scal_G$ with respective measurable functions $m$ and $\widetilde m$, the convex combination $c s + (1 - c) \widetilde s$ lies in $\scal_G$ by taking the corresponding measurable functions to be the convex combination $c m + (1 - c) \widetilde m$. Hence, $\scal_G$ is convex. It also has free disposal since, for any $s \in \scal_G$ and $0 \leq \widetilde s \leq s$, the measurable functions $m$ corresponding to $s$ also satisfy the inequalities in \eqref{eqn:pooling-feasible-outputs} for $\widetilde s$.

    It only remains to prove compactness. Let
    \[
        \ccal_G := \cbr{m \in \br{L^\infty\br{dG}}^K: m_k\br{\xi} \geq 0 \;\text{ and } \;\sum_{k \in K}m_k\br{\xi} \leq 1 \quad dG\text{-almost everywhere.}}
    \]
    The set $\ccal_G$ is contained in the unit ball of $\br{L^\infty\br{dG}}^K$ and is weak$^*$-closed because its constraints are weak$^*$-closed. It is therefore weak$^*$-compact by the Banach--Alaoglu theorem. Next define the linear map $A:\ccal_G \to \rbb_+^K$ by
    \[
        A_k\br{m} := \int_{\sbr{0,1}^K} \xi_km_k\br{\xi}\,dG\br{\xi} \quad \text{for all } m \in \ccal_G.
    \]
    Since each coordinate function $\xi\mapsto\xi_k$ belongs to $L^1\br{dG}$, $A$ is weak$^*$-continuous. By definition of $\scal_G$, $A\br{\ccal_G} \subseteq \scal_G$. Conversely, for any $s \in \scal_G$, choose $m \in \ccal_G$ with $s_k \leq A_k\br{m}$ for every $k$. Then multiplying each $m_k$ by $s_k/A_k\br{m}$ when $A_k\br{m} > 0$, and setting it to zero otherwise, gives shares in $\ccal_G$ with image $s$. Hence, $\scal_G = A\br{\ccal_G}$, so $\scal_G$ is compact as the continuous image of a compact set, concluding the proof.
\end{proof}

\begin{lemma}
    \label{lem:support-function}
    Suppose $G$ has a strictly positive density on $(0,1)^K$, and fix $\lambda \in \rbb_+^K\setminus\cbr{\mathbf{0}}$. Then:
    \begin{enumerate}[label=(\roman*), ref=\roman*]
        \item
        \label{lem:support-function:value}
        The support function satisfies
        \[
            \sigma\br{\lambda}
            = \E_G\sbr{\max_{k \in K}\cbr{\lambda_k\xi_k}}.
        \]

        \item
        \label{lem:support-function:supply}
        The value $\sigma\br{\lambda}$ is uniquely attained at $s^\lambda$ defined in \eqref{eqn:optimal-s-frontier}, and the corresponding customer shares are uniquely determined $dG$-almost everywhere by
        \[
            m_k^\lambda\br{\xi}
            = \mathbf{1}\cbr{k = \argmax_{j \in K}\lambda_j\xi_j}
            \quad \text{for every }k \in K.
        \]

        \item
        \label{lem:support-function:positive-supply}
        For every $k \in K$, $s_k^\lambda > 0$ if and only if $\lambda_k > 0$.
    \end{enumerate}
    The identity in part \textup{(\ref{lem:support-function:value})} holds for any distribution $G$ on $\sbr{0,1}^K$ and every $\lambda \in \rbb_+^K$.
\end{lemma}
\begin{proof}[Proof of \cref{lem:support-function}]
    Fix $\lambda \in \rbb_+^K$.

    \paragraph{Part \textup{(\ref{lem:support-function:value})}: Value of Support Function.} Let $s \in \scal_G$ and $m = \br{m_k}_{k \in K}$ be its corresponding measurable functions. Then,
    \begin{align*}
        \lambda \cdot s
        &\leq
        \int_{\sbr{0,1}^K}
        \sum_{k \in K}
        \lambda_k\xi_km_k\br{\xi}\,dG\br{\xi}
        \leq
        \int_{\sbr{0,1}^K}
        \max_{k \in K}\cbr{\lambda_k\xi_k}
        \br{\sum_{k \in K}m_k\br{\xi}}\,dG\br{\xi}\\
        &\leq \int_{\sbr{0,1}^K}
        \max_{k \in K}\cbr{\lambda_k\xi_k} \,dG\br{\xi} =
        \E_G\sbr{\max_{k \in K}\cbr{\lambda_k\xi_k}},
    \end{align*}
    where the first and third inequalities follow by definition of $m$. Thus, taking the maximum over $s \in \scal_G$ gives
    \[
        \sigma\br{\lambda} \leq \E_G\sbr{\max_{k \in K}\cbr{\lambda_k\xi_k}}.
    \]

    For the reverse inequality, define a tie-breaking rule $\kappa^\lambda: \sbr{0,1}^K \to K$ by
    \begin{equation}
        \label{eqn:kappa}
        \kappa^\lambda\br{\xi}
        :=
        \min\argmax_{j \in K}\cbr{\lambda_j\xi_j}.
    \end{equation}

    Because $K$ is finite, $\kappa^\lambda$ is well defined and measurable with respect to $G$. Then define for each $k \in K$ the measurable functions $m_k^\lambda$ by
    \[
        m_k^\lambda\br{\xi} := \mathbf{1}\cbr{\kappa^\lambda\br{\xi} = k} \quad dG\text{-almost everywhere.}
    \]
    Since
    \[
        \sum_{k \in K} m^\lambda_k\br{\xi} = 1 \quad dG\text{-almost everywhere,}
    \]
    the supply vector $s^\lambda \in \rbb_+^K$ defined by
    \[
        s_k^\lambda := \int_{\sbr{0,1}^K} \xi_km_k^\lambda\br{\xi}\,dG\br{\xi}
    \]
    lies in $\scal_G$ and attains the value on the support since
    \[
        \lambda \cdot s^\lambda
        =
        \int_{\sbr{0,1}^K}
        \sum_{k \in K}
        \lambda_k\xi_k
        \mathbf{1}\cbr{\kappa^\lambda\br{\xi} = k}
        \,dG\br{\xi}
        =
        \E_G\sbr{
            \max_{k \in K}\cbr{\lambda_k\xi_k}
        }.
    \]

    \paragraph{Part \textup{(\ref{lem:support-function:supply})}: Uniqueness.} Fix $\lambda \neq \mathbf{0}$, and let $\widetilde s \in \scal_G$ attain $\sigma\br{\lambda}$, with corresponding shares $\widetilde m$. Define
    \[
        M^\lambda\br{\xi} := \max_{j \in K}\cbr{\lambda_j\xi_j}.
    \]
    By the support bound above,
    \[
        \sigma\br{\lambda}
        = \lambda \cdot \widetilde s
        \leq \E_G\sbr{\sum_{k \in K}\lambda_k\xi_k\widetilde m_k\br{\xi}}
        \leq \E_G\sbr{M^\lambda\br{\xi}}
        = \sigma\br{\lambda}.
    \]
    Thus, equality holds throughout, and
    \begin{align*}
        0
        &= \E_G\sbr{M^\lambda\br{\xi} - \sum_{k \in K}\lambda_k\xi_k\widetilde m_k\br{\xi}}\\
        &= \E_G\sbr{
            M^\lambda\br{\xi}\br{1 - \sum_{k \in K}\widetilde m_k\br{\xi}}
            + \sum_{k \in K}\br{M^\lambda\br{\xi} - \lambda_k\xi_k}\widetilde m_k\br{\xi}
        }.
    \end{align*}
    Every term inside the last expectation is nonnegative. Hence, we get the identities
    \[
        M^\lambda\br{\xi}\br{1 - \sum_{k \in K}\widetilde m_k\br{\xi}}
        = 0
        \quad dG\text{-almost everywhere},
    \]
    and, for every $k \in K$,
    \[
        \br{M^\lambda\br{\xi} - \lambda_k\xi_k}\widetilde m_k\br{\xi}
        = 0
        \quad dG\text{-almost everywhere}.
    \]
    Since $G$ has a density, every CTR is strictly positive $dG$-almost everywhere. Together with $\lambda \neq \mathbf{0}$, this implies $M^\lambda > 0$ $dG$-almost everywhere. The first identity therefore gives
    \[
        \sum_{k \in K}\widetilde m_k\br{\xi} = 1
        \quad dG\text{-almost everywhere}.
    \]

    Moreover, since $G$ has a density, $\kappa^\lambda$ defined in \eqref{eqn:kappa} is the unique maximizing tag $dG$-almost everywhere. For every $k \neq \kappa^\lambda\br{\xi}$, we therefore have
    \[
        M^\lambda\br{\xi} - \lambda_k\xi_k > 0.
    \]
    The second identity then implies
    \[
        \widetilde m_k\br{\xi} = 0
        \quad \text{for every }k \neq \kappa^\lambda\br{\xi},
        \quad dG\text{-almost everywhere}.
    \]
    Since the shares sum to one, the unique maximizing tag must receive share one. Thus, for every $k \in K$,
    \[
        \widetilde m_k\br{\xi}
        = \mathbf{1}\cbr{\kappa^\lambda\br{\xi} = k}
        = m_k^\lambda\br{\xi}
        \quad dG\text{-almost everywhere}.
    \]

    Next, by feasibility of $\widetilde m$ for $\widetilde s$ and the definition of $s^\lambda$,
    \begin{equation}
        \label{eqn:supply-inequality}
        \widetilde s_k
        \leq \E_G\sbr{\xi_k\widetilde m_k\br{\xi}}
        = \E_G\sbr{\xi_km_k^\lambda\br{\xi}}
        = s_k^\lambda
        \quad \text{for every }k \in K.
    \end{equation}
    Since both $s^\lambda$ and $\widetilde s$ attain the value of the support function,
    \[
        0 = \lambda \cdot s^\lambda - \lambda \cdot \widetilde s
        = \sum_{k \in K}\lambda_k\br{s_k^\lambda - \widetilde s_k}.
    \]
    Each term in the sum is nonnegative, so $\widetilde s_k = s_k^\lambda$ whenever $\lambda_k > 0$. If $\lambda_k = 0$, then $\lambda_k\xi_k = 0 < M^\lambda\br{\xi}$ almost everywhere, so $m_k^\lambda = 0$ almost everywhere and $s_k^\lambda = 0$. The preceding inequality \eqref{eqn:supply-inequality} then gives $\widetilde s_k = 0$ as well. Therefore, $\widetilde s = s^\lambda$.

    \paragraph{Part \textup{(\ref{lem:support-function:positive-supply})}: Positive Supply.} Suppose $\lambda_k = 0$. Since $\lambda \neq \mathbf{0}$ and every CTR is strictly positive $dG$-almost everywhere, $\lambda_k\xi_k = 0 < M^\lambda\br{\xi}$ $dG$-almost everywhere, so $m_k^\lambda = 0$ $dG$-almost everywhere and $s_k^\lambda = 0$.

    Conversely, suppose $\lambda_k > 0$. The set
    \[
        \cbr{
            \xi \in (0,1)^K:
            \lambda_k\xi_k > \lambda_j\xi_j
            \text{ for every }j \neq k
        }
    \]
    has positive probability since $G$ has a strictly positive density. Hence every customer in this set is assigned to tag $k$ and has $\xi_k > 0$, so $s_k^\lambda > 0$.
\end{proof}

\paragraph{Reduced-Form Feasibility.}

\begin{lemma}
    \label{lem:implementable}
    Under the density assumption on $G$, an allocation $q = \br{q_k}_{k \in K}$ is implementable if and only if $d\br{q} \in \scal_G$.
\end{lemma}
\begin{proof}[Proof of \cref{lem:implementable}]
    $\br{\Rightarrow}$ Suppose $x$ implements $q$ at every type, and define the aggregate targeted ads for tag $k$ by
    \[
        \overline x_k\br{B} := \beta_k\int_{\Theta_k}x_k\br{\theta;B}\,dF_k\br{\theta}.
    \]
    By \eqref{eqn:aggregate-feasibility-limit} we have $\sum_k\overline x_k \leq G$. Since $0 \leq \overline x_k \leq G$, each $\overline x_k$ is absolutely continuous with respect to $G$. By the Radon--Nikodym theorem, there exist nonnegative measurable functions $m_k := d\overline x_k/dG$ such that
    \[
        \overline x_k\br{B}
        = \int_B m_k\br{\xi}\,dG\br{\xi}
        \quad \text{for every Borel set }B \subseteq \sbr{0,1}^K.
    \]
    \eqref{eqn:aggregate-feasibility-limit} then gives
    \[
        \int_B\sum_{k \in K}m_k\br{\xi}\,dG\br{\xi}
        = \sum_{k \in K}\overline x_k\br{B}
        \leq G\br{B}
        = \int_B1\,dG\br{\xi} \quad \text{for every Borel set }B \subseteq \sbr{0,1}^K,
    \]
    hence $\sum_{k \in K}m_k\br{\xi} \leq 1$ $dG\text{-almost everywhere}$.

    Finally, for every $k \in K$,
    \begin{align*}
        d_k\br{q}
        &= \beta_k\E_{F_k}\sbr{q_k\br{\theta}}\\
        &= \beta_k\int_{\Theta_k}
        \int_{\sbr{0,1}^K}\xi_k\,x_k\br{\theta;d\xi}
        \,dF_k\br{\theta}\\
        &= \int_{\sbr{0,1}^K}\xi_k\,d\overline x_k\br{\xi}\\
        &= \int_{\sbr{0,1}^K}\xi_km_k\br{\xi}\,dG\br{\xi}
        = \E_G\sbr{\xi_km_k\br{\xi}},
    \end{align*}
    where the second equality follows by definition of implementable $q$, the third follows from Fubini's theorem and the definition of $\overline x_k$, and the fourth follows from its Radon--Nikodym representation. Hence, $d\br{q} \in \scal_G$.

    $\br{\Leftarrow}$ Suppose $d\br{q} \in \scal_G$, and choose shares $m$ satisfying \eqref{eqn:pooling-feasible-outputs}. Scale them down to
    \[
        \widetilde m_k\br{\xi} :=
        \begin{cases}
            \displaystyle\frac{d_k\br{q}}{\E_G\sbr{\xi_km_k\br{\xi}}}\,m_k\br{\xi}
            &\text{if }\E_G\sbr{\xi_km_k\br{\xi}} > 0,\\[2mm]
            0&\text{otherwise.}
        \end{cases}
    \]
    Since $0 \leq d_k\br{q} \leq \E_G\sbr{\xi_km_k\br{\xi}}$, these shares satisfy $\sum_k\widetilde m_k \leq 1$ $dG$-almost everywhere and $\E_G\sbr{\xi_k\widetilde m_k\br{\xi}} = d_k\br{q}$.

    Since $\E_G\sbr{\xi_k} > 0$ under the density assumption, define
    \begin{equation}
        \label{eqn:allocation-density}
        \rho_k\br{\theta,\xi} :=
        \begin{cases}
            \displaystyle\frac{q_k\br{\theta}\widetilde m_k\br{\xi}}{d_k\br{q}}
            &\text{if }d_k\br{q} > 0,\\[2mm]
            \displaystyle\frac{q_k\br{\theta}}{\E_G\sbr{\xi_k}}
            &\text{if }d_k\br{q} = 0,
        \end{cases}
    \end{equation}
    and let $x_k\br{\theta;B} := \int_B\rho_k\br{\theta,\xi}\,dG\br{\xi}$. Since $d_k\br{q} = \beta_k\E_{F_k}\sbr{q_k\br{\theta}}$,
    \[
        \beta_k\int_{\Theta_k}\rho_k\br{\theta,\xi}\,dF_k\br{\theta}
        = \widetilde m_k\br{\xi}.
    \]
    Then, by Fubini's theorem
    \[
        \sum_k\beta_k\int_{\Theta_k}x_k\br{\theta;B}\,dF_k\br{\theta}
        = \int_B\sum_k\widetilde m_k\br{\xi}\,dG\br{\xi}
        \leq G\br{B} \quad \text{for all Borel set } B \subseteq \sbr{0,1}^K
    \]
    hence $x$ satisfies \eqref{eqn:aggregate-feasibility-limit}.

    Now, if $d_k\br{q} > 0$, then
    \[
        \int_{\sbr{0,1}^K}\xi_k\,x_k\br{\theta;d\xi}
        = \frac{q_k\br{\theta}}{d_k\br{q}}
        \E_G\sbr{\xi_k\widetilde m_k\br{\xi}}
        = q_k\br{\theta} \quad \text{for all } \theta \in \Theta_k.
    \]
    If instead $d_k\br{q} = 0$, then
    \[
        \int_{\sbr{0,1}^K}\xi_k\,x_k\br{\theta;d\xi}
        = \frac{q_k\br{\theta}}{\E_G\sbr{\xi_k}}
        \E_G\sbr{\xi_k}
        = q_k\br{\theta} \quad \text{for all } \theta \in \Theta_k.
    \]
    Hence, $x$ implements $q$.
\end{proof}

\begin{proof}[Proof of \cref{prop:reduced-form-feasibility}]
    If the allocation $q$ is implementable, then $d\br{q} \in \scal_G$ by \cref{lem:implementable}. Hence, for every $\lambda \in \rbb_+^K$,
    \[
        \sum_{k \in K}\lambda_k\beta_k\E_{F_k}\sbr{q_k\br{\theta}}
        = \lambda \cdot d\br{q}
        \leq \sigma\br{\lambda}
        = \E_G\sbr{\max_{k \in K}\cbr{\lambda_k\xi_k}},
    \]
    where the last equality follows from \cref{lem:support-function}\textup{(\ref{lem:support-function:value})}. This proves necessity.

    Conversely, suppose $q$ satisfies \eqref{eqn:R-Fsb} but $d\br{q} \notin \scal_G$. Since $\scal_G$ is nonempty, convex, and compact by \cref{lem:feasible-set}, the strict separating hyperplane theorem gives $a \in \rbb^K$ such that
    \[
        a \cdot d\br{q} > \max_{s \in \scal_G}a \cdot s.
    \]
    Let $a_+ := \br{\max\cbr{a_k,0}}_{k \in K}$. By free disposal, the maximum over $\scal_G$ is attained with $s_k = 0$ whenever $a_k < 0$, so
    \[
        \max_{s \in \scal_G}a \cdot s
        = \max_{s \in \scal_G}a_+\cdot s
        = \sigma\br{a_+}.
    \]
    Since $d\br{q} \geq 0$, it follows that
    \[
        a_+\cdot d\br{q}
        \geq a \cdot d\br{q}
        > \sigma\br{a_+},
    \]
    contradicting \eqref{eqn:R-Fsb} at $\lambda = a_+$. Thus, $d\br{q} \in \scal_G$, and by \cref{lem:implementable} $q$ is implementable.
\end{proof}

\subsection{Proof of \texorpdfstring{\cref{thm:optimal-limit-mechanism}}{Theorem \ref*{thm:optimal-limit-mechanism}}}
\label{app:optimal-limit}

We provide the omitted details in the proof sketch of \cref{thm:optimal-limit-mechanism}.

\subsubsection{Virtual Objective and Weak Duality}
\label{app:available-output-screening}

The reduction of the designer's problem in \eqref{eqn:limit-reduced-problem} follows from the same arguments as in the proof of \cref{thm:optimal-finite-mechanism}. Since $\eta_r > 0$, optimality requires the minimum payoff in each tag to be zero: otherwise, increasing every transfer in that tag by its minimum payoff preserves incentive compatibility and individual rationality and strictly increases the designer's objective.

To prove \eqref{eqn:pseudo-lagrangian-equality}, observe that, for every $q \in \qcal$,
\[
    \inf_{\lambda \in \rbb_+^K}
    \cbr{\sigma\br{\lambda} - \lambda \cdot d\br{q}}
    =
    \begin{cases}
        0 & \text{if }q \in \qcal_R,\\
        - \infty & \text{otherwise.}
    \end{cases}
\]
This holds because, when $q$ is feasible, $\sigma\br{\lambda} - \lambda \cdot d\br{q} \geq 0$ at every $\lambda \geq 0$ and equals zero at $\lambda = \mathbf{0}$. Otherwise, if $q$ is not feasible, that is, $q \notin \qcal_R$, then there exists $\bar\lambda \geq 0$ such that $\sigma\br{\bar\lambda} - \bar\lambda \cdot d\br{q} < 0$. But then, since $\sigma\br{\lambda}$ is homogeneous of degree one,
\[
    \sigma\br{c\bar\lambda} - c\bar\lambda \cdot d\br{q} = c\br{\sigma\br{\bar\lambda} - \bar\lambda \cdot d\br{q}} \longrightarrow - \infty \quad \text{as } c \longrightarrow + \infty.
\]
Since this added term is independent of $\boldtheta^{\crt}$, the infimum over $\lambda$ and the minimum over critical types commute, and together with \eqref{eqn:limit-pseudo-lagrangian} and \eqref{eqn:limit-reduced-problem} we get \eqref{eqn:pseudo-lagrangian-equality}.

Lastly, for every $q \in \qcal_R$ and $\lambda \geq 0$, \eqref{eqn:screening-value} gives
\[
    \min_{\boldtheta^{\crt} \in \Theta_K}
    \psi_\eta\br{q,\boldtheta^{\crt}}
    \leq \min_{\boldtheta^{\crt} \in \Theta_K}
    \lcal\br{q,\boldtheta^{\crt};\lambda}
    \leq \sigma\br{\lambda} + \ell\br{\lambda}.
\]
Taking the supremum over $q \in \qcal_R$ and then the infimum over $\lambda \geq 0$ gives \eqref{eqn:dual-upper-bound}.

\subsubsection{The Screening Problem}
\label{app:shadow-price-screening}

\paragraph{Minimizing over Critical Types.} Fix an allocation $q \in \qcal$. For every $\boldtheta^{\crt} \in \Theta_K$,
\[
    \psi_\eta\br{q,\boldtheta^{\crt}}
    \geq
    \sum_{k \in K}\beta_k\E_{F_k}\sbr{
        \min\cbr{
            \br{q_k\br{\theta} - \mu_k}\phi_{\eta,k}^S\br{\theta},\,
            \br{q_k\br{\theta} - \mu_k}\phi_{\eta,k}^B\br{\theta}
        }
    }.
\]
To attain this bound, let
\[
    \theta_k^{\crt}
    := \inf\br{
        \cbr{\theta \in \Theta_k:q_k\br{\theta} \geq \mu_k}
        \cup\cbr{\btheta_k}
    }.
\]
Since $q_k$ is nondecreasing, $q_k\br{\theta} \leq \mu_k$ for $\theta < \theta_k^{\crt}$ and $q_k\br{\theta} \geq \mu_k$ for $\theta > \theta_k^{\crt}$. Moreover, since $\phi_{\eta,k}^S \geq \phi_{\eta,k}^B$,
\begin{align*}
    \br{q_k\br{\theta} - \mu_k}
    \phi_{\eta,k}\br{\theta;\theta_k^{\crt}}
    &=
    \begin{cases}
        \br{q_k\br{\theta} - \mu_k}\phi_{\eta,k}^S\br{\theta}
        & \text{if }\theta < \theta_k^{\crt},\\
        \br{q_k\br{\theta} - \mu_k}\phi_{\eta,k}^B\br{\theta}
        & \text{if }\theta \geq \theta_k^{\crt}
    \end{cases}\\
    &= \min\cbr{
        \br{q_k\br{\theta} - \mu_k}\phi_{\eta,k}^S\br{\theta},\,
        \br{q_k\br{\theta} - \mu_k}\phi_{\eta,k}^B\br{\theta}
    }.
\end{align*}
Integrating and summing across tags gives equality in the bound above, proving \eqref{eqn:reduced-screening-objective}.

\paragraph{Maximizing over Allocations.} For $\lambda \in \rbb_+^K$, let $p_k^S\br{\lambda_k} := \br{\phi_{\eta,k}^S}^{-1}\br{\lambda_k}$ and $p_k^B\br{\lambda_k} := \br{\phi_{\eta,k}^B}^{-1}\br{\lambda_k}$, using the inverse conventions in \cref{subsec:optimal-mechanism-finite}.

\begin{lemma}[Optimal Screening at Given Shadow Prices]
    \label{lem:screening}
    For every $\lambda \in \rbb_+^K$, the screening value defined in \eqref{eqn:screening-value} is given by \eqref{eqn:pooling-dual-decomposition} and attained by the allocation $q^\lambda$ in \eqref{eqn:screening-clicks}. Every maximizing allocation $\widetilde q$ satisfies $\widetilde q_k = q_k^\lambda$ $dF_k$-almost everywhere for every $k \in K$.
\end{lemma}

\begin{proof}[Proof of \cref{lem:screening}]
    Fix $\lambda \in \rbb_+^K$. For each tag $k$, define
    \begin{align*}
        h_k^\lambda\br{\theta,r}
        &:=
        \min\cbr{
            \br{r - \mu_k}\phi_{\eta,k}^S\br{\theta},\,
            \br{r - \mu_k}\phi_{\eta,k}^B\br{\theta}
        }
        - \lambda_kr\\
        &=
        \begin{cases}
            \br{\phi_{\eta,k}^S\br{\theta} - \lambda_k}r
            - \mu_k\phi_{\eta,k}^S\br{\theta}
            & \text{if }0 \leq r \leq \mu_k,\\
            \br{\phi_{\eta,k}^B\br{\theta} - \lambda_k}r
            - \mu_k\phi_{\eta,k}^B\br{\theta}
            & \text{if }\mu_k < r \leq 1.
        \end{cases}
    \end{align*}
    Since $\phi_{\eta,k}^S > \phi_{\eta,k}^B$, at each $\theta \in \Theta_k$ the function $h_k^\lambda\br{\theta,\cdot}$ is piecewise linear and concave on $\sbr{0,1}$ with a kink at $\mu_k$. It is therefore maximized at $0$, $\mu_k$, or $1$ according to the signs of its slopes $\phi_{\eta,k}^S\br{\theta} - \lambda_k$ and $\phi_{\eta,k}^B\br{\theta} - \lambda_k$, so a maximizer $r^* \in \sbr{0,1}$ is given by
    \[
        r^* =
        \begin{cases}
            0
            & \text{if }\phi_{\eta,k}^S\br{\theta} \leq \lambda_k,\\
            \mu_k
            & \text{if }\phi_{\eta,k}^B\br{\theta} \leq \lambda_k
            < \phi_{\eta,k}^S\br{\theta},\\
            1
            & \text{if }\lambda_k < \phi_{\eta,k}^B\br{\theta}.
        \end{cases}
    \]
    This is exactly $q_k^\lambda\br{\theta}$ in \eqref{eqn:screening-clicks}, so $q^\lambda$ is a pointwise maximizing allocation. Hence,
    \begin{align*}
        h_k^\lambda\br{\theta,q_k^\lambda\br{\theta}}
        &= \max\cbr{
            h_k^\lambda\br{\theta,0},\,
            h_k^\lambda\br{\theta,\mu_k},\,
            h_k^\lambda\br{\theta,1}
        }\\
        &= \max\cbr{
            - \mu_k\phi_{\eta,k}^S\br{\theta},\,
            - \mu_k\lambda_k,\,
            \br{1 - \mu_k}\phi_{\eta,k}^B\br{\theta} - \lambda_k
        }.
    \end{align*}
    Thus, $q^\lambda$ attains the screening value, which proves \eqref{eqn:pooling-dual-decomposition}.

    This pointwise maximizer $q^\lambda$ is unique except possibly where $\phi_{\eta,k}^S\br{\theta} = \lambda_k$ or $\phi_{\eta,k}^B\br{\theta} = \lambda_k$. By regularity, there are at most two such cutoff types, which have probability zero since $F_k$ has a density. Hence, $q^\lambda$ is the unique maximizer $dF_k$-almost everywhere.
\end{proof}

\subsubsection{The Targeting Problem}

By \cref{lem:support-function}\textup{(\ref{lem:support-function:supply})}, at every $\lambda \in \rbb_+^K\setminus\cbr{\mathbf{0}}$ the supply vector in \eqref{eqn:optimal-s-frontier} uniquely attains the targeting value in \eqref{eqn:pooling-support-function}.

\subsubsection{Market Clearing and Optimality}
\label{app:market-clearing-optimality}

Let $\overline\lambda_k := \phi_{\eta,k}^S\br{\btheta_k}$ for each $k \in K$, and define the compact product set
\[
    \Lambda := \prod_{k \in K}\sbr{0,\overline\lambda_k}.
\]

\begin{lemma}
    \label{lem:dual-compact}
    The sum $\sigma + \ell$ is convex and continuous on $\rbb_+^K$ and satisfies
    \[
        \inf_{\lambda \in \rbb_+^K}\cbr{\sigma\br{\lambda} + \ell\br{\lambda}}
        = \min_{\lambda \in \Lambda\vphantom{\lambda \in \rbb_+^K}}\cbr{\sigma\br{\lambda} + \ell\br{\lambda}}.
    \]
    In particular, $\sigma + \ell$ attains its minimum over $\rbb_+^K$ at some $\lambda^* \in \Lambda$.
\end{lemma}
\begin{proof}[Proof of \cref{lem:dual-compact}]
    We prove each claim separately.

    \paragraph{Convexity.} By \eqref{eqn:pooling-dual-decomposition}, $\ell$ is convex since each term inside the expectation is the maximum of affine functions of $\lambda_k$. Moreover, $\sigma$ is convex by definition of a support function, so $\sigma + \ell$ is convex on $\rbb_+^K$.

    \paragraph{Restriction to a Compact Set.} Fix any $\lambda \in \rbb_+^K$. If $\lambda_k \geq \overline\lambda_k = \phi_{\eta,k}^S\br{\btheta_k}$, then \eqref{eqn:screening-clicks} gives $q_k^\lambda\equiv0$. Tag $k$'s contribution to the screening value in \eqref{eqn:pooling-dual-decomposition} is therefore $ - \beta_k\mu_k\E_{F_k}\sbr{\phi_{\eta,k}^S\br{\theta}}$.

    Now, let $\widetilde\lambda_k := \min\cbr{\lambda_k,\overline\lambda_k}$. Then $\ell\br{\widetilde\lambda} = \ell\br{\lambda}$ and $\sigma\br{\widetilde\lambda} \leq \sigma\br{\lambda}$, and hence $ \sigma\br{\widetilde\lambda} + \ell\br{\widetilde\lambda} \leq \sigma\br{\lambda} + \ell\br{\lambda}$. Therefore,
    \[
        \inf_{\lambda \in \rbb_+^K} \cbr{\sigma\br{\lambda} + \ell\br{\lambda}} = \inf_{\lambda \in \Lambda\vphantom{\lambda \in \rbb_+^K}} \cbr{\sigma\br{\lambda} + \ell\br{\lambda}}.
    \]

    \paragraph{Existence.} By \eqref{eqn:pooling-dual-decomposition} and \cref{lem:support-function}\textup{(\ref{lem:support-function:value})}, both $\ell$ and $\sigma$ are continuous in $\lambda$. Since $\Lambda$ is compact, the Weierstrass theorem implies that $\sigma + \ell$ attains its minimum on $\Lambda$ at some $\lambda^* \in \Lambda$.
\end{proof}

The following definition for directional derivatives is used in \cref{lem:market-clearing} to rule out zero shadow prices.

\begin{definition}
    \label{def:directional-derivative}
    Fix $\lambda \in \rbb_+^K$. We say that a direction $h \in \rbb^K$ is \textit{feasible at $\lambda$} if $h_k \geq 0$ whenever $\lambda_k = 0$, or equivalently, if $\lambda + th \in \rbb_+^K$ for all sufficiently small $t > 0$. For a function $f:\rbb_+^K \to \rbb$, we define its \textit{right directional derivative} at $\lambda$ in a feasible direction $h$ by
    \[
        D_h^+f\br{\lambda}
        := \lim_{t \longrightarrow 0^+}
        \frac{f\br{\lambda + th} - f\br{\lambda}}{t},
    \]
    whenever the limit exists.
\end{definition}

\begin{lemma}[Virtual Prices and Market Clearing]
    \label{lem:market-clearing}
    Suppose $G$ has a strictly positive density on $(0,1)^K$ and \cref{asm:scarcity-limit} holds.
    \begin{enumerate}[label=(\roman*), ref=\roman*]
        \item
        \label{lem:market-clearing:foc}
        A vector $\lambda \in \rbb_{++}^K$ minimizes $\sigma + \ell$ if and only if it satisfies the market-clearing conditions \eqref{eqn:market-clearing-limit}.

        \item
        \label{lem:market-clearing:bounds}
        Every minimizer $\lambda^*$ satisfies
        \[
            0 < \lambda_k^* < \overline\lambda_k = \phi_{\eta,k}^S\br{\btheta_k},
            \qquad
            s_k^{\lambda^*} = d_k\br{q^{\lambda^*}} > 0
            \quad \text{for all }k \in K.
        \]
    \end{enumerate}
\end{lemma}

\begin{proof}[Proof of \cref{lem:market-clearing}]
    We prove each part in turn.

    \paragraph{Part \textup{(\ref{lem:market-clearing:foc})}:} Fix $\lambda \in \rbb_+^K$ and $k \in K$. By \cref{lem:screening}, $q_k^\lambda\br{\theta}$ is the unique maximizer except possibly at the cutoff types, so
    \[
        \frac{\partial}{\partial\lambda_k}
        \max\cbr{
            - \mu_k\phi_{\eta,k}^S\br{\theta},\,
            - \mu_k\lambda_k,\,
            \br{1 - \mu_k}\phi_{\eta,k}^B\br{\theta} - \lambda_k
        }
        = - q_k^\lambda\br{\theta}
        \quad dF_k\text{-almost everywhere.}
    \]
    Hence,
    \begin{align*}
        \frac{\partial\ell}{\partial\lambda_k}\br{\lambda}
        &= \frac{\partial}{\partial\lambda_k}
        \br{\sum_{j \in K}\beta_j\E_{F_j}\sbr{
                \max\cbr{
                    - \mu_j\phi_{\eta,j}^S\br{\theta},\,
                    - \mu_j\lambda_j,\,
                    \br{1 - \mu_j}\phi_{\eta,j}^B\br{\theta} - \lambda_j
                }
        }}\\
        &= - \beta_k\E_{F_k}\sbr{q_k^\lambda\br{\theta}}
        = - d_k\br{q^\lambda},
    \end{align*}
    where the dominated convergence theorem (DCT) justifies interchanging expectation and differentiation.

    Next, suppose $\lambda \neq \mathbf{0}$. By \cref{lem:feasible-set}, $\scal_G$ is nonempty and compact, and by \cref{lem:support-function}\textup{(\ref{lem:support-function:supply})}, $s^\lambda$ uniquely maximizes $\lambda \cdot s$ over $s \in \scal_G$. The objective $\lambda \cdot s$ and its partial derivative $s_k$ are jointly continuous in $(s,\lambda)$. Danskin's theorem\footnote{Corollary~4 of \cite{milgrom2002envelope} provides a generalized version of Danskin's theorem, which gives the envelope formula for the maximum of a parametrized family of functions over a fixed compact choice set.} therefore gives
    \begin{equation}
        \label{eqn:appendix-support-partial-derivative}
        \frac{\partial\sigma}{\partial\lambda_k}\br{\lambda}
        = s_k^\lambda.
    \end{equation}

    If $\lambda \in \rbb^K_{++}$ minimizes $\sigma + \ell$, the first-order conditions give
    \[
        0 = \frac{\partial\br{\sigma + \ell}}{\partial\lambda_k}\br{\lambda}
        = s_k^\lambda - d_k\br{q^\lambda}
        \quad \text{for all }k \in K.
    \]
    By \cref{lem:support-function}\textup{(\ref{lem:support-function:supply})} and \eqref{eqn:screening-demand}, these are the market-clearing conditions \eqref{eqn:market-clearing-limit}. Conversely, for any $\lambda \in \rbb_{++}^K$ that satisfies the first-order conditions, convexity of $\sigma + \ell$ implies global optimality.

    \paragraph{Part \textup{(\ref{lem:market-clearing:bounds})}:} Let $\lambda^* \in \rbb_+^K$ be a minimizer of $\sigma + \ell$, which exists by \cref{lem:dual-compact}. At zero shadow prices, $p_k^S\br{0} = \ltheta_k$ and $p_k^B\br{0} = \br{\phi_{\eta,k}^B}^{-1}\br{0}$, so \eqref{eqn:screening-demand} gives
    \[
        d_k\br{q^{\mathbf{0}}}
        = \beta_k\sbr{1 - \br{1 - \mu_k}F_k\br{p_k^B\br{0}}}.
    \]
    Since $\sigma\br{\mathbf{0}} = 0$ and $\sigma\br{t\mathbf{1}} = t\sigma\br{\mathbf{1}}$ for every $t > 0$,
    \[
        D_{\mathbf{1}}^+\sigma\br{\mathbf{0}}
        = \lim_{t \longrightarrow 0^+}
        \frac{\sigma\br{t\mathbf1} - \sigma\br{\mathbf0}}{t}
        = \lim_{t \longrightarrow 0^+}
        \frac{t\sigma\br{\mathbf1}}{t}
        = \sigma\br{\mathbf{1}}
        = \E_G\sbr{\max_{k \in K}\xi_k}.
    \]
    Since $\ell$ is separable across tags, writing $e_k$ for the $k$th coordinate vector gives
    \[
        D_{\mathbf{1}}^+\ell\br{\mathbf{0}}
        = \lim_{t \longrightarrow 0^+}
        \sum_{k \in K}\frac{\ell\br{te_k} - \ell\br{\mathbf0}}{t}
        = - \sum_{k \in K}d_k\br{q^{\mathbf{0}}},
    \]
    where the last equality uses $\partial\ell/\partial\lambda_k = - d_k\br{q^\lambda}$ from part \textup{(\ref{lem:market-clearing:foc})}, which holds as a right derivative at $\lambda_k = 0$.

    Combining the two derivatives gives
    \[
        D_{\mathbf{1}}^+\br{\sigma + \ell}\br{\mathbf{0}}
        = \E_G\sbr{\max_{k \in K}\xi_k}
        - \sum_{k \in K}d_k\br{q^{\mathbf{0}}}
        < 0,
    \]
    where the strict inequality follows from \cref{asm:scarcity-limit}. Consequently, $\sigma\br{t\mathbf{1}} + \ell\br{t\mathbf{1}} < \sigma\br{\mathbf{0}} + \ell\br{\mathbf{0}}$ for all sufficiently small $t > 0$, so $\mathbf{0}$ cannot minimize $\sigma + \ell$. Thus, $\lambda^* \neq \mathbf{0}$.

    Now, the first-order conditions for optimality (including endpoints) give
    \[
        \begin{cases}
            \displaystyle
            \frac{\partial\br{\sigma + \ell}}{\partial\lambda_k}
            \br{\lambda^*} = 0
            & \text{if }\lambda_k^* > 0,\\[2mm]
            \displaystyle
            \frac{\partial\br{\sigma + \ell}}{\partial\lambda_k}
            \br{\lambda^*} \geq 0
            & \text{if }\lambda_k^* = 0,
        \end{cases}
    \]
    where the derivative is understood from the right when $\lambda_k^* = 0$.

    If $\lambda_k^* = 0$, \cref{lem:support-function}\textup{(\ref{lem:support-function:positive-supply})} gives $s_k^{\lambda^*} = 0$, whereas demand depends only on $\lambda_k$ and satisfies
    \[
        d_k\br{q^{\lambda^*}}
        = d_k\br{q^{\mathbf{0}}}
        \geq \beta_k\mu_k > 0.
    \]
    Consequently, $\frac{\partial\br{\sigma + \ell}}{\partial\lambda_k}\br{\lambda^*} = - d_k\br{q^{\mathbf{0}}} < 0$, contradicting the second condition above. Hence, $\lambda^* \in \rbb_{++}^K$.

    By part \textup{(\ref{lem:market-clearing:foc})}, $\lambda^*$ satisfies the market-clearing conditions \eqref{eqn:market-clearing-limit} for all $k \in K$. By \cref{lem:support-function}\textup{(\ref{lem:support-function:positive-supply})}, strictly positive shadow prices give $s_k^{\lambda^*} > 0$ for every tag. Market clearing therefore gives $d_k\br{q^{\lambda^*}} > 0$. Since demand is zero whenever $\lambda_k \geq \overline\lambda_k$, we obtain $\lambda_k^* < \overline\lambda_k$.
\end{proof}

\paragraph{Optimality.} Let $\lambda^* \in \Lambda$ minimize $\sigma + \ell$, and let $q^*$ be the optimal allocation in \cref{thm:optimal-limit-mechanism} at the corresponding posted prices. This allocation agrees with $q^{\lambda^*}$ $dF_k$-almost everywhere in each tag. By \cref{lem:market-clearing},
\[
    d\br{q^*} = s^{\lambda^*} \in \scal_G.
\]
Thus, $q^* \in \qcal_R$ and is implementable by \cref{lem:implementable}. By \cref{lem:screening}, $q^*$ attains the screening value. Then market clearing and \cref{lem:support-function}\textup{(\ref{lem:support-function:supply})} give
\[
    \begin{aligned}
        \min_{\boldtheta^{\crt} \in \Theta_K}
        \psi_\eta\br{q^*,\boldtheta^{\crt}}
        &=
        \min_{\boldtheta^{\crt} \in \Theta_K}
        \cbr{
            \psi_\eta\br{q^*,\boldtheta^{\crt}}
            - \lambda^* \cdot d\br{q^*}
        }
        + \lambda^* \cdot d\br{q^*}\\
        &= \ell\br{\lambda^*} + \lambda^* \cdot d\br{q^*} = \ell\br{\lambda^*} + \lambda^* \cdot s^{\lambda^*}\\
        &= \ell\br{\lambda^*} + \sigma\br{\lambda^*}.
    \end{aligned}
\]
Thus, $q^*$ is feasible and attains the weak duality bound \eqref{eqn:dual-upper-bound}, so it is optimal and
\begin{equation}
    \label{eqn:pooling-dual-value}
    \Pi_\eta^\infty
    = \sigma\br{\lambda^*} + \ell\br{\lambda^*}
    = \min_{\lambda \in \rbb_+^K}
    \cbr{\sigma\br{\lambda} + \ell\br{\lambda}}.
\end{equation}

\paragraph{Payoffs and Transfers.} For ease of notation, let $p_k^j := p_k^j\br{\lambda_k^*}$ for $j \in \cbr{S,B}$ and $k \in K$. Let $u_k^*\br{\theta} := \theta\br{q_k^*\br{\theta} - \mu_k} - t_k^*\br{\theta}$ denote the payoff of a type $\theta$ with tag $k$, relative to his outside option.

By the worst-off-type characterization, every worst-off type under $q_k^*$ satisfies $\htheta_k \in \sbr{p_k^S,p_k^B}$. Since $\eta_r > 0$, optimal transfers are such that $u_k^*\br{\htheta_k} = 0$. The envelope formula then gives, for any $\htheta_k \in \sbr{p_k^S,p_k^B}$,
\[
    u_k^*\br{\theta}
    = \int_{\htheta_k}^{\theta}\br{q_k^*\br{y} - \mu_k}\,dy
    =
    \begin{cases}
        \mu_k\br{p_k^S - \theta}
        & \text{if }\theta \leq p_k^S,\\
        0
        & \text{if }p_k^S < \theta \leq p_k^B,\\
        \br{1 - \mu_k}\br{\theta - p_k^B}
        & \text{if }\theta > p_k^B.
    \end{cases}
\]
Hence, the corresponding transfers are
\[
    t_k^*\br{\theta}
    = \theta\br{q_k^*\br{\theta} - \mu_k} - u_k^*\br{\theta}
    =
    \begin{cases}
        - \mu_kp_k^S
        & \text{if }\theta \leq p_k^S,\\
        0
        & \text{if }p_k^S < \theta \leq p_k^B,\\
        \br{1 - \mu_k}p_k^B
        & \text{if }\theta > p_k^B,
    \end{cases}
\]
as in \cref{thm:optimal-limit-mechanism}. Since $q_k^*$ is nondecreasing and $u_k^*$ is nonnegative, these transfers satisfy \eqref{eqn:ICk} and \eqref{eqn:IRk}.

\subsubsection{(Essential) Uniqueness}
\label[appendix]{app:limit-uniqueness}

\paragraph{Allocation and Transfers.}

Let $\widetilde q$ and $\widetilde t$ be the allocation and transfers induced by any optimal mechanism. Then
\[
    \begin{aligned}
        \Pi_\eta^\infty
        &= \sum_{k \in K}\beta_k\E_{F_k}\sbr{
            \eta_r\widetilde t_k\br{\theta}
            + \br{\eta_v\theta + \eta_w}
            \br{\widetilde q_k\br{\theta} - \mu_k}
        }\\
        &= \min_{\boldtheta^{\crt} \in \Theta_K}
        \psi_\eta\br{\widetilde q,\boldtheta^{\crt}}\\
        &\leq \sigma\br{\lambda^*} - \lambda^* \cdot d\br{\widetilde q}
        + \min_{\boldtheta^{\crt} \in \Theta_K}
        \psi_\eta\br{\widetilde q,\boldtheta^{\crt}}\\
        &\leq \sigma\br{\lambda^*} + \ell\br{\lambda^*}
        = \Pi_\eta^\infty.
    \end{aligned}
\]
The first equality follows from optimality of $\br{\widetilde q,\widetilde t}$. The second equality follows from the reduction of the designer's objective to the virtual objective. The first and second inequalities follow from \eqref{eqn:R-Fsb} and \eqref{eqn:screening-value}, respectively, and the last equality follows from \eqref{eqn:pooling-dual-value}.

Both inequalities therefore hold with equality. In particular, equality in the second inequality gives, after canceling $\sigma\br{\lambda^*}$,
\[
    \min_{\boldtheta^{\crt} \in \Theta_K}
    \cbr{
        \psi_\eta\br{\widetilde q,\boldtheta^{\crt}}
        - \lambda^* \cdot d\br{\widetilde q}
    }
    = \ell\br{\lambda^*} = \adjustlimits\max_{q \in \qcal}
    \min_{\boldtheta^{\crt} \in \Theta_K}
    \cbr{
        \psi_\eta\br{q,\boldtheta^{\crt}}
        - \lambda^* \cdot d\br{q}
    },
\]
where the last equality follows from \eqref{eqn:screening-value}. Thus, $\widetilde q$ solves the screening problem at $\lambda^*$. By \cref{lem:screening}, its maximizer is essentially unique, so $\widetilde q_k = q_k^*$ $dF_k$-almost everywhere for every $k \in K$.

Now, let $\widetilde u_k\br{\theta} := \theta\br{\widetilde q_k\br{\theta} - \mu_k} - \widetilde t_k\br{\theta}$ denote the corresponding payoffs. By the envelope formula, both payoff functions are absolutely continuous, and
\[
    \widetilde u_k'\br{\theta} - \br{u_k^*}'\br{\theta}
    = \widetilde q_k\br{\theta} - q_k^*\br{\theta}
    = 0
    \quad dF_k\text{-almost everywhere.}
\]
Since $f_k > 0$, the two payoff functions differ only by a constant. Moreover, since worst-off types earn zero rents, $\widetilde u_k \equiv u_k^*$. Therefore, for all $k \in K$,
\[
    \widetilde t_k\br{\theta} - t_k^*\br{\theta}
    = \theta\br{\widetilde q_k\br{\theta} - q_k^*\br{\theta}}
    = 0
    \quad dF_k\text{-almost everywhere}.
\]

\paragraph{Virtual Prices.} Let $\lambda$ and $\widetilde\lambda$ be two minimizers of $\sigma + \ell$. By \cref{lem:market-clearing}, both are strictly positive and satisfy \eqref{eqn:market-clearing-limit}. We show that $\lambda = \widetilde\lambda$, proving uniqueness of the virtual prices.

By \cref{lem:support-function}\textup{(\ref{lem:support-function:supply})}, $s^\lambda$ and $s^{\widetilde\lambda}$ attain the respective support values, so $\lambda \cdot s^\lambda \geq \lambda \cdot s^{\widetilde\lambda}$ and $\widetilde\lambda \cdot s^{\widetilde\lambda} \geq \widetilde\lambda \cdot s^\lambda$. Adding these inequalities and using \eqref{eqn:market-clearing-limit} gives
\[
    0 \leq
    \br{\lambda - \widetilde\lambda} \cdot \br{s^\lambda - s^{\widetilde\lambda}}
    = \sum_{k \in K}\br{\lambda_k - \widetilde\lambda_k}
    \br{d_k\br{q^\lambda} - d_k\br{q^{\widetilde\lambda}}}
    \leq 0,
\]
where the last inequality follows from \eqref{eqn:screening-demand}, since $ \lambda_k \leq \widetilde\lambda_k$ implies
\[
    d_k\br{q^\lambda} \geq d_k\br{q^{\widetilde\lambda}}
    \quad \text{for every }k \in K.
\]
Every term in the sum is therefore zero, so demands coincide in every tag, and market clearing gives $s^\lambda = s^{\widetilde\lambda}$.

The posted prices corresponding to the allocations $q^\lambda$ and $q^{\widetilde\lambda}$ also coincide. Fix $k$ and suppose without loss of generality that $\widetilde\lambda_k \geq \lambda_k$. Both posted prices are nondecreasing in the shadow price, so $d_k\br{q^\lambda} = d_k\br{q^{\widetilde\lambda}}$ and \eqref{eqn:screening-demand} give
\[
    0
    = (1 - \mu_k)\sbr{
        F_k\br{p_k^B\br{\widetilde\lambda_k}}
        - F_k\br{p_k^B\br{\lambda_k}}
    }
    + \mu_k\sbr{
        F_k\br{p_k^S\br{\widetilde\lambda_k}}
        - F_k\br{p_k^S\br{\lambda_k}}
    },
\]
which is the sum of two nonnegative terms with strictly positive weights. Both terms therefore are zero, and since $F_k$ is strictly increasing,
\[
    p_k^S\br{\lambda_k} = p_k^S\br{\widetilde\lambda_k}
    \quad \text{and} \quad
    p_k^B\br{\lambda_k} = p_k^B\br{\widetilde\lambda_k}
    \quad \text{for every }k \in K.
\]
Write $p_k^j := p_k^j\br{\lambda_k} = p_k^j\br{\widetilde\lambda_k}$ for these common posted prices, where $j \in \cbr{S,B}$ and $k \in K$.

Next, since $s^{\widetilde\lambda} = s^\lambda$ attains $\sigma\br{\lambda}$, \cref{lem:support-function}\textup{(\ref{lem:support-function:supply})} gives $m^{\widetilde\lambda} = m^\lambda$ $dG$-almost everywhere. If $\lambda$ and $\widetilde\lambda$ were not proportional, then there exist tags $k,j$ satisfying $\lambda_k/\lambda_j < \widetilde\lambda_k/\widetilde\lambda_j$. Then, consider customers whose CTRs satisfy
\[
    \frac{\lambda_k}{\lambda_j}
    < \frac{\xi_j}{\xi_k}
    < \frac{\widetilde\lambda_k}{\widetilde\lambda_j},
\]
with all remaining CTRs sufficiently small that no other tag wins under either price vector. These customers are assigned to tag $j$ under $m^\lambda$ and to tag $k$ under $m^{\widetilde\lambda}$. The set of such CTRs is nonempty and open in $(0,1)^K$, so it has positive probability under the density assumption, contradicting $m^\lambda = m^{\widetilde\lambda}$ $dG$-almost everywhere. Hence, $\widetilde\lambda = c\lambda$ for some $c > 0$.

Lastly, $\sum_ks_k^\lambda \leq \sigma\br{\mathbf{1}} = \E_G\sbr{\max_k\xi_k} < 1 = \sum_k\beta_k$, so some tag $k$ has $d_k\br{q^\lambda} = s_k^\lambda < \beta_k$, and $d_k\br{q^\lambda} = s_k^\lambda > 0$ by \cref{lem:support-function}\textup{(\ref{lem:support-function:positive-supply})}. Then $d_k\br{q^\lambda} > 0$ rules out $p_k^S = \btheta_k$, whereas $d_k\br{q^\lambda} < \beta_k$ rules out $p_k^B = \ltheta_k$. If neither posted price were interior, we would therefore have $p_k^S = \ltheta_k$ and $p_k^B = \btheta_k$, which requires $\phi_{\eta,k}^B\br{\btheta_k} \leq \lambda_k \leq \phi_{\eta,k}^S\br{\ltheta_k}$, whereas
\[
    \phi_{\eta,k}^S\br{\ltheta_k}
    = (\eta_r + \eta_v)\ltheta_k + \eta_w
    < (\eta_r + \eta_v)\btheta_k + \eta_w
    = \phi_{\eta,k}^B\br{\btheta_k}.
\]
Hence, some posted price $p_k^j$ common to both $\lambda$ and $\widetilde\lambda$ lies in $(\ltheta_k,\btheta_k)$, so $\lambda_k = \phi_{\eta,k}^j\br{p_k^j} = \widetilde\lambda_k > 0$. Since $\widetilde\lambda = c\lambda$, it follows that $c = 1$.

Thus, $\sigma + \ell$ has a unique minimizer, proving uniqueness of the vector of virtual prices. By \cref{lem:market-clearing}, it satisfies \eqref{eqn:market-clearing-limit} and $0 < \lambda_k < \overline\lambda_k$ for every $k \in K$. The posted prices are then uniquely determined by \eqref{eqn:posted-price-cutoffs}, with $p_k^S < \btheta_k$ in every tag. This completes the proof of \cref{thm:optimal-limit-mechanism}.

\section{Proofs for Finite-to-Limit Market Convergence}
\label[appendix]{app:symmetric-large-markets}

The consistency argument and the proof of part~I) of \cref{thm:finite-limit-convergence} apply to arbitrary distributions $G$ defined by \eqref{eqn:convergence-frontier}, so no density assumption is required.

\subsection{Consistency of the Frontier Distribution}
\label{app:convergence-ownership}

\begin{proof}[Proof of \cref{prop:consistency}]
    For each $i \in \mathbb N$ and $k \in K$, conditional on a realization $\bP$, each $\omega_{i,k}$ has measure $P_k$, and so by the definition of the essential supremum $\xi_k\br{P_k}$ we get
    \[
        \p_\gamma\br{\omega_{i,k} > \xi_k\br{P_k} \mid \bP}
        = P_k\br{\rsbr{\xi_k\br{P_k},1}}
        = 0
        \quad \Gamma\text{-almost everywhere}.
    \]
    Taking expectations over $\bP$ under $\Gamma$ gives $\p_\gamma\br{\omega_{i,k} > \xi_k\br{P_k}} = 0$. Hence, $\omega_{i,k} \leq \xi_k\br{P_k}$ $\gamma$-almost everywhere.

    Now fix $\lambda \in \rbb_+^K$. Then,
    \begin{align*}
        \sum_{k \in K}\lambda_k\beta_k^n\mu_k
        &= \frac{1}{n}\sum_{i \in N_k,\,k \in K}\lambda_k\mu_k
        = \frac{1}{n}\sum_{i \in N_k,\,k \in K}
        \int_{\sbr{0,1}^n}\lambda_k\omega^n_{i,k}\,d\gamma^n_{i,k}\br{\omega^n}\\
        &\leq \frac{1}{n}\sum_{i \in N_k,\,k \in K}
        \int_{\sbr{0,1}^n}\max_{j \in N_\ell,\,\ell \in K}\cbr{\lambda_\ell\omega^n_{j,\ell}}\,d\gamma^n_{i,k}\br{\omega^n}\\
        &= \int_{\sbr{0,1}^n}\max_{j \in N_\ell,\,\ell \in K}\cbr{\lambda_\ell\omega^n_{j,\ell}}\,d\gamma^n\br{\omega^n},
    \end{align*}
    where the first and second equalities follow from $\beta_k^n = n_k/n$ and \eqref{eqn:convergence-ownership}, and the last equality follows from \eqref{eqn:convergence-mixture-dist}. Then using the law of iterated expectations,
    \begin{align*}
        \int_{\sbr{0,1}^n}\max_{j \in N_\ell,\,\ell \in K}\cbr{\lambda_\ell\omega^n_{j,\ell}}\,d\gamma^n\br{\omega^n}
        &= \E_{\gamma^n}\sbr{\max_{j \in N_\ell,\,\ell \in K}\cbr{\lambda_\ell\omega^n_{j,\ell}}}
        = \E_\Gamma\sbr{\E_\gamma\sbr{\max_{j \in N_\ell,\,\ell \in K}\cbr{\lambda_\ell\omega^n_{j,\ell}}\;\middle|\;\bP}}\\
        &\leq \E_\Gamma\sbr{\max_{k \in K}\lambda_k\xi_k\br{P_k}}
        = \E_G\sbr{\max_{k \in K}\lambda_k\xi_k},
    \end{align*}
    where the inequality uses $\lambda_k \geq 0$ and that $\omega^n_{i,k} \leq \xi_k$ $\gamma$-almost everywhere, and the last equality follows by definition $G = \Gamma\circ\xi^{-1}$. Combining the two preceding inequalities and letting $n \longrightarrow \infty$ gives
    \[
        \sum_{k \in K}\lambda_k\beta_k\mu_k
        = \lim_{n \longrightarrow \infty}\sum_{k \in K}\lambda_k\beta_k^n\mu_k
        \leq \E_G\sbr{\max_{k \in K}\lambda_k\xi_k},
    \]
    where we used $\beta_k^n \longrightarrow \beta_k$ and interchanged the order of the limit with the sum since $K$ is finite. Thus $G$ satisfies \eqref{eqn:G-consistency}.
\end{proof}

\subsection{Proof of \texorpdfstring{\cref{thm:finite-limit-convergence}}{Theorem \ref*{thm:finite-limit-convergence}}}
\label{app:limit-market-convergence}

\subsubsection{Limits of Feasible Finite-Market Allocations}
\label{app:convergence-finite-to-limit}

We show that the sets of feasible and nondecreasing normalized interim allocations in symmetric finite markets converge to the set of feasible and nondecreasing allocations in the limit market.

\paragraph{Finite-to-Limit Feasibility.} Fix a sequence $\br{q^n}_{n \geq 1}$ in $\qcal_{\gamma^n}$ and $\lambda \in \rbb_+^K$. For any type report $\boldtheta \in \Theta^n$, \eqref{eqn:convergence-feasibility} implies
\[
    \sum_{i \in N_k,k \in K}\lambda_k q_{i,k}^n\br{\boldtheta}
    \leq \int\max_{j \in N_\ell,\,\ell \in K}\cbr{\lambda_\ell\omega^n_{j,\ell}}
    \,d\gamma^n\br{\omega^n}.
\]
Taking expectations over $\boldtheta \in \Theta^n$ and conditioning on $\bP$ as in the proof of \cref{prop:consistency} gives
\begin{equation}
    \label{eqn:feasibility-qn}
    \sum_{k \in K}\lambda_k\beta_k^n\E_{F_k} \sbr{q_k^n\br{\theta}}
    \leq \E_G\sbr{\max_{k \in K}\lambda_k\xi_k}.
\end{equation}

Since $0 \leq q_k^n \leq 1$ and each $q_k^n$ is nondecreasing, Helly's selection theorem implies the existence of a subsequence $\br{n_j}_{j \geq 1}$ along which $q_k^{n_j} \longrightarrow q_k^\infty$ $dF_k$-almost everywhere for every $k \in K$, where each $q_k^\infty$ is nondecreasing and takes values in $\sbr{0,1}$. Hence, $q^\infty \in \qcal$.

Lastly, using \eqref{eqn:feasibility-qn} and $\beta_k^{n_j} \longrightarrow \beta_k$ for each $k \in K$ we get
\[
    \sum_{k \in K}\lambda_k\beta_k\E_{F_k}\sbr{q^\infty_k\br{\theta}}
    = \lim_{j \longrightarrow \infty}\sum_{k \in K}\lambda_k\beta_k^{n_j}\E_{F_k}\sbr{q_k^{n_j}\br{\theta}}
    \leq \E_G\sbr{\max_{k \in K}\lambda_k\xi_k}.
\]
Hence, $q^\infty$ satisfies \eqref{eqn:R-Fsb}, so $q^\infty \in \qcal_R$. This proves the first direction in part~I) of \cref{thm:finite-limit-convergence}.

\subsubsection{Constructing Finite-Market Approximations}
\label{app:convergence-approximation}

\paragraph{Tentative Expected Clicks.} Fix $q \in \qcal_R$ and define the nonnegative constants
\[
    \bar q_k := \E_{F_k}\sbr{q_k\br{\theta}} \quad \text{for all } k \in K.
\]
Recall that $d_k\br{q} = \beta_k\bar q_k$ gives the total demand for clicks in tag $k$. Then, by \cref{lem:implementable}, we can choose measurable functions $m = \br{m_k}_{k \in K}$\textemdash referred to as \textit{limit} audience shares\textemdash such that, for all $k \in K$, $m_k \equiv 0$ if $\bar q_k = 0$ and $d_k\br{q} = \E_G\sbr{\xi_km_k\br{\xi}}$ for some
\begin{equation}
    \label{eqn:convergence-reserved-supply}
    m_k\br{\xi} \geq 0 \quad \text{and} \quad \sum_{k \in K}m_k\br{\xi} \leq 1 \quad dG\text{-almost everywhere.}
\end{equation}

Now fix the $n$th market for some $n \in \mathbb N$. Define the \textit{tentative} audience shares $m^n = \br{m^n_k}_{k \in K}$ with each $m^n_k:[0,1]^n \to \rbb_+$ defined by\footnote{Conditional on $\bP$, CTRs $\omega_{j,k}$ are i.i.d.\ with measure $P_k$, so by the strong law of large numbers, $\frac{1}{h}\sum_{j = 1}^{h}\mathbf{1}_{\{\omega_{j,k} \leq a\}} \longrightarrow P_k\br{\sbr{0,a}} \quad \text{for every } a \in \sbr{0,1}$, $P_k^{\otimes \mathbb N}$-almost everywhere. Thus, $\bP$ is $\gamma$-almost surely a measurable function of $\omega$, and the conditional expectation in \eqref{eqn:convergence-observable} is well defined under $\gamma$.}
\begin{equation}
    \label{eqn:convergence-observable}
    m_k^n\br{\omega^n} := \E_{\gamma}\sbr{m_k\br{\xi} \mid \omega^n}
    \quad d\gamma^n\text{-almost everywhere},
\end{equation}
where by \eqref{eqn:convergence-reserved-supply}, they satisfy
\begin{equation}
    \label{eqn:convergence-observable-finite}
    m^n_k\br{\omega^n} \geq 0 \quad \text{and} \quad \sum_{k \in K}m^n_k\br{\omega^n} \leq 1 \quad d\gamma^n\text{-almost everywhere.}
\end{equation}

In the finite market, the designer observes the realized CTRs $\omega^n$, which need not determine the frontier $\xi$ or the limit audience shares $m_k\br{\xi}$. We therefore assign each tag its conditional expected share $m_k^n\br{\omega^n}$ in \eqref{eqn:convergence-observable}. As $n$ grows, $\omega^n$ reveals more about the latent CTR distributions $\bP$, and hence about $\xi$, so these shares converge to the limit shares.\footnote{Precisely, $m_k^n\br{\omega^n} \longrightarrow m_k\br{\xi}$ in $L^1\br{\gamma}$, and hence in probability, for every $k \in K$.} We call the resulting assignments \textit{tentative} because capacity constraints have yet to be imposed and the shares may change as more CTRs are observed. We show that the tentative total expected clicks converge to $d_k\br{q}$ in each tag as the number of eligible merchants grows.

Fix a nonempty set $B \subseteq N_k$ and assign the tentative audience share $m_k^n\br{\omega^n}$ to the merchant in $B$ with the highest CTR, with ties broken uniformly. Formally, the probability $\pi_{i,k}^B\br{\omega^n}$ that merchant $i$ wins is given by
\begin{equation}
    \label{eqn:convergence-tie-breaking}
    \pi_{i,k}^B\br{\omega^n}
    :=
    \begin{cases}
        \displaystyle
        \frac{1}{\left|\cbr{j \in B:\omega^n_{j,k} = \max_{\ell \in B}\omega^n_{\ell,k}}\right|}
        & \text{if } i \in B \text{ and } \omega^n_{i,k} = \max_{j \in B}\omega^n_{j,k},\\[1em]
        0 & \text{otherwise}.
    \end{cases}
\end{equation}

Next, let $S^n_{i,k}\br{B}$ denote the tentative expected clicks assigned to merchant $i$ in $B$, defined by
\[
    S^n_{i,k}\br{B} := \E_{\gamma^n}\sbr{\omega^n_{i,k}m_k^n\br{\omega^n}\pi_{i,k}^B\br{\omega^n}},
\]
and let $S^n_k\br{B}$ denote the total expected clicks supplied to $B$, given by
\[
    S^n_k\br{B} := \E_{\gamma^n}\sbr{m_k^n\br{\omega^n}\max_{i \in B}\omega^n_{i,k}}.
\]
Moreover, for each $k \in K$ and $h \in \mathbb N$, define
\[
    S_k\br{h} := \E_{\gamma}\sbr{m_k\br{\xi}\max_{1 \leq j \leq h}\omega_{j,k}},
\]
where the expectation is under the infinite CTR measure $\gamma$, and the maximum is taken over the merchants indexed by $i = 1,\ldots,h$ with tag $k$.

\begin{lemma}[Tentative Expected Clicks]
    \label{lem:tentative-clicks}
    Fix $k \in K$.
    \begin{enumerate}[label=\roman*)]
        \item For every $n$ and nonempty set $B \subseteq N_k$, the tentative total expected clicks supplied to $B$ depend only on its size:
        \[
            S^n_k\br{B} = S_k\br{|B|}.
        \]
        \item For every $n$ and nonempty set $B \subseteq N_k$, each merchant $i \in B$ receives an equal share of the tentative expected clicks:
        \[
            S^n_{i,k}\br{B} = \frac{S_k\br{|B|}}{|B|}.
        \]
        \item The sequence $\br{S_k\br{h}}_{h \geq 1}$ converges to $d_k\br{q}$:
        \[
            S_k\br{h} \longrightarrow \E_G\sbr{\xi_km_k\br{\xi}} = d_k\br{q}.
        \]
    \end{enumerate}
\end{lemma}

\begin{proof}[Proof of \cref{lem:tentative-clicks}, part~i)]
    Fix $n \in \mathbb N$, $k \in K$, and a nonempty set $B \subseteq N_k$ with size $h := |B|$. By the definitions of $m_k^n\br{\omega^n}$ and $S_k^n\br{B}$, and the law of iterated expectations,
    \[
        S_k^n\br{B}
        = \E_{\gamma^n}\sbr{m_k^n\br{\omega^n}\max_{i \in B}\omega^n_{i,k}}
        = \E_{\gamma^n}\sbr{\E_\gamma\sbr{m_k\br{\xi} \mid \omega^n}\max_{i \in B}\omega^n_{i,k}}
        = \E_\gamma\sbr{m_k\br{\xi}\max_{i \in B}\omega^n_{i,k}}.
    \]
    Moreover, by de Finetti's representation theorem (\cref{prop:de-finetti}), conditional independence and identical distributions within each tag $k$ imply that, for every $a \in \sbr{0,1}$,
    \[
        \p_\gamma\br{\max_{i \in B}\omega^n_{i,k} \leq a\;\middle|\;\bP}
        = P_k\br{\sbr{0,a}}^h
        = \p_\gamma\br{\max_{1 \leq j \leq h}\omega_{j,k} \leq a\;\middle|\;\bP}.
    \]
    Then, since $\xi$ is a measurable function of $\bP$, the law of iterated expectations yields
    \begin{align*}
        \E_\gamma\sbr{m_k\br{\xi}\max_{i \in B}\omega^n_{i,k}}
        &= \E_\Gamma\sbr{m_k\br{\xi\br{\bP}}\E_\gamma\sbr{\max_{i \in B}\omega^n_{i,k}\;\middle|\;\bP}}
        = \E_\Gamma\sbr{m_k\br{\xi\br{\bP}}\E_\gamma\sbr{\max_{1 \leq j \leq h}\omega_{j,k}\;\middle|\;\bP}}\\
        &= \E_\gamma\sbr{m_k\br{\xi}\max_{1 \leq j \leq h}\omega_{j,k}}
        = S_k\br{h}.
    \end{align*}
    Combining the first and last identities in the proof gives $S_k^n\br{B} = S_k\br{|B|}$.
\end{proof}

\begin{proof}[Proof of \cref{lem:tentative-clicks}, part~ii)]
    Fix $n \in \mathbb N$, $k \in K$, and a nonempty set $B \subseteq N_k$. By de Finetti's representation theorem (\cref{prop:de-finetti}), conditional on $\bP$, the CTRs of merchants in $B$ are independent and identically distributed according to $P_k$. Then, together with the uniform tie-breaking rule defined in \eqref{eqn:convergence-tie-breaking}, this implies
    \[
        \E_\gamma\sbr{\omega^n_{i,k}\pi_{i,k}^B\br{\omega^n}\;\middle|\;\bP}
        = \E_\gamma\sbr{\omega^n_{j,k}\pi_{j,k}^B\br{\omega^n}\;\middle|\;\bP}
        \quad \text{for all }i,j \in B.
    \]

    Then, for every $i \in B$,
    \begin{equation}
        \label{eqn:convergence-uniform-ties}
        \E_\gamma\sbr{\omega^n_{i,k}\pi_{i,k}^B\br{\omega^n}\;\middle|\;\bP}
        = \frac{1}{|B|}\E_\gamma\sbr{\sum_{j \in B}\omega^n_{j,k}\pi_{j,k}^B\br{\omega^n}\;\middle|\;\bP}
        = \frac{1}{|B|}\E_\gamma\sbr{\max_{j \in B}\omega^n_{j,k}\;\middle|\;\bP}.
    \end{equation}
    Therefore, merchant $i \in B$ receives tentative expected clicks
    \begin{align*}
        S^n_{i,k}\br{B}
        &= \E_{\gamma^n}\sbr{\omega^n_{i,k}m_k^n\br{\omega^n}\pi_{i,k}^B\br{\omega^n}}
        = \E_\gamma\sbr{m_k\br{\xi}\omega^n_{i,k}\pi_{i,k}^B\br{\omega^n}}\\
        &= \E_\Gamma\sbr{m_k\br{\xi\br{\bP}}\E_\gamma\sbr{\omega^n_{i,k}\pi_{i,k}^B\br{\omega^n}\;\middle|\;\bP}}
        = \frac{1}{|B|}\E_\Gamma\sbr{m_k\br{\xi\br{\bP}}\E_\gamma\sbr{\max_{j \in B}\omega^n_{j,k}\;\middle|\;\bP}}\\
        &= \frac{1}{|B|}\E_\gamma\sbr{m_k\br{\xi}\max_{j \in B}\omega^n_{j,k}}
        = \frac{S_k\br{|B|}}{|B|},
    \end{align*}
    where the first equality follows by definition of $S^n_{i,k}\br{B}$, the second follows by definition \eqref{eqn:convergence-observable} and the law of iterated expectations, the third and fifth again use the law of iterated expectations conditional on $\bP$, the fourth follows by \eqref{eqn:convergence-uniform-ties}, and the last follows from part~i). This concludes the proof of part~ii).
\end{proof}

\begin{proof}[Proof of \cref{lem:tentative-clicks}, part~iii)]
    Fix $k \in K$. We have already shown in the proof of \cref{prop:consistency} that $\omega_{i,k} \leq \xi_k$ $\gamma$-almost everywhere. We claim that
    \[
        \max_{1 \leq j \leq h}\omega_{j,k} \longrightarrow \xi_k
        \quad \gamma\text{-almost everywhere, as }h \longrightarrow \infty.
    \]
    For each $h \geq 1$, define $y_h := \max_{1 \leq j \leq h}\omega_{j,k}$. The sequence $\br{y_h}_{h \geq 1}$ is nondecreasing and bounded above by $\xi_k$ $\gamma$-almost everywhere, so it converges to some $\bar y \leq \xi_k$. Then, for every $M \in \mathbb N$ and $h \geq 1$, since $y_h \leq \bar y$ $\gamma$-almost everywhere,
    \begin{equation}
        \label{eqn:convergence-frontier-tail}
        \p_\gamma\br{\bar y \leq \xi_k - 1/M}
        \leq \p_\gamma\br{y_h \leq \xi_k - 1/M}
        = \E_\Gamma\sbr{\p_\gamma\br{y_h \leq \xi_k - 1/M \mid \bP}}.
    \end{equation}
    By de Finetti's representation theorem (\cref{prop:de-finetti}), conditional on $\bP$, the CTRs $\br{\omega_{j,k}}_{j \in \mathbb N}$ are independent and identically distributed according to $P_k$. Since
    \[
        \p_\gamma\br{\omega_{1,k} \leq \xi_k - 1/M \mid \bP} < 1
    \]
    by definition of the essential supremum $\xi_k$,
    \[
        \p_\gamma\br{y_h \leq \xi_k - 1/M \mid \bP}
        = \p_\gamma\br{\omega_{1,k} \leq \xi_k - 1/M \mid \bP}^{h}
        \longrightarrow 0
        \quad \text{as }h \longrightarrow \infty,
    \]
    $\Gamma$-almost everywhere. Then, by the DCT,
    \[
        \E_\Gamma\sbr{\p_\gamma\br{y_h \leq \xi_k - 1/M \mid \bP}} \longrightarrow 0
        \quad \text{as }h \longrightarrow \infty,
    \]
    so \eqref{eqn:convergence-frontier-tail} implies
    \[
        \p_\gamma\br{\bar y \leq \xi_k - 1/M} = 0
        \quad \text{for every }M \in \mathbb N.
    \]
    Since
    \[
        \cbr{\bar y < \xi_k} = \bigcup_{M \geq 1}\cbr{\bar y \leq \xi_k - 1/M},
    \]
    countable subadditivity of the measure gives $\p_\gamma\br{\bar y < \xi_k} = 0$. Together with $\bar y \leq \xi_k$ $\gamma$-almost everywhere, this proves the claim that $\bar y = \xi_k$ $\gamma$-almost everywhere.

    Finally, since $0 \leq m_k \leq 1$ $dG$-almost everywhere, applying DCT gives
    \[
        S_k\br{h} = \E_\gamma\sbr{m_k\br{\xi}y_h}
        \longrightarrow \E_G\sbr{\xi_km_k\br{\xi}} = d_k\br{q},
    \]
    concluding the proof of part~iii).
\end{proof}

We next construct random eligibility sets $B_k^n$ as functions of the limit allocation $q$, with each merchant $i \in N_k$ included independently with probability $q_k\br{\theta_{i,k}}$. We show that under this construction the mass of eligible merchants converges in probability to $d_k\br{q}$, the aggregate expected clicks in tag $k$.

\paragraph{Random Eligible Sets.} For each $k \in K$ and type report $\boldtheta \in \Theta^n$, let $B_k^n\br{\boldtheta} \subseteq N_k$ denote the random set of merchants eligible to receive targeted ads. We construct $B_k^n\br{\boldtheta}$ by including each merchant $j \in N_k$ with probability $q_k\br{\theta_{j,k}}$, independently across merchants and of the observed CTRs.

Specifically, for each $j \in N_k$ and $k \in K$, draw $U_{j,k}^n\sim\operatorname{Uniform}[0,1]$ independently of types and observed CTRs, and write $U_k^n = \br{U_{j,k}^n}_{j \in N_k}$. Then, for each $\boldtheta \in \Theta^n$, define the random eligible set $B_k^n\br{\boldtheta} \subseteq N_k$ by
\[
    B_k^n\br{\boldtheta}
    := \cbr{j \in N_k:U_{j,k}^n \leq q_k\br{\theta_{j,k}}}.
\]
Thus, for each fixed type report $\boldtheta \in \Theta^n$, merchant $j$'s eligibility probability depends only on his own report:
\[
    \p\br{j \in B_k^n\br{\boldtheta}}
    = \p\br{U_{j,k}^n \leq q_k\br{\theta_{j,k}}}
    = q_k\br{\theta_{j,k}}.
\]
Let $\nu_k^n\br{B \mid \boldtheta}$ denote the probability that $B^n_k = B$ at a given type report $\boldtheta$, defined by
\begin{equation}
    \label{eqn:convergence-eligible-law}
    \nu_k^n\br{B \mid \boldtheta}
    := \p\br{B_k^n\br{\boldtheta} = B}
    =
    \prod_{j \in B}q_k\br{\theta_{j,k}}
    \prod_{j \in N_k\setminus B}\sbr{1 - q_k\br{\theta_{j,k}}}
    \quad \text{for all } B \subseteq N_k.
\end{equation}

Now, let $B_k^n$ denote the ex ante random eligible set obtained when types are drawn from $F^n$, which has conditional distribution $\nu_k^n\br{\cdot \mid \boldtheta}$. Then, by the law of iterated expectations, merchant $i$'s ex ante eligibility probability is
\[
    \p\br{i \in B_k^n}
    = \E_{F^n}\sbr{\p\br{i \in B_k^n\br{\boldtheta}}}
    = \E_{F_k}\sbr{q_k\br{\theta}}
    = \bar q_k.
\]
Since types and eligibility draws are independent across merchants, the size $b_k^n := |B_k^n|$ is the sum of $n_k$ independent $\Bernoulli\br{\bar q_k}$ variables:
\[
    b_k^n \sim \Binomial\br{n_k,\bar q_k}.
\]

Next, for each $i \in N_k$, define
\[
    b^n_{i,k} := 1 + |B_k^n\setminus\cbr{i}|.
\]
By construction, $b^n_{i,k}$ is independent of merchant $i$'s type $\theta_{i,k}$ and uniform draw $U_{i,k}^n$, and hence of his eligibility. It follows that
\[
    b^n_{i,k} \sim 1 + \Binomial\br{n_k - 1,\bar q_k}.
\]
In particular, $b^n_{i,k} = b_k^n$ whenever $i \in B_k^n$.

Now, for each $n \in \mathbb N$ and realized eligible set $B \subseteq N_k$ with $h := |B| \geq 1$, using part~ii) of \cref{lem:tentative-clicks}, define the normalized tentative expected clicks for each eligible merchant $i \in B$ by
\[
    s_k^n\br{h} := nS^n_{i,k}\br{B} = \frac{S_k\br{h}}{h/n},
\]
where we set $s_k^n\br{0} := 0$. In the next lemma, we give convergence results for the mass of eligible merchants and their normalized tentative expected clicks.

\begin{lemma}[Convergence under Eligibility]
    \label{lem:eligibility-convergence}
    Suppose $\beta_k^n \longrightarrow \beta_k > 0$ for all $k \in K$, and fix $k \in K$.
    \begin{enumerate}[label=\roman*)]

        \item The mass of eligible merchants converges in probability to $d_k\br{q}$:
        \[
            \frac{b_k^n}{n}\xlongrightarrow{\p}d_k\br{q}.
        \]
        Consequently, $b^n_{i,k}/n\xlongrightarrow{\p}d_k\br{q}$ for all $i \in N_k$ as well.

        \item If $d_k\br{q} > 0$, normalized tentative expected clicks for every merchant $i \in N_k$, conditional on being eligible, converge in probability to one:
        \[
            s_k^n\br{b^n_{i,k}}\xlongrightarrow{\p}1 \quad \text{for all } i \in N_k.
        \]
    \end{enumerate}
\end{lemma}

\begin{proof}[Proof of \cref{lem:eligibility-convergence}, part~i)]
    Fix $k \in K$. Since $b_k^n\sim\Binomial\br{n_k,\bar q_k}$,
    \begin{align*}
        \E\sbr{\frac{b_k^n}{n}}
        &= \beta_k^n\bar q_k \longrightarrow d_k\br{q} \quad \text{as }n \longrightarrow \infty,\\
        \Var\br{\frac{b_k^n}{n}}
        &= \frac{n_k\bar q_k\br{1 - \bar q_k}}{n^2}
        \leq \frac{1}{n} \longrightarrow 0 \quad \text{as }n \longrightarrow \infty.
    \end{align*}
    Chebyshev's inequality then gives convergence in probability
    \[
        \frac{b_k^n}{n}\xlongrightarrow{\p}d_k\br{q}.
    \]
    Moreover, for every $i \in N_k$, since $ 0 \leq b^n_{i,k} / n - b_k^n/n \leq 1/n$, $b^n_{i,k}/n\xlongrightarrow{\p}d_k\br{q}$ as well, thus proving part~i).
\end{proof}

\begin{proof}[Proof of \cref{lem:eligibility-convergence}, part~ii)]
    Fix $i \in N_k$ and $k \in K$, and suppose $d_k\br{q} > 0$. By part~i), $b^n_{i,k}/n\xlongrightarrow{\p}d_k\br{q} > 0$, implying that $b^n_{i,k}\xlongrightarrow{\p}\infty$. Part~iii) of \cref{lem:tentative-clicks} then gives $S_k\br{b^n_{i,k}}\xlongrightarrow{\p}d_k\br{q}$. Since $d_k\br{q} > 0$, the continuous mapping theorem gives
    \[
        s_k^n\br{b^n_{i,k}}
        = \frac{S_k\br{b^n_{i,k}}}{b^n_{i,k}/n}
        \xlongrightarrow{\p}\frac{d_k\br{q}}{d_k\br{q}} = 1,
    \]
    proving part~ii).
\end{proof}

We now use these convergence results to construct feasible normalized interim allocations $q^n$ that converge to $q$, which then completes the proof of the second direction in part~I) of \cref{thm:finite-limit-convergence}.

\paragraph{Limit-to-Finite Approximation.} To complete the construction of feasible targeted ads $x^n$, we use the tentative audience shares $m_k^n$ defined in \eqref{eqn:convergence-observable}, the tie-breaking probabilities $\pi_{i,k}^B\br{\omega^n}$ defined in \eqref{eqn:convergence-tie-breaking}, and the random eligible sets $B_k^n\br{\boldtheta}$ with conditional law $\nu_k^n\br{\cdot \mid \boldtheta}$ defined in \eqref{eqn:convergence-eligible-law}. Specifically, for all $\boldtheta \in \Theta^n$, $i \in N_k$, and $k \in K$, define targeted ads $x_{i,k}^n$ by
\begin{equation}
    \label{eqn:convergence-targeted-ads}
    x_{i,k}^n\br{\boldtheta;d\omega^n}
    :=
    \E_{\nu_k^n\br{\cdot \mid \boldtheta}}\sbr{
        \frac{m_k^n\br{\omega^n}\pi_{i,k}^{B_k^n\br{\boldtheta}}\br{\omega^n}}{\max\cbr{1,nS^n_{i,k}\br{B_k^n\br{\boldtheta}}}}
    }
    \,d\gamma^n\br{\omega^n} \quad d\gamma^n\text{-almost everywhere,}
\end{equation}
where we set $\pi_{i,k}^{\varnothing} := 0$ and $S^n_{i,k}\br{B} := 0$ whenever $i \notin B$ (including $B = \varnothing$). Scaling by $1/\max\cbr{1,nS^n_{i,k}\br{B_k^n\br{\boldtheta}}}$ caps each eligible merchant's expected clicks at $1/n$, thus ensuring \eqref{eqn:convergence-capacity}.

The following proposition concludes the proof of part~I) in \cref{thm:finite-limit-convergence}.

\begin{proposition}[Limit-to-Finite Approximation]
    \label{prop:limit-to-finite}
    Fix an allocation $q \in \qcal_R$ and define targeted ads $x^n$ by \eqref{eqn:convergence-targeted-ads}. Then $x^n \in \xcal^n$, their normalized interim allocation satisfies $q^n \in \qcal_{\gamma^n}$, and, for all $i \in N_k$ and $k \in K$,
    \[
        q_k^n\br{\theta}
        = \E_{b^n_{i,k}}\sbr{\min\cbr{1,s_k^n\br{b^n_{i,k}}}}q_k\br{\theta}
        \longrightarrow q_k\br{\theta}
        \quad dF_k\text{-almost everywhere}.
    \]
\end{proposition}

\begin{proof}[Proof of \cref{prop:limit-to-finite}]
    The densities defined in \eqref{eqn:convergence-targeted-ads} are measurable and nonnegative. We show that the resulting targeted ads $x^n$ are symmetric and satisfy the feasibility constraint \eqref{eqn:convergence-feasibility} and the capacity constraint \eqref{eqn:convergence-capacity}. Lastly, we prove convergence of the interim allocations to the limit allocation $q$.

    \paragraph{Feasibility.} For all $\boldtheta \in \Theta^n$ and Borel sets $A \subseteq [0,1]^n$,
    \begin{align*}
        \sum_{i \in N_k,\,k \in K}x_{i,k}^n\br{\boldtheta;A}
        &= \int_A\sum_{k \in K}m_k^n\br{\omega^n}
        \E_{\nu_k^n\br{\cdot \mid \boldtheta}}\sbr{
            \sum_{i \in N_k}
            \frac{\pi_{i,k}^{B_k^n\br{\boldtheta}}\br{\omega^n}}{\max\cbr{1,nS^n_{i,k}\br{B_k^n\br{\boldtheta}}}}
        }\,d\gamma^n\br{\omega^n}\\
        &\leq \int_A\sum_{k \in K}m_k^n\br{\omega^n}
        \E_{\nu_k^n\br{\cdot \mid \boldtheta}}\sbr{
            \sum_{i \in N_k}\pi_{i,k}^{B_k^n\br{\boldtheta}}\br{\omega^n}
        }\,d\gamma^n\br{\omega^n}\\
        &\leq \int_A\sum_{k \in K}m_k^n\br{\omega^n}
        \,d\gamma^n\br{\omega^n}
        \leq \gamma^n\br{A},
    \end{align*}
    where the second inequality follows from $\sum_{i \in N_k}\pi_{i,k}^{B_k^n\br{\boldtheta}}\br{\omega^n} = \mathbf{1}_{\cbr{B_k^n\br{\boldtheta} \neq \varnothing}} \leq 1$ and the last uses $\sum_{k \in K}m_k^n \leq 1$ $d\gamma^n$-almost everywhere from \eqref{eqn:convergence-observable-finite}. Thus, $x^n$ satisfies \eqref{eqn:convergence-feasibility}.

    \paragraph{Capacity.} For all $\boldtheta \in \Theta^n$, $i \in N_k$ and $k \in K$, the corresponding expected clicks $q_{i,k}^n\br{\boldtheta}$ are given by
    \begin{align}
        q_{i,k}^n\br{\boldtheta}
        &= \int_{[0,1]^n}\omega^n_{i,k}
        x_{i,k}^n\br{\boldtheta;d\omega^n}\notag\\
        &= \sum_{i \in B \subseteq N_k}
        \frac{\nu_k^n\br{B \mid \boldtheta}}{\max\cbr{1,nS^n_{i,k}\br{B}}}
        \underbrace{
            \E_{\gamma^n}\sbr{
                \omega^n_{i,k}m_k^n\br{\omega^n}\pi_{i,k}^B\br{\omega^n}
            }
        }_{S^n_{i,k}\br{B}}\notag\\
        &= \frac{1}{n}
        \sum_{i \in B \subseteq N_k}
        \nu_k^n\br{B \mid \boldtheta}
        \frac{s_k^n\br{|B|}}{\max\cbr{1,s_k^n\br{|B|}}}\notag\\
        &= \frac{1}{n}\,
        \E_{\nu_k^n\br{\cdot \mid \boldtheta}}\sbr{
            \mathbf{1}_{\cbr{i \in B_k^n\br{\boldtheta}}}
            \min\cbr{1,s_k^n\br{|B_k^n\br{\boldtheta}|}}
        }\notag\\
        &= \frac{1}{n}\,
        \E_{\nu_k^n\br{\cdot \mid \boldtheta}}\sbr{
            \mathbf{1}_{\cbr{i \in B_k^n\br{\boldtheta}}}
            \min\cbr{1,s_k^n\br{b^n_{i,k}}}
        }\notag\\
        &= \frac{q_k\br{\theta_{i,k}}}{n}\,
        \E_{\nu_k^n\br{\cdot \mid \boldtheta}}\sbr{
            \min\cbr{1,s_k^n\br{b^n_{i,k}}}
        }
        \leq \frac{1}{n},
        \label{eqn:convergence-expost-clicks}
    \end{align}
    where the second equality uses \eqref{eqn:convergence-targeted-ads} and the independence of eligibility draws from CTRs $\omega^n$, the third uses part~ii) of \cref{lem:tentative-clicks} and the definition of $s_k^n$, the fourth uses $s/\max\cbr{1,s} = \min\cbr{1,s}$ for $s \geq 0$, the fifth uses $b^n_{i,k} = |B_k^n\br{\boldtheta}|$ whenever $i \in B_k^n\br{\boldtheta}$. The last equality uses that, given $\boldtheta$, merchant $i$'s eligibility is independent of $b^n_{i,k}$ and has probability $q_k\br{\theta_{i,k}}$. The last inequality follows since $q_k\br{\theta_{i,k}} \leq 1$ and $\min\cbr{1,s} \leq 1$, so $x^n$ satisfies \eqref{eqn:convergence-capacity}.

    \paragraph{Symmetry.} Fix $i \in N_k$, $k \in K$, and a type report $\br{\theta,\boldtheta_{-\br{i,k}}} \in \Theta^n$, and define
    \begin{equation}
        \label{eqn:convergence-delivery-factor}
        c_{i,k}^n\br{\boldtheta_{-\br{i,k}}}
        := \E_{\nu_k^n\br{\cdot \mid \boldtheta}}\sbr{
            \min\cbr{1,s_k^n\br{b^n_{i,k}}}
        },
    \end{equation}
    which depends on $\boldtheta$ only through $\boldtheta_{-\br{i,k}}$ because, given $\boldtheta$, the law of $b^n_{i,k} = 1 + |B_k^n\br{\boldtheta}\setminus\cbr{i}|$ involves only the other merchants' reports. By \eqref{eqn:convergence-expost-clicks} and \eqref{eqn:convergence-delivery-factor},
    \begin{equation}
        \label{eqn:convergence-profile-factorization}
        nq_{i,k}^n\br{\theta,\boldtheta_{-\br{i,k}}}
        = c_{i,k}^n\br{\boldtheta_{-\br{i,k}}}q_k\br{\theta}.
    \end{equation}
    Taking expectations over other merchants' types gives
    \begin{align*}
        nQ_{i,k}^n\br{\theta}
        &= n\E_{F^n_{-\br{i,k}}}\sbr{
            q_{i,k}^n\br{\theta,\boldtheta_{-\br{i,k}}}
        }\\
        &= q_k\br{\theta}\,
        \E_{F^n_{-\br{i,k}}}\sbr{
            c_{i,k}^n\br{\boldtheta_{-\br{i,k}}}
        }\\
        &= q_k\br{\theta}\,
        \E_{b^n_{i,k}}\sbr{\min\cbr{1,s_k^n\br{b^n_{i,k}}}},
    \end{align*}
    where the last equality uses the law of iterated expectations, with the expectation taken over $b^n_{i,k}\sim1 + \Binomial\br{n_k - 1,\bar q_k}$. Since $b^n_{i,k}$ is identically distributed for all $i \in N_k$, merchants with the same tag have identical interim allocation rules, proving symmetry.

    Therefore, along with the preceding steps, we conclude that $x^n \in \xcal^n$.

    \paragraph{Convergence.} By the preceding analysis, the normalized expected clicks $q_k^n\br{\theta}$ satisfy
    \begin{equation}
        \label{eqn:normalized-clicks-proof}
        q_k^n\br{\theta} = nQ_{i,k}^n\br{\theta} = q_k\br{\theta}\,
        \E_{b^n_{i,k}}\sbr{\min\cbr{1,s_k^n\br{b^n_{i,k}}}}.
    \end{equation}

    Since $q_k$ is nondecreasing, $q_k^n$ is also nondecreasing. Therefore, $q^n \in \qcal_{\gamma^n}$.

    Now, if $d_k\br{q} > 0$, \cref{lem:eligibility-convergence} gives
    \[
        \min\cbr{1,s_k^n\br{b^n_{i,k}}}
        \xlongrightarrow{\p}1.
    \]
    The preceding convergence in probability and $0 \leq \min\cbr{1,s_k^n\br{b^n_{i,k}}} \leq 1$ imply
    \[
        \E_{b^n_{i,k}}\sbr{\min\cbr{1,s_k^n\br{b^n_{i,k}}}}
        \longrightarrow 1.
    \]
    Hence, $q_k^n\br{\theta} \longrightarrow q_k\br{\theta}$ for every $\theta \in \Theta_k$. On the other hand, if $d_k\br{q} = 0$, then $\beta_k > 0$ implies $\bar q_k = 0$, hence $q_k = 0$ $dF_k$-almost everywhere. Therefore, by \eqref{eqn:normalized-clicks-proof}, it follows that $q_k^n = 0$ $dF_k$-almost everywhere, concluding the proof.
\end{proof}

\subsubsection{Optimal Values and Transfers}
\label{subsec:convergence-part-ii}

For part~II), we impose the assumptions of \cref{thm:optimal-limit-mechanism}, including a strictly positive density of $G$ on $\br{0,1}^K$ and audience scarcity. These assumptions give the optimal limit-market mechanism and its essential uniqueness used below.

By symmetry and following similar steps as in the proof of \cref{thm:optimal-finite-mechanism}, we can write the optimal value of \eqref{eqn:designer-problem-finite-n} as
\begin{equation}
    \label{eqn:convergence-finite-reduced-value}
    \Pi_\eta^n = \adjustlimits
    \sup_{q \in \qcal_{\gamma^n}}
    \min_{\boldtheta^{\crt} \in \Theta_K}
    \psi_\eta^n\br{q,\boldtheta^{\crt}}.
\end{equation}
Here, let $\psi_\eta^n$ be the virtual objective in \eqref{eqn:virtual-objective-limit} with $\beta_k^n$ in place of $\beta_k$:
\[
    \psi_\eta^n\br{q,\boldtheta^{\crt}}
    := \sum_{k \in K}\beta_k^n\E_{F_k}\sbr{\phi_{\eta,k}\br{\theta;\theta^{\crt}_k}\br{q_k\br{\theta} - \mu_k}}.
\]

\paragraph{Upper Bound on Asymptotic Optimal Value.} We show that the optimal limit value $\Pi_\eta^\infty$ gives an asymptotic upper bound on the optimal value in \eqref{eqn:convergence-finite-reduced-value}, that is, $\limsup_n\Pi_\eta^n \leq \Pi_\eta^\infty$.

To see this, we first show that the value of the reduced objective in the finite market converges to the limit-market value whenever nondecreasing normalized interim allocation rules $q_k^n$ converge to nondecreasing limit allocation rules $q_k^\infty$ $dF_k$-almost everywhere in each tag.

\begin{lemma}[Convergence of Reduced Objective Values]
    \label{lem:objective-convergence}
    Let $q^n,q^\infty \in \qcal$ satisfy $q_k^n \longrightarrow q_k^\infty$ $dF_k$-almost everywhere for every $k \in K$. Then,
    \[
        \min_{\boldtheta^{\crt} \in \Theta_K}\psi_\eta^n\br{q^n,\boldtheta^{\crt}}
        \longrightarrow
        \min_{\boldtheta^{\crt} \in \Theta_K}\psi_\eta\br{q^\infty,\boldtheta^{\crt}}.
    \]
\end{lemma}

\begin{proof}[Proof of \cref{lem:objective-convergence}]
    Define
    \[
        M := \adjustlimits\max_{k \in K}\sup_{\theta,\theta_k^{\crt} \in \Theta_k}
        \left|\phi_{\eta,k}\br{\theta;\theta_k^{\crt}}\right|
        \quad \text{and} \quad
        \delta_n := M\sum_{k \in K}\left|\beta_k^n - \beta_k\right|,
    \]
    where $M < \infty$ since for all $k \in K$, $\Theta_k$ is a compact set and each density $f_k$ is strictly positive and continuous. Moreover, $\delta_n < \infty$ since $0 < \beta_k^n,\beta_k \leq 1$.

    Then, for any critical type $\boldtheta^{\crt} \in \Theta_K$, using the triangle inequality we get
    \begin{align*}
        \left|\psi_\eta^n\br{q^n,\boldtheta^{\crt}} - \psi_\eta\br{q^\infty,\boldtheta^{\crt}}\right|
        &\leq \sum_{k \in K}\left|\beta_k^n - \beta_k\right|
        \E_{F_k}\sbr{\left|\phi_{\eta,k}\br{\theta;\theta_k^{\crt}}\br{q_k^n\br{\theta} - \mu_k}\right|}\\
        & \quad + \sum_{k \in K}\beta_k
        \E_{F_k}\sbr{\left|\phi_{\eta,k}\br{\theta;\theta_k^{\crt}}\br{q_k^n\br{\theta} - q_k^\infty\br{\theta}}\right|}\\
        &\leq \delta_n + M\sum_{k \in K}\beta_k\int_{\Theta_k}\left|q_k^n - q_k^\infty\right|dF_k,
    \end{align*}
    where the last inequality uses $\left|q_k^n - \mu_k\right| \leq 1$ and the definitions for $M$ and $\delta_n$. Similarly,
    \begin{align*}
        \left|\min_{\boldtheta^{\crt} \in \Theta_K}\psi_\eta^n\br{q^n,\boldtheta^{\crt}}
        - \min_{\boldtheta^{\crt} \in \Theta_K}\psi_\eta\br{q^\infty,\boldtheta^{\crt}}\right|
        &\leq \sup_{\boldtheta^{\crt} \in \Theta_K}
        \left|\psi_\eta^n\br{q^n,\boldtheta^{\crt}} - \psi_\eta\br{q^\infty,\boldtheta^{\crt}}\right|\\
        &\leq \delta_n + M\sum_{k \in K}\beta_k\int_{\Theta_k}\left|q_k^n - q_k^\infty\right|dF_k.
    \end{align*}
    Since $q_k^n \longrightarrow q_k^\infty$ $dF_k$-almost everywhere and $\left|q_k^n - q_k^\infty\right| \leq 1$, by the DCT we have
    \[
        \int_{\Theta_k}\left|q_k^n - q_k^\infty\right|dF_k \longrightarrow 0.
    \]
    Since $\delta_n \longrightarrow 0$, we get
    \[
        \left|\min_{\boldtheta^{\crt} \in \Theta_K}\psi_\eta^n\br{q^n,\boldtheta^{\crt}}
        - \min_{\boldtheta^{\crt} \in \Theta_K}\psi_\eta\br{q^\infty,\boldtheta^{\crt}}\right|
        \longrightarrow 0,
    \]
    concluding the proof.
\end{proof}

Then, by definition of the supremum in \eqref{eqn:convergence-finite-reduced-value} choose $q^n \in \qcal_{\gamma^n}$ for each $n$ such that
\[
    \Pi_\eta^n - \frac{1}{n}
    \leq \min_{\boldtheta^{\crt} \in \Theta_K}\psi_\eta^n\br{q^n,\boldtheta^{\crt}}
    \leq \Pi_\eta^n.
\]
Moreover, since \eqref{eqn:convergence-finite-reduced-value} is bounded above, we can choose a subsequence $\br{n_j}_{j \geq 1}$ along which $\Pi_\eta^{n_j} \longrightarrow \limsup_n\Pi_\eta^n$. Finally, by part~I), we pass to a further subsequence $q^{n_j}$ (keeping the same notation for ease of exposition) along which $q_k^{n_j}$ converges to $q_k^\infty$ $dF_k$-almost everywhere for all $k \in K$, for some $q^\infty \in \qcal_R$. Thus, by construction of subsequences  $\br{q_k^{n_j}}_{j \geq 1}$ and \cref{lem:objective-convergence} we get
\begin{equation}
    \label{eqn:upper-bound}
    \limsup_{n \to \infty}\Pi_\eta^n
    = \lim_{j \to \infty}\br{\min_{\boldtheta^{\crt} \in \Theta_K}
        \psi_\eta^{n_j}\br{q^{n_j},\boldtheta^{\crt}}}
    = \min_{\boldtheta^{\crt} \in \Theta_K}
    \psi_\eta\br{q^\infty,\boldtheta^{\crt}}
    \leq \Pi_\eta^\infty.
\end{equation}

\paragraph{Lower Bound on Asymptotic Optimal Value.} We construct feasible symmetric mechanisms in the finite market whose values converge to $\Pi_\eta^\infty$, thus proving the lower bound $\liminf_n\Pi_\eta^n \geq \Pi_\eta^\infty$.

Fix the allocation $q^*$ and transfers $t^*$ of the optimal limit-market mechanism characterized in \cref{thm:optimal-limit-mechanism}, and construct targeted ads $x^n$ as in the proof for part~I) in \cref{thm:finite-limit-convergence} using \eqref{eqn:convergence-targeted-ads}. Then, for all $\br{\theta,\boldtheta_{-\br{i,k}}} \in \Theta^n$, $i \in N_k$ and $k \in K$, define transfers $t^n_{i,k}$ by
\begin{equation}
    \label{eqn:convergence-expost-transfer}
    t_{i,k}^n\br{\theta,\boldtheta_{-\br{i,k}}} := \frac{1}{n}\br{
        c_{i,k}^n\br{\boldtheta_{-\br{i,k}}} t_k^*\br{\theta}
        - \br{1 - c_{i,k}^n\br{\boldtheta_{-\br{i,k}}}}
        \mu_k\btheta_k},
\end{equation}
where $c_{i,k}^n\br{\boldtheta_{-\br{i,k}}} \in \sbr{0,1}$ is defined in \eqref{eqn:convergence-delivery-factor}.

\begin{lemma}
    \label{lem:approx-transfers}
    For all $n \in \mathbb N$, the targeted ads $x^n$ defined in \eqref{eqn:convergence-targeted-ads} and transfers $t^n$ defined in \eqref{eqn:convergence-expost-transfer} constitute symmetric mechanisms satisfying \eqref{eqn:convergence-IC} and \eqref{eqn:convergence-IR}, where the normalized interim transfers satisfy, for all $k \in K$,
    \[
        t_k^n\br{\theta} \longrightarrow t_k^*\br{\theta}
        \quad dF_k\text{-almost everywhere.}
    \]
\end{lemma}

\begin{proof}[Proof of \cref{lem:approx-transfers}]

    Fix $i \in N_k$ and $k \in K$, and a type report $\boldtheta_{-\br{i,k}}$, and let
    \[
        c := c_{i,k}^n\br{\boldtheta_{-\br{i,k}}} \in \sbr{0,1}.
    \]

    First, using the envelope theorem and \eqref{eqn:IRk} for the optimal limit-market allocation and transfers $\br{q^*,t^*}$, the payoff $u_k^*\br{\theta}$ satisfies
    \begin{equation}
        \label{eqn:payoff-nonnegative-limit}
        u_k^*\br{\theta}
        := \theta\br{q_k^*\br{\theta} - \mu_k} - t_k^*\br{\theta} \geq 0.
    \end{equation}

    \paragraph{IR.} By \eqref{eqn:convergence-expost-transfer} and \eqref{eqn:convergence-profile-factorization}, the normalized payoffs $u^n_k\br{\theta}$ satisfy
    \begin{align*}
        u^n_k\br{\theta} &= n\br{\theta\br{q_{i,k}^n\br{\theta,\boldtheta_{-\br{i,k}}} - \mu_k/n}
            - t_{i,k}^n\br{\theta,\boldtheta_{-\br{i,k}}}}\\
        &= \theta\br{cq_k^*\br{\theta} - \mu_k}
        - ct_k^*\br{\theta} + \br{1 - c}\mu_k\btheta_k\\
        &= c\,u_k^*\br{\theta}
        + \br{1 - c}\mu_k\br{\btheta_k - \theta}
        \geq 0,
    \end{align*}
    where the last inequality uses \eqref{eqn:payoff-nonnegative-limit}. Hence, $\br{x^n,t^n}$ satisfies \eqref{eqn:convergence-IR}.

    \paragraph{IC.} For any deviation $\widetilde\theta \in \Theta_k$, $c$ does not depend on merchant $i$'s report and the second term in \eqref{eqn:convergence-expost-transfer} that defines transfers $t^n_{i,k}$ is independent of his report. Then
    \[
        c\sbr{
            \theta q_k^*\br{\theta} - t_k^*\br{\theta}
            - \br{\theta q_k^*\br{\widetilde\theta} - t_k^*\br{\widetilde\theta}}
        } \geq 0,
    \]
    where the inequality follows from \eqref{eqn:ICk} of the optimal limit-market mechanism and $c \geq 0$. Thus, $\br{x^n,t^n}$ satisfies \eqref{eqn:convergence-IC}.

    \paragraph{Symmetry.} Taking expectations in \eqref{eqn:convergence-expost-transfer} over the other merchants' types gives
    \begin{align*}
        nT_{i,k}^n\br{\theta}
        &= n\E_{F^n_{-\br{i,k}}}\sbr{t_{i,k}^n\br{\theta,\boldtheta_{-\br{i,k}}}}\\
        &= \E_{b^n_{i,k}}\sbr{\min\cbr{1,s_k^n\br{b^n_{i,k}}}}
        t_k^*\br{\theta}
        - \br{1 - \E_{b^n_{i,k}}\sbr{
                \min\cbr{1,s_k^n\br{b^n_{i,k}}}
        }}\mu_k\btheta_k,
    \end{align*}
    where the last equality uses \eqref{eqn:convergence-delivery-factor} and the law of iterated expectations. Since $b^n_{i,k}$ is identically distributed across $i \in N_k$, the interim expected transfers are equal within each tag, proving symmetry.

    The mechanism $\br{x^n,t^n}$ is therefore symmetric.

    \paragraph{Convergence.} By the preceding analysis, the normalized expected transfers $t_k^n\br{\theta}$ satisfy
    \begin{equation}
        \label{eqn:normalized-transfers-proof}
        t_k^n\br{\theta} = nT_{i,k}^n\br{\theta} = \E_{b^n_{i,k}}\sbr{\min\cbr{1,s_k^n\br{b^n_{i,k}}}}
        t_k^*\br{\theta}
        - \br{1 - \E_{b^n_{i,k}}\sbr{
                \min\cbr{1,s_k^n\br{b^n_{i,k}}}
        }}\mu_k\btheta_k.
    \end{equation}
    If $d_k\br{q^*} > 0$, the same argument as in the proof of \cref{prop:limit-to-finite} gives
    \[
        \E_{b^n_{i,k}}\sbr{\min\cbr{1,s_k^n\br{b^n_{i,k}}}} \longrightarrow 1.
    \]
    Hence, $t_k^n\br{\theta} \longrightarrow t_k^*\br{\theta}$ for every $\theta \in \Theta_k$. On the other hand, if $d_k\br{q^*} = 0$, then $q_k^* = 0$ $dF_k$-almost everywhere. Since $\mu_k > 0$, the worst-off type is $\htheta_k^* = \btheta_k$, hence $t_k^* = - \mu_k\btheta_k$ $dF_k$-almost everywhere. Then by \eqref{eqn:normalized-transfers-proof} it follows that $t_k^n = t_k^*$ $dF_k$-almost everywhere for every $n$, concluding the proof.
\end{proof}

By \cref{prop:limit-to-finite,lem:approx-transfers}, normalized interim allocations $q^n$ and transfers $t^n$ converge $dF_k$-almost everywhere in each tag and are uniformly bounded. Since $\beta_k^n \longrightarrow \beta_k$, by DCT we get
\begin{equation}
    \label{eqn:convergence-implemented-objective-bound}
    \begin{aligned}
        \Psi_\eta^n\br{x^n,t^n}
        &= \sum_{k \in K}\beta_k^n\E_{F_k}\sbr{
            \eta_r t_k^n\br{\theta}
            + \br{\eta_v\theta + \eta_w}\br{q_k^n\br{\theta} - \mu_k}}\\
        & \quad \longrightarrow \sum_{k \in K}\beta_k\E_{F_k}\sbr{
            \eta_r t_k^*\br{\theta}
            + \br{\eta_v\theta + \eta_w}\br{q_k^*\br{\theta} - \mu_k}}
        = \Pi_\eta^\infty.
    \end{aligned}
\end{equation}
Since $\br{x^n,t^n}$ are feasible by construction,
\[
    \liminf_n\Pi_\eta^n
    \geq \lim_n\Psi_\eta^n\br{x^n,t^n}
    = \Pi_\eta^\infty.
\]
Therefore, combining the preceding inequality with the upper bound in \eqref{eqn:upper-bound} gives $\Pi_\eta^n \longrightarrow \Pi_\eta^\infty$. Since $\Psi_\eta^n\br{x^n,t^n} \longrightarrow \Pi_\eta^\infty$ by \eqref{eqn:convergence-implemented-objective-bound}, we also obtain $\Pi_\eta^n - \Psi_\eta^n\br{x^n,t^n} \longrightarrow 0$.

\subsubsection{Convergence of Optimal Finite-Market Mechanisms}
\label{app:convergence-optimal-sequences}

\begin{proof}[Completion of part~II) of \cref{thm:finite-limit-convergence}]
    We have already proved that optimal values converge and that the constructed mechanisms are asymptotically optimal, with normalized interim allocations and transfers converging to $\br{q^*,t^*}$. It remains to prove the same allocation and transfer convergence for every sequence of symmetric optimal finite-market mechanisms.

    Fix any sequence of symmetric optimal finite-market mechanisms. For each market size $n$ and tag $k$, denote the normalized interim allocations and transfers by $\br{\widehat q_k^n,\widehat t_k^n}$. Since $\eta_r > 0$, optimality sets the payoff of the worst-off type in each tag to zero. Hence,
    \[
        \min_{\boldtheta^{\crt} \in \Theta_K}\psi_\eta^n\br{\widehat q^n,\boldtheta^{\crt}} = \Pi_\eta^n.
    \]

    Take any subsequence $\br{\widehat q^{n_\ell}}_{\ell \geq 1}$ of the allocation sequence $\br{\widehat q^n}_{n \geq 1}$. By part~I), it has a further subsequence $\br{\widehat q^{n_{\ell_j}}}_{j \geq 1}$ along which $\widehat q_k^{n_{\ell_j}}$ converges $dF_k$-almost everywhere to $\bar q_k$ for every $k$, with $\bar q \in \qcal_R$. By \cref{lem:objective-convergence} and convergence of optimal values,
    \[
        \min_{\boldtheta^{\crt} \in \Theta_K}\psi_\eta\br{\bar q,\boldtheta^{\crt}}
        = \lim_{j \longrightarrow \infty}\Pi_\eta^{n_{\ell_j}}
        = \Pi_\eta^\infty.
    \]
    Thus, essential uniqueness in \cref{thm:optimal-limit-mechanism} gives $\bar q_k = q_k^*$ $dF_k$-almost everywhere.

    Now fix an interior continuity point $\theta$ of $q_k^*$ and choose types $a < \theta < b$ where $\widehat q_k^{n_{\ell_j}}$ converges to $q_k^*$. Monotonicity gives
    \[
        q_k^*\br{a}
        \leq \liminf_j\widehat q_k^{n_{\ell_j}}\br{\theta}
        \leq \limsup_j\widehat q_k^{n_{\ell_j}}\br{\theta}
        \leq q_k^*\br{b}.
    \]
    Since $f_k > 0$, these points can be chosen arbitrarily close to $\theta$, so continuity gives $\widehat q_k^{n_{\ell_j}}\br{\theta} \longrightarrow q_k^*\br{\theta}$. The initial subsequence $\br{\widehat q^{n_\ell}}_{\ell \geq 1}$ was arbitrary, so every subsequence admits a further subsequence converging at $\theta$. Hence, the full sequence converges at $\theta$. Since $q_k^*$ has at most countably many discontinuities and $F_k$ has a density, $\widehat q_k^n \longrightarrow q_k^*$ $dF_k$-almost everywhere. This convergence, together with the envelope formula and zero payoff for the worst-off type, implies $\widehat t_k^n \longrightarrow t_k^*$ $dF_k$-almost everywhere, proving \eqref{eqn:convergence-optimal-allocations-transfers}.
\end{proof}

\section{Proofs for Efficiency and Sequential Targeting}
\label[appendix]{app:extensions}

\subsection{Proof of \texorpdfstring{\cref{lem:efficient-clicks}}{Lemma \ref*{lem:efficient-clicks}}}

We closely follow the proof of \cref{thm:optimal-limit-mechanism}. With $\eta_r = 0$, information rents no longer enter the designer's objective, and solving for the efficient allocation also pins down the critical types as the cutoff types $p_k^E$.

\paragraph{Audience Scarcity.} With $\eta_r = 0$, $\MR_{\eta,k}^{-1}\br{0} = \ltheta_k$ for every tag, so demand at zero shadow prices is one expected click for almost every merchant. The left-hand side of \cref{asm:scarcity-limit} then equals $\sum_{k \in K}\beta_k = 1$. Because $G$ has a density on $[0,1]^K$, $\max_k\xi_k < 1$ almost surely implies
\[
    \E_G\sbr{\max_{k \in K}\xi_k} < 1,
\]
hence \cref{asm:scarcity-limit} holds.

\paragraph{Efficient Objective.} Let $\phi_\eta^E\br{\theta} := \eta_v\theta + \eta_w$. Up to a constant, the efficient objective is
\[
    \psi_\eta^E\br{q}
    := \sum_{k \in K}\beta_k\E_{F_k}\sbr{\phi_\eta^E\br{\theta}q_k\br{\theta}}.
\]
For $\lambda \in \rbb_+^K$, define
\[
    \ell^E\br{\lambda}
    := \sum_{k \in K}\beta_k\E_{F_k}\sbr{\br{\phi_\eta^E\br{\theta} - \lambda_k}_+}.
\]
Then \eqref{eqn:R-Fsb} and \eqref{eqn:Ck} give the upper bound
\begin{equation}
    \label{eqn:appendix-efficient-dual-bound}
    \psi_\eta^E\br{q}
    = \lambda \cdot d\br{q}
    + \sum_{k \in K}\beta_k\E_{F_k}\sbr{\br{\phi_\eta^E\br{\theta} - \lambda_k}q_k\br{\theta}}
    \leq \sigma\br{\lambda} + \ell^E\br{\lambda}
\end{equation}
for every feasible allocation $q$. Since $\eta_v > 0$, $\phi_\eta^E$ is strictly increasing. Hence, the (essentially) unique pointwise maximizer $q_k^{E,\lambda}\br{\theta} \in [0,1]$ of $\br{\phi_\eta^E\br{\theta} - \lambda_k}q_k\br{\theta}$ is given $dF_k$-almost everywhere by
\[
    q_k^{E,\lambda}\br{\theta}
    := \mathbf{1}\cbr{\theta > p_k^E\br{\lambda_k}},
\]
where, using the inverse convention in \cref{subsec:optimal-mechanism-finite}, cutoff type $\theta = p_k^E\br{\lambda_k}$ is given by
\[
    p_k^E\br{\lambda_k}
    := \br{\phi_\eta^E}^{-1}\br{\lambda_k}
    = \min\cbr{\btheta_k,\max\cbr{\ltheta_k,\frac{\lambda_k - \eta_w}{\eta_v}}}.
\]
The aggregate demand $d_k^E\br{\lambda_k}$ by $q_k^{E,\lambda}$ in tag $k$ is given by
\[
    d_k^E\br{\lambda_k}
    := d_k\br{q^{E,\lambda}}
    = \beta_k\sbr{1 - F_k\br{p_k^E\br{\lambda_k}}}.
\]

We next characterize the minimizers of $\sigma + \ell^E$ and show that the corresponding allocations are feasible and attain the bound in \eqref{eqn:appendix-efficient-dual-bound}, which proves efficiency.

\paragraph{Market Clearing.} We now show that minimizing $\sigma + \ell^E$ yields the market-clearing conditions \eqref{eqn:market-clearing-efficiency}. Since tag $k$'s term in $\ell^E$ is zero once $\lambda_k \geq \phi_\eta^E\br{\btheta_k}$ and $\sigma$ is nondecreasing in each coordinate, minimization can be restricted to the compact set $\prod_{k \in K}\sbr{0,\phi_\eta^E\br{\btheta_k}}$. As in the proof of \cref{lem:dual-compact}, $\sigma + \ell^E$ is continuous and convex, so a minimizer exists by the Weierstrass theorem.

Let $\lambda^E$ be any minimizer, and write $p_k^E := p_k^E\br{\lambda_k^E}$. To obtain market clearing from the first-order conditions, we first show that every coordinate of $\lambda^E$ is positive. Applying the derivative argument in the proof of \cref{lem:market-clearing}\textup{(\ref{lem:market-clearing:foc})} together with \eqref{eqn:appendix-support-partial-derivative} gives
\[
    \frac{\partial\ell^E}{\partial\lambda_k}\br{\lambda}
    = - d_k^E\br{\lambda_k}
    \quad \text{and} \quad
    \frac{\partial\br{\sigma + \ell^E}}{\partial\lambda_k}\br{\lambda}
    = s_k^\lambda - d_k^E\br{\lambda_k},
\]
where the first identity holds for every $\lambda \in \rbb_+^K$ and the second for $\lambda \neq \mathbf{0}$.

First, we rule out $\lambda^E = \mathbf{0}$. Since $d_k^E\br{0} = \beta_k$, the same calculation as in the proof of \cref{lem:market-clearing}\textup{(\ref{lem:market-clearing:bounds})} gives
\[
    D_{\mathbf{1}}^+\br{\sigma + \ell^E}\br{\mathbf{0}}
    = \E_G\sbr{\max_{k \in K}\xi_k} - 1 < 0,
\]
where the inequality follows from the audience-scarcity argument above. Hence, $\mathbf{0}$ cannot minimize $\sigma + \ell^E$.

Next, suppose $\lambda_k^E = 0$ for some $k$. Then $d_k^E\br{0} = \beta_k\sbr{1 - F_k\br{\ltheta_k}} = \beta_k$. Since in addition $s_k^{\lambda^E} = 0$ by \cref{lem:support-function}\textup{(\ref{lem:support-function:positive-supply})},
\[
    \frac{\partial\br{\sigma + \ell^E}}{\partial\lambda_k}\br{\lambda^E}
    = s_k^{\lambda^E} - d_k^E\br{0}
    = - \beta_k < 0,
\]
which contradicts $\lambda^E$ being a minimizer.

Thus, $\lambda_k^E > 0$ for all $k \in K$, so convexity implies that the first-order conditions are necessary and sufficient:
\[
    s_k^{\lambda^E} = d_k^E\br{\lambda_k^E}
    \quad \text{for all }k \in K,
\]
which give the market-clearing conditions \eqref{eqn:market-clearing-efficiency}.

Lastly, $\lambda_k^E > 0$ implies $s_k^{\lambda^E} > 0$ by \cref{lem:support-function}\textup{(\ref{lem:support-function:positive-supply})}, hence $d_k^E\br{\lambda_k^E} > 0$ as well. So let $\overline\lambda_k^E := \phi_\eta^E\br{\btheta_k} = \eta_v\btheta_k + \eta_w$ for each $k \in K$. Since demand is zero whenever $\lambda_k \geq \overline\lambda_k^E$, $d_k^E\br{\lambda_k^E} > 0$ implies $p_k^E < \btheta_k$, and therefore $0 < \lambda_k^E < \overline\lambda_k^E$ for all $k \in K$.

\paragraph{Optimality.} Let $q^E := q^{E,\lambda^E}$. Since $q_k^E\br{\theta} = 1$ when $\phi_\eta^E\br{\theta} > \lambda_k^E$ and $q_k^E\br{\theta} = 0$ when $\phi_\eta^E\br{\theta} < \lambda_k^E$, we have
\[
    \br{\phi_\eta^E\br{\theta} - \lambda_k^E}q_k^E\br{\theta}
    = \br{\phi_\eta^E\br{\theta} - \lambda_k^E}_+\quad dF_k\text{-almost everywhere.}
\]
Integrating and summing across tags gives
\[
    \begin{aligned}
        \ell^E\br{\lambda^E}
        &= \sum_{k \in K}\beta_k\E_{F_k}\sbr{\br{\phi_\eta^E\br{\theta} - \lambda_k^E}_+}\\
        &= \sum_{k \in K}\beta_k\E_{F_k}\sbr{\br{\phi_\eta^E\br{\theta} - \lambda_k^E}q_k^E\br{\theta}}\\
        &= \psi_\eta^E\br{q^E} - \lambda^E \cdot d\br{q^E}.
    \end{aligned}
\]

Next, since market clearing gives $d\br{q^E} = s^{\lambda^E} \in \scal_G$, the allocation $q^E$ is implementable by \cref{lem:implementable}. Moreover, each $q_k^E$ is nondecreasing and satisfies capacity constraints. Using the preceding identity and market clearing, we obtain
\[
    \psi_\eta^E\br{q^E}
    = \ell^E\br{\lambda^E} + \lambda^E \cdot d\br{q^E}
    = \ell^E\br{\lambda^E} + \lambda^E \cdot s^{\lambda^E}
    = \ell^E\br{\lambda^E} + \sigma\br{\lambda^E},
\]
where the last equality follows from \cref{lem:support-function}\textup{(\ref{lem:support-function:supply})}. Thus, $q^E$ is feasible and attains the bound in \eqref{eqn:appendix-efficient-dual-bound}, so it is efficient. Essential uniqueness of the efficient allocation follows by the same argument as in the proof of \cref{thm:optimal-limit-mechanism}.

\paragraph{Virtual Prices.} Let $\widetilde\lambda^E$ be another minimizer of $\sigma + \ell^E$. Its allocation is also efficient, so essential uniqueness of the efficient allocation gives the same cutoff types, and hence the same demand, in every tag. Market clearing therefore gives $s^{\widetilde\lambda^E} = s^{\lambda^E}$. By \cref{lem:support-function}\textup{(\ref{lem:support-function:supply})} and the proportionality argument in \cref{app:limit-uniqueness}, $\widetilde\lambda^E = c\lambda^E$ for some $c > 0$.

Next, observe that
\[
    \sum_{k \in K}d_k\br{q^E}
    \leq \E_G\sbr{\max_{k \in K}\xi_k} < 1
    = \sum_{k \in K}\beta_k.
\]
By \cref{lem:support-function}\textup{(\ref{lem:support-function:positive-supply})}, every tag receives positive expected clicks, so some tag has $0 < d_k\br{q^E} < \beta_k$ and therefore an interior cutoff. Its unique cutoff satisfies
\[
    \lambda_k^E = \eta_vp_k^E + \eta_w = \widetilde\lambda_k^E.
\]
Since $\lambda_k^E > 0$, we obtain $c = 1$. Thus, $\sigma + \ell^E$ has a unique minimizer, proving uniqueness of the vector of efficient virtual prices.

\subsection{Proof of \texorpdfstring{\cref{prop:cm-efficient-design}}{Proposition \ref*{prop:cm-efficient-design}}}
\label{app:efficient-transfers}

By the worst-off-type characterization, $p_k^E$ is a worst-off type in tag $k$. To maximize net payments subject to incentive compatibility and individual rationality, we set its payoff to zero. The same steps as in the proof of \cref{thm:optimal-limit-mechanism}, with $p_k^S = p_k^B = p_k^E$, give

\begin{equation}
    \label{eqn:efficient-transfers}
    t_k^E\br{\theta}
    =
    \begin{cases}
        - \mu_kp_k^E & \text{if }\theta \leq p_k^E,\\
        \br{1 - \mu_k}p_k^E & \text{if }\theta > p_k^E,
    \end{cases}
\end{equation}
which, together with $q_k^E$, satisfy \eqref{eqn:ICk} and \eqref{eqn:IRk}, and also maximize expected net payments among all \eqref{eqn:ICk} and \eqref{eqn:IRk} mechanisms that implement efficient expected clicks. For each tag $k$, the net payments are
\begin{align*}
    \beta_k\E_{F_k}\sbr{t_k^E\br{\theta}}
    &= \beta_k \sbr{\br{1 - \mu_k}p_k^E \sbr{1 - F_k\br{p_k^E}}
        - \mu_kp_k^E F_k\br{p_k^E}}\\
    &= \beta_k p_k^E \sbr{1 - \mu_k - F_k\br{p_k^E}}.
\end{align*}
By summing across tags, we conclude that the customer- and merchant-efficient platform design can be achieved without an outside subsidy if and only if \eqref{eqn:efficiency-no-deficit-condition} holds.

\subsection{Proof of \texorpdfstring{\cref{prop:m-efficient-design}}{Proposition \ref*{prop:m-efficient-design}}}
\label{app:merchant-surplus-efficiency}

We show that the transfers in \eqref{eqn:efficient-transfers} yield nonnegative aggregate net payments. Let $d_k^E := \beta_k\sbr{1 - F_k\br{p_k^E}}$ and $b_k := \beta_k\mu_k$ denote efficient demand and total baseline expected clicks in tag $k$, respectively. Market clearing \eqref{eqn:market-clearing-efficiency} and \cref{lem:support-function}\textup{(\ref{lem:support-function:supply})} give $\lambda^E \cdot d^E = \sigma\br{\lambda^E}$, while consistency \eqref{eqn:G-consistency} gives $\lambda^E \cdot b \leq \sigma\br{\lambda^E}$. Hence,
\[
    \lambda^E \cdot \br{d^E - b}
    = \sigma\br{\lambda^E} - \lambda^E \cdot b \geq 0.
\]

Moreover, because $\eta_v = 1$, \cref{lem:efficient-clicks} gives
\[
    p_k^E = \min\cbr{\btheta_k,\max\cbr{\ltheta_k,\lambda_k^E}},
\]
with $p_k^E < \btheta_k$. We claim that
\[
    \br{p_k^E - \lambda_k^E}\br{d_k^E - b_k} \geq 0
    \quad \text{for all }k \in K.
\]
If $p_k^E \in \br{\ltheta_k,\btheta_k}$, then $p_k^E = \lambda_k^E$, so the product is zero. If $p_k^E = \ltheta_k$, then $p_k^E \geq \lambda_k^E$ and $d_k^E = \beta_k > \beta_k\mu_k = b_k$. Thus, the inequality holds in every case.

The maximal expected transfers generated by the efficient allocation therefore satisfy
\[
    \begin{aligned}
        \sum_{k \in K}\beta_kp_k^E\sbr{1 - F_k\br{p_k^E} - \mu_k}
        &= p^E \cdot \br{d^E - b}\\
        &= \lambda^E \cdot \br{d^E - b}
        + \br{p^E - \lambda^E} \cdot \br{d^E - b}\\
        &\geq 0,
    \end{aligned}
\]
where the inequality follows from the two preceding inequalities. Hence, the transfers in \eqref{eqn:efficient-transfers} satisfy \eqref{eqn:no-deficit}, and the efficient allocation can be implemented without an outside subsidy.

\subsection{Proof of \texorpdfstring{\cref{prop:sequential-targeting}}{Proposition \ref*{prop:sequential-targeting}}}
\label{app:sequential-targeting-value}

The proof follows that of \cref{thm:optimal-limit-mechanism}, with the targeting value $\sigma^{(2)}$ in place of $\sigma$. The screening problem, the weak duality bound, and the existence arguments are unchanged, so the optimal mechanism is characterized by virtual prices that solve
\begin{equation}
    \label{eqn:sequential-dual}
    \min_{\lambda \in \rbb_+^K}\cbr{\sigma^{(2)}\br{\lambda} + \ell\br{\lambda}}.
\end{equation}

We omit the details of these steps and only verify that $\sigma^{(2)}$ is the targeting value. Specifically, let $m_k\br{\xi}$ be the share of customers with CTRs $\xi$ assigned only an ad from tag $k$, and let $m_{kj}\br{\xi}$ be the share assigned tag $k$ first and tag $j$ after no click. These shares are measurable, nonnegative, and satisfy
\begin{equation}
    \label{eqn:sequential-shares}
    \sum_{k \in K}m_k\br{\xi} + \sum_{k \neq j}m_{kj}\br{\xi} \leq 1
    \quad dG\text{-almost everywhere}.
\end{equation}

Let $s\br{m} := \br{s_k\br{m}}_{k \in K}$ denote the vector of aggregate expected clicks, where
\[
    s_k\br{m}
    := \E_G\sbr{
        \xi_k m_k\br{\xi}
        + \sum_{j \neq k}\xi_k m_{kj}\br{\xi}
        + \sum_{j \neq k}\br{1 - \xi_j}\xi_k m_{jk}\br{\xi}
    }.
\]
The first term counts clicks on tag $k$'s ad when it is shown alone, the second when it is shown first in a pair, and the third when it is shown second. For shadow prices $\lambda \in \rbb_+^K$, we have
\begin{equation}
    \label{eqn:sequential-targeting-bound}
    \lambda \cdot s\br{m}
    = \E_G\sbr{
        \sum_k\lambda_k\xi_km_k\br{\xi}
        + \sum_{k \neq j}\sbr{\lambda_k\xi_k + \br{1 - \xi_k}\lambda_j\xi_j}m_{kj}\br{\xi}
    }
    \leq \sigma^{(2)}\br{\lambda},
\end{equation}
with equality whenever each customer is assigned an ordered pair attaining the maximum in \eqref{eqn:sequential-support}. Thus, $\sigma^{(2)}\br{\lambda}$ is the targeting value under sequential targeting, and the screening and weak-duality arguments in the proof of \cref{thm:optimal-limit-mechanism} apply with $\sigma^{(2)}$ in place of $\sigma$.

\end{document}

%% file: figures/price-discrimination-classic.tex
\begin{tikzpicture}[
    x=1cm,
    y=1cm,
    line cap=round,
    line join=round
]
    \path[use as bounding box]
        (-0.25,-0.20) rectangle (9.05,5.85);

    \draw[
        -{Triangle[scale=0.8]},
        line width=1.2pt
    ]
    (0.35,0.80) -- (8.85,0.80);

    \draw[
        -{Triangle[scale=0.8]},
        line width=1.2pt
    ]
    (0.75,0.38) -- (0.75,5.35);

    \node[
        font=\large
    ] at (4.75,6)
    {Tag-$k$ Market};

    \draw[
        densely dashed,
        line width=0.9pt
    ]
    (0.75,2.82) -- (4.60,2.82);

    \draw[
        densely dashed,
        line width=0.9pt
    ]
    (4.60,0.80) -- (4.60,2.82);

    \draw[
        line width=1.3pt
    ]
    (1.20,1.95) -- (8.20,3.75);

    \node[
        font=\large,
        anchor=south east
    ] at (8.15,3.8)
    {$\mathrm{MR}_k$};


    \node[
        font=\large,
        align=center
    ] at (6.40,1.40)
    {Buy};

    \node[
        font=\large,
        anchor=north east,
        inner sep=1pt
    ] at (0.64,0.42)
    {$\underline{\theta}_k$};

    \node[
        font=\large,
        anchor=north,
        inner sep=1pt
    ] at (4.60,0.5)
    {$p_k^B$};

    \draw[
        line width=1.2pt
    ]
    (8.20,0.58) -- (8.20,1.02);

    \node[
        font=\large,
        anchor=north,
        inner sep=1pt
    ] at (8.20,0.42)
    {$\overline{\theta}_k$};
\end{tikzpicture}

%% file: figures/price-discrimination-platform.tex
\begin{tikzpicture}[
    x=1cm,
    y=1cm,
    line cap=round,
    line join=round
]
    \path[use as bounding box]
        (-0.25,-0.20) rectangle (9.05,5.85);

    \draw[
        -{Triangle[scale=0.8]},
        line width=1.2pt
    ]
    (0.35,0.80) -- (8.85,0.80);

    \draw[
        -{Triangle[scale=0.8]},
        line width=1.2pt
    ]
    (0.75,0.38) -- (0.75,5.35);

    \node[
        font=\large
    ] at (4.75,6)
    {Tag-$k$ Market};

    \draw[
        densely dashed,
        line width=0.9pt
    ]
    (0.75,3.20) -- (6.06,3.20);

    \draw[
        densely dashed,
        line width=0.9pt
    ]
    (3.16,0.80) -- (3.16,3.20);

    \draw[
        densely dashed,
        line width=0.9pt
    ]
    (6.06,0.80) -- (6.06,3.20);

    \draw[
        line width=1.3pt
    ]
    (1.00,2.45) -- (7.90,4.85);

    \node[
        font=\large,
        anchor=south east
    ] at (7.85,4.8)
    {$\mathrm{MC}_{\eta,k}$};

    \draw[
        line width=1.3pt
    ]
    (1.20,1.95) -- (8.20,3.75);

    \node[
        font=\large,
        anchor=south east
    ] at (8.15,3.7)
    {$\mathrm{MR}_{\eta,k}$};

    \node[
        font=\large,
        anchor=east
    ] at (0.57,3.20)
    {$\lambda_k$};

    \node[
        font=\large,
        align=center
    ] at (1.95,1.40)
    {Sell};

    \node[
        font=\large,
        align=center
    ] at (4.60,1.40)
    {Exchange};

    \node[
        font=\large,
        align=center
    ] at (7.20,1.40)
    {Buy};

    \node[
        font=\large,
        anchor=north east,
        inner sep=1pt
    ] at (0.64,0.42)
    {$\underline{\theta}_k$};

    \node[
        font=\large,
        anchor=north,
        inner sep=1pt
    ] at (3.16,0.5)
    {$p_k^S$};

    \node[
        font=\large,
        anchor=north,
        inner sep=1pt
    ] at (6.06,0.5)
    {$p_k^B$};

    \draw[
        line width=1.2pt
    ]
    (8.20,0.58) -- (8.20,1.02);

    \node[
        font=\large,
        anchor=north,
        inner sep=1pt
    ] at (8.20,0.42)
    {$\overline{\theta}_k$};
\end{tikzpicture}

%% file: figures/scoring-negative-intercept.tex
\begin{tikzpicture}[
    x=10cm,y=7cm,
    >=Latex,
    axis/.style={-{Triangle[scale=1]},very thick},
    guide/.style={densely dotted,gray!65,line width=.9pt},
    guideblack/.style={densely dotted,black,line width=.9pt},
    main/.style={black,very thick,line cap=round,line join=round},
    aux/.style={gray!60,very thick,line cap=round,line join=round},
    counterfactual/.style={gray!35,very thick,dash pattern=on 4pt off 5pt,line cap=round,line join=round},
    vdash/.style={gray!60,very thick,dash pattern=on 4pt off 5pt},
    labgray/.style={gray!55},
    every node/.style={font=\Large}
]
    \def\xS{0.30}
    \def\xhat{0.50}
    \def\xB{0.72}
    \def\yBzero{-0.13}
    \def\ySw{0.28}
    \def\yz{0.50}
    \def\yhatBottom{0.30}
    \def\yTopMid{0.70}
    \def\yBone{0.75}
    \def\ySone{0.97}

    \draw[axis] (-0.10,0) -- (1.14,0) node[below right] {$\theta_i$};
    \draw[axis] (0,-0.23) -- (0,1.12) node[above left] {$h_i$};

    \draw[guide] (0,\ySone) -- (1,\ySone) -- (1,0);
    \draw[guide] (0,\yBone) -- (1,\yBone);

    \draw[guideblack] (0,\yz) -- (\xS,\yz);
    \draw[guideblack] (\xS,0) -- (\xS,\yz);
    \draw[guideblack] (\xB,0) -- (\xB,\yz);
    \draw[guideblack] (\xhat,0) -- (\xhat,\yhatBottom);
    \draw[vdash] (\xhat,\yhatBottom) -- (\xhat,\yTopMid);

    \draw[counterfactual]
        (0,\yBzero)
        .. controls (0.12,-0.03) and (0.18,0.09) .. (0.29,0.12)
        .. controls (0.41,0.16) and (0.44,0.25) .. (\xhat,\yhatBottom);
    \draw[aux]
        (\xhat,\yhatBottom)
        .. controls (0.56,0.39) and (0.55,0.40) .. (0.60,0.45)
        .. controls (0.67,0.50) and (0.68,0.47) .. (\xB,\yz)
        .. controls (0.79,0.54) and (0.84,0.60) .. (0.90,0.66)
        .. controls (0.94,0.70) and (0.98,0.73) .. (1.00,\yBone);

    \draw[aux]
        (0,\ySw)
        .. controls (0.12,0.30) and (0.18,0.35) .. (\xS,\yz)
        .. controls (0.36,0.58) and (0.41,0.55) .. (0.47,0.66)
        .. controls (0.49,0.70) and (0.50,0.70) .. (\xhat,\yTopMid);
    \draw[counterfactual]
        (\xhat,\yTopMid)
        .. controls (0.57,0.76) and (0.62,0.74) .. (0.69,0.79)
        .. controls (0.82,0.84) and (0.89,0.93) .. (1.00,\ySone);

    \draw[main]
        (0,\ySw)
        .. controls (0.12,0.30) and (0.18,0.35) .. (\xS,\yz);
    \draw[main] (\xS,\yz) -- (\xB,\yz);
    \draw[main]
        (\xB,\yz)
        .. controls (0.79,0.54) and (0.84,0.60) .. (0.90,0.66)
        .. controls (0.94,0.70) and (0.98,0.73) .. (1.00,\yBone);

    \node[left] at (0,\ySone) {$\phi^S_{\eta,i}(1)$};
    \node[left] at (0,\yBone) {$1$};
    \node[left] at (0,\yz) {$z_i$};
    \node[left] at (0,\ySw) {$\eta_w$};
    \node[left] at (0,\yBzero) {$\phi^B_{\eta,i}(0)$};

    \node[below] at (-0.03,0) {$0$};
    \node[below] at (\xS,0) {$\theta^S_i$};
    \node[below] at (\xhat,0) {$\hat{\theta}_i(z_i)$};
    \node[below] at (\xB,0) {$\theta^B_i$};
    \node[below] at (1,0) {$1$};

    \node[labgray,rotate=23] at (0.16,0.13) {$\phi^B_{\eta,i}(\theta_i)$};
    \node[labgray,rotate=25] at (0.7,0.9) {$\phi^S_{\eta,i}(\theta_i)$};
    \node[rotate=15] at (0.67,0.63) {$\overline\phi_{\eta,i}(\theta_i;z_i)$};
\end{tikzpicture}

%% file: figures/scoring-positive-intercept.tex
\begin{tikzpicture}[
    x=10cm,y=7cm,
    >=Latex,
    axis/.style={-{Triangle[scale=1]},very thick},
    guide/.style={densely dotted,gray!65,line width=.9pt},
    guideblack/.style={densely dotted,black,line width=.9pt},
    main/.style={black,very thick,line cap=round,line join=round},
    aux/.style={gray!60,very thick,line cap=round,line join=round},
    counterfactual/.style={gray!35,very thick,dash pattern=on 4pt off 5pt,line cap=round,line join=round},
    vdash/.style={gray!60,very thick,dash pattern=on 4pt off 5pt},
    labgray/.style={gray!55},
    every node/.style={font=\Large}
]
    \def\xhat{0.22}
    \def\xB{0.50}
    \def\yBzero{0.10}
    \def\ySw{0.37}
    \def\yz{0.27}
    \def\yhatBottom{0.20}
    \def\yhatTop{0.45}
    \def\yBone{0.68}
    \def\ySone{0.95}

    \draw[axis] (-0.10,0) -- (1.14,0) node[below right] {$\theta_i$};
    \draw[axis] (0,-0.20) -- (0,1.12) node[above left] {$h_i$};

    \draw[guide] (0,\ySone) -- (1,\ySone) -- (1,0);
    \draw[guide] (0,\yBone) -- (1,\yBone);

    \draw[guideblack] (\xB,0) -- (\xB,\yz);
    \draw[guideblack] (\xhat,0) -- (\xhat,\yhatBottom);
    \draw[vdash] (\xhat,\yhatBottom) -- (\xhat,\yhatTop);

    \draw[counterfactual]
        (0,\yBzero)
        .. controls (0.06,0.12) and (0.13,0.14) .. (0.19,0.18)
        .. controls (0.2,0.18) and (0.2,0.19) .. (\xhat,\yhatBottom);
    \draw[aux]
        (\xhat,\yhatBottom)
        .. controls (0.29,0.225) and (0.34,0.245) .. (0.40,0.248)
        .. controls (0.44,0.250) and (0.47,0.255) .. (0.485,0.262)
        .. controls (0.492,0.265) and (0.497,0.268) .. (\xB,\yz);

    \draw[aux]
        (0,\ySw)
        .. controls (0.08,0.38) and (0.15,0.39) .. (\xhat,\yhatTop);
    \draw[counterfactual]
        (\xhat,\yhatTop)
        .. controls (0.34,0.54) and (0.48,0.60) .. (0.66,0.68)
        .. controls (0.80,0.76) and (0.92,0.92) .. (1.00,\ySone);

    \draw[main] (0,\yz) -- (\xB,\yz);
    \draw[main]
        (\xB,\yz)
        .. controls (0.60,0.31) and (0.67,0.36) .. (0.71,0.45)
        .. controls (0.74,0.53) and (0.80,0.55) .. (0.87,0.59)
        .. controls (0.94,0.63) and (0.98,0.66) .. (1.00,\yBone);

    \node[left] at (0,\ySone) {$\phi^S_{\eta,i}(1)$};
    \node[left] at (0,\yBone) {$1$};
    \node[left] at (0,\ySw) {$\eta_w$};
    \node[left] at (0,\yz) {$z_i$};
    \node[left] at (0,\yBzero) {$\phi^B_{\eta,i}(0)$};

    \node[below] at (-0.03,0) {$0$};
    \node[below] at (\xhat,0) {$\hat{\theta}_i(z_i)$};
    \node[below] at (\xB,0) {$\theta^B_i$};
    \node[below] at (1,0) {$1$};

    \node[labgray,rotate=18] at (0.36,0.09) {$\phi^B_{\eta,i}(\theta_i)$};
    \node[labgray,rotate=26] at (0.7,0.82) {$\phi^S_{\eta,i}(\theta_i)$};
    \node[rotate=15] at (0.50,0.4) {$\overline\phi_{\eta,i}(\theta_i;z_i)$};
\end{tikzpicture}

%% file: figures/ironing-equal-audiences.tex
\begin{tikzpicture}
    \begin{axis}[
        width=\linewidth,
        height=0.76\linewidth,
        xmin=0.5,
        xmax=1.015,
        ymin=0.05,
        ymax=0.65,
        axis lines=left,
        axis line style={-{Triangle[scale=0.8]},thick},
        xlabel={\(\varepsilon\)},
        xtick={0.5,1},
        xticklabels={\(0.5\),\(1\)},
        ytick={0.1,0.2,0.3,0.4,0.5,0.6},
        yticklabels={
            \(0.1\),\(0.2\),\(0.3\),
            \(0.4\),\(0.5\),\(0.6\)
        },
        tick style={black,thick},
        legend style={
            draw=none,
            fill=none,
            font=\small,
            at={(0.5,0.4)},
            anchor=north west
        }
    ]
        \addplot[
            black,
            densely dotted,
            thick,
            forget plot
        ] coordinates {
            (0.50,0.50)
            (1.00,0.50)
        };

        \addplot[
            black,
            very thick,
            smooth
        ] coordinates {
            (0.50,0.1570)
            (0.55,0.2222)
            (0.60,0.2805)
            (0.65,0.3319)
            (0.70,0.3765)
            (0.75,0.4143)
            (0.80,0.4451)
            (0.85,0.4691)
            (0.90,0.4863)
            (0.95,0.4966)
            (0.99,0.4999)
            (1.00,0.5000)
        };

        \addlegendentry{\(z_1(\varepsilon)=z_2(\varepsilon)\)}

    \addplot[
        only marks,
        mark=o,
        mark size=2.8pt,
        mark options={
            draw=black,
            fill=none,
            line width=0.9pt
        },
        forget plot
    ] coordinates {
        (1.00,0.50)
    };

    \node[
        font=\scriptsize,
        align=center,
        anchor=south east,
        inner sep=1pt
    ] (pdlabel) at (axis cs:0.950,0.550)
        {Partnership\\Dissolution};
    \draw[
        thick,
        -{Triangle[scale=0.7]},
        shorten <=1.5pt
    ]
    (pdlabel.south east)
    --
    (axis cs:0.992,0.508);
    \end{axis}
\end{tikzpicture}

%% file: figures/ironing-unequal-audiences.tex
\begin{tikzpicture}
    \begin{axis}[
        width=\linewidth,
        height=0.76\linewidth,
        xmin=0.5,
        xmax=1.015,
        ymin=0.05,
        ymax=0.65,
        axis lines=left,
        axis line style={-{Triangle[scale=0.8]},thick},
        xlabel={\(\varepsilon\)},
        xtick={0.5,1},
        xticklabels={\(0.5\),\(1\)},
        ytick={0.1,0.2,0.3,0.4,0.5,0.6},
        yticklabels={
            \(0.1\),\(0.2\),\(0.3\),
            \(0.4\),\(0.5\),\(0.6\)
        },
        tick style={black,thick},
        legend style={
            draw=none,
            fill=none,
            font=\small,
            at={(0.7,0.4)},
            anchor=north west
        }
    ]
        \addplot[
            black,
            densely dotted,
            line width=0.9pt,
            forget plot
        ] coordinates {
            (0.50,0.60)
            (1.00,0.60)
        };

        \addplot[
            black,
            densely dotted,
            line width=0.9pt,
            forget plot
        ] coordinates {
            (0.50,0.40)
            (1.00,0.40)
        };

        \addplot[
            black,
            line width=1.3pt,
            smooth,
            tension=0.6,
            line cap=round,
            line join=round,
            mark=none
        ] coordinates {
            (0.50,0.4230)
            (0.55,0.4640)
            (0.60,0.4987)
            (0.65,0.5275)
            (0.70,0.5507)
            (0.75,0.5687)
            (0.80,0.5821)
            (0.85,0.5913)
            (0.90,0.5968)
            (0.95,0.5994)
            (0.99,0.6000)
            (1.00,0.6000)
        };
        \addlegendentry{\(z_1(\varepsilon)\)}

        \addplot[
            black,
            line width=1.3pt,
            dash pattern=on 4.5pt off 2.8pt,
            smooth,
            tension=0.6,
            line cap=round,
            line join=round,
            mark=none
        ] coordinates {
            (0.50,0.1000)
            (0.55,0.1695)
            (0.60,0.2283)
            (0.65,0.2770)
            (0.70,0.3164)
            (0.75,0.3470)
            (0.80,0.3696)
            (0.85,0.3852)
            (0.90,0.3946)
            (0.95,0.3991)
            (0.99,0.4000)
            (1.00,0.4000)
        };
        \addlegendentry{\(z_2(\varepsilon)\)}

        \addplot[
            only marks,
            mark=o,
            mark size=2.8pt,
            mark options={
                draw=black,
                fill=none,
                line width=0.9pt
            },
            forget plot
        ] coordinates {
            (1.00,0.60)
            (1.00,0.40)
        };

        \node[
            font=\scriptsize,
            align=center,
            anchor=east,
            inner sep=1pt
        ] (pdlabel) at (axis cs:0.900,0.500)
        {Partnership\\Dissolution};

        \draw[
            thick,
            -{Triangle[scale=0.7]},
            shorten <=1.5pt
        ]
        (pdlabel.north east)
        --
        (axis cs:0.992,0.592);

        \draw[
            thick,
            -{Triangle[scale=0.7]},
            shorten <=1.5pt
        ]
        (pdlabel.south east)
        --
        (axis cs:0.992,0.408);
    \end{axis}
\end{tikzpicture}

%% file: figures/feasible-set-uniform-ctrs.tex
\begin{tikzpicture}[font=\large]
    \begin{axis}[
        width=0.4\textwidth,
        height={0.4\textwidth * 0.54/0.58},
        scale only axis,
        xmin=0,
        xmax=0.58,
        ymin=0,
        ymax=0.54,
        axis lines=left,
        axis line style={-{Triangle[scale=0.8]}, thick},
        ticks=none,
        clip=false,
        xlabel={$s_1$},
        ylabel={$s_2$},
        xlabel style={
            at={(axis description cs:1,0)},
            anchor=north east,
            xshift=18pt
        },
        ylabel style={
            at={(axis description cs:0,1)},
            anchor=north east,
            rotate=-90,
            yshift=22pt
        }
    ]
        \addplot[
            draw=none,
            fill=gray!15,
            domain=0:1/3,
            samples=100
        ]
        {1/2-3*x^2/2}
        \closedcycle;

        \addplot[
            draw=none,
            fill=gray!15,
            domain=1/3:0.5,
            samples=100
        ]
        {sqrt((1-2*x)/3)}
        \closedcycle;

        \addplot[
            black,
            very thick,
            domain=0:1/3,
            samples=100
        ]
        {1/2-3*x^2/2};

        \addplot[
            black,
            very thick,
            domain=1/3:0.5,
            samples=100
        ]
        {sqrt((1-2*x)/3)};

        \addplot[
            black,
            dashed,
            thick,
            domain=0.13:0.55,
            samples=2
        ]
        {2/3-x};

        \addplot[
            black,
            only marks,
            mark=*,
            mark size=2.4pt,
            mark options={draw=black,fill=black}
        ]
        coordinates {
            (0,0.5)
            (1/3,1/3)
            (0.5,0)
            (3/15,3/15)
        };

        \node at (axis cs:0.05,0.05)
            {$\scal_G$};

        \node[
            anchor=north east,
            inner sep=1.5pt
        ]
        at (axis cs:0.43,0.54)
            {$\lambda\cdot s=\sigma\br{\lambda}$};

        \node[
            anchor=north east
        ]
        at (axis cs:{3/15},{4/16})
            {$s^0$};

        \node[
            anchor=north east
        ]
        at (axis cs:0.32,0.36)
            {$s^\lambda$};

        \draw[->,thick]
            (axis cs:{1/3},{1/3})
            --
            (axis cs:0.42,0.42);

        \node[
            anchor=north west
        ]
        at (axis cs:0.42,0.415)
            {$\lambda$};

        \node[anchor=north east]
            at (axis cs:0,0) {$0$};

        \node[anchor=north, font=\small]
            at (axis cs:0.5,0) {$\mathbb{E}[\xi_1]$};
        
        \node[anchor=east, font=\small]
            at (axis cs:0,0.5) {$\mathbb{E}[\xi_2]$};

        \draw[semithick]
            (axis cs:{1/3+0.015},{1/3+0.015})
            --
            (axis cs:{1/3+0.030},{1/3})
            --
            (axis cs:{1/3+0.015},{1/3-0.015});
    \end{axis}
\end{tikzpicture}